\documentclass[12pt]{article}

\usepackage{lmodern}
\usepackage[T1]{fontenc}
\usepackage[utf8]{inputenc}
\usepackage{graphicx}
\usepackage{amssymb}
\usepackage{times}
\usepackage{amsmath}
\usepackage[bookmarks=false]{hyperref} 

\newcounter{lastnote}

\usepackage[colorinlistoftodos,prependcaption,textsize=footnotesize]{todonotes}

\title{\textbf{Introduction to Loop Quantum Gravity} \\
       Rovelli's lectures on LQG} 

\author
{Transcribed by Pietropaolo Frisoni \\
}

 \date{}

\begin{document} 

\baselineskip18pt

\maketitle 

\begin{abstract}
These notes are a transcript of Carlo Rovelli's lectures on Loop Quantum Gravity, given in Marseille in 2018, which (at present) can be entirely found on YouTube. I transcribed them in LaTeX in early 2020 as an exercise to get ready for my Ph.D. in LQG at Western University. This transcript is meant to be a (hopefully helpful) integration for the video version. I reported the order of the topics and the chronological structure exactly as presented by Rovelli throughout the course, primarily to facilitate the comparison. Each Section corresponds to a different Lecture. The parts written in \textit{textit} are my additions. Sometimes in the text, I report references, which specify precisely the minute and the second of the corresponding video on YouTube, to very short historical digressions or excursus made during the lectures by Rovelli that I have not explicitly transcribed in these notes. Where appropriate, I took some figures from the book "Covariant Loop Quantum Gravity - An elementary introduction to Quantum Gravity and Spinfoam Theory" by Carlo Rovelli and Francesca Vidotto, to which I always refer by the term "the book" in the following. For what concerns the equations, where possible, I tried to write down the "correct" versions present within the book. Finally, I thank Carlo Rovelli himself for reviewing these notes. I apologize in advance for any errors, and I wish everyone a lot of fun! 
\end{abstract}

\tableofcontents

\newpage

\section{The empirical basis of quantum gravity}
\label{Lecture_1_empirical_basis_LQG}
LQG is a theory that attempts to describe the quantum behavior of gravitational fields. This means, according to Einstein, to describe spacetime. We have (yet) no solid empirical support for this theory. \textit{Rovelli says he hopes to see some tests applicable to this theory before dying}. The theory can be described using different formalisms (canonical and covariant). It is beginning to be approximately thirty years old. The goal of the theory is not to have a theory of everything but "just" to describe the quantum behavior of gravity (or spacetime). We need such a theory since we still do not know what happens in certain universe regions, such as objects that fall into black holes.

\medskip

Concerning the experimental discoveries, what have we recently learned from the Universe?
\begin{enumerate}
\item A few months ago, a merging between neutron stars was observed. It was measured that the gravitational and electromagnetic signals arrived practically simultaneously (with a relative accuracy of $10^{- 15}$), so the signals had traveled at the same speed. In other words, the gravitational waves travel at the speed of light: $\frac{v_g}{c} \simeq 1$. Many quantum theories of gravity based on GR modifications predicted this was not the case. Therefore, they have been (in this sense) refuted.
\item Another observation that surprised everyone a few years ago was that supersymmetry was not observed at LHC. However, some physicists were sure to find it already at LEP, even before such experiments were performed at CERN. This is relevant as it doesn't directly rule out any theory of QG, but it constrains the direction of research that somehow led to the supersymmetry prediction. \textit{Rovelli starts discussing some aspects concerning the philosophy of science}.
\item Writing a consistent theory of quantum gravity, from a mathematical point of view, is much easier if one admits the possibility of breaking the Lorentz invariance at the Planck scale. Nowadays, there are limits to this violation up to the order $10^{-6}$. \textit{See lecture for further considerations regarding this point}.
\end{enumerate}

\medskip

What do we know about the word? The best theories that we have to describe nature are:
\begin{itemize}
\item \textbf{Quantum theory}
\item \textbf{General relativity}
\item \textbf{Standard model}
\item \textbf{Thermodynamics}
\end{itemize}
The shocking feature regarding these theories is that nobody initially took them seriously. A few years ago \textit{(when Rovelli was a student in Bologna)}, nobody took the Standard model seriously, and everybody expected it to be wrong and violated very soon. \textit{Rovelli tells the class an exciting anecdote with Carlo Rubbia}. People like Fermi, Landau, and Einstein used to say that thermodynamics is the theory that we know much better than others. We will see why and how this is relevant for QG.
Furthermore \textit{(when Rovelli was a kid)}, nobody believed in black holes: it was a belief that these entities were only a mathematical solution of Einstein's equations. The discovery of gravitational waves also confirmed General Relativity in the strong field regime. General Relativity certainly goes wrong when $\hbar \neq 0$, and Quantum Mechanics has to be adapted to consider it in such a regime. Finally, there is a subtle question regarding the "Quantum mechanics" concept: if we mean the "core" of the theory, we have no reason to believe that the latter doesn't work in QG, but if we are referring to the Schroedinger equation, then the latter does not fit with GR, since there is no an external time parameter $t$ in Einstein's theory. Dirac was the first who started trying to adapt QM to GR. Regarding the Standard Model, we don't know what happens when energies are higher than $15$ TeV: maybe there are new particles, black holes, etc. 

\medskip

To build a consistent QG theory, it seems reasonable to try to base it on theories that we consider correct. The Standard model is not so relevant for QG because the energies for its regime are much higher, but GR and QM turn out to be the core. LQG does not consider strings, supersymmetry, and other stuff: it just tries to assemble things that work well. \textit{Rovelli shares interesting considerations regarding the philosophy of science, see the lecture for details}. In the following lecture, we discuss what we learned about "space" and "time" and what we mean by these definitions. 
\section{Space}
\label{Lecture_2_space}
This lecture is important because a lot of confusion in QG arises from mixing the space and time concepts. When introducing LQG, it is easier to treat first the space and, later on, time. There is a useful review \textit{(written by Rovelli)} called "Space and Time in LQG" on arXiv, which explains these concepts in every detail. When we say "space," we mean different things, creating confusion. 
\subsection{Concepts of Space}
\label{subsec_Lec_2_concepts_of_space}
The first thing we mean when we say space, or \textbf{relational space}, is the following. Space is the thing we refer to when we are talking about "where" and when we are referring to the position of an object to other objects, for example: "I was in Iceland." So we answer the question "where" essentially by saying "next to what." Aristotle's concept of location defined an object as the inner boundary of other objects, for example: "The air surrounds me." Therefore, we can say that \underline{"space" is a relation between things}. Another intellectual who made a remarkable disquisition regarding space was Descartes, with his concept of "contiguity."
\begin{equation}
\text{Relational space} \longleftrightarrow \text{Where} \longleftrightarrow \text{Space is a relation between things} \ .
\end{equation}
Two corollaries descend from this concept of space:
\begin{enumerate}
\item \textbf{There's no vacuum, i.e., no empty space}. If we remove all the things, there is no space since the latter is defined as a relation between things. This seems in contradiction with the modern way of thinking. We'll see why in a while.
\item \textbf{Motion is "relative" to other objects}. Motion is defined as the change of location, and location is "what is next to me." Therefore, during a motion, we are changing the contiguity of an object to other objects.
\end{enumerate}
This is a straightforward way to think about space, but we immediately realize it is not what we find in physics textbooks. 

\medskip

Books usually adopt the second notion of space, which is surprisingly modern, namely the \textbf{Newtonian space}. It was first described by Newton, explicitly disputing Descartes. Newton pointed out that, with the previous relational notion of space, it is impossible to do physics. We emphasize that Newton makes a distinction, in the "Philosophiae Naturalis Principia Mathematica," between notions of space: the "common" (or "apparent") and the "true" space. The first one is Aristotle's and Descartes' notions of space. It is fundamental to remember that Newton does not say that the first notion is wrong or doesn't exist, but only that there's another one. His revolution is to establish that \underline{space is an entity}, so it exists even if there are no things. Besides, it has a (geometrical) structure. We use a modern description for it, specifying that space is a three-dimensional vector space endowed with a metric, so it turns out to be a metric space (with Euclidean geometry):
$$
\mathbb{E}^3 = \left( \mathbb{R}^3, \hspace{1mm} \text{distance} \right) \ .
$$
Therefore, space is a container of the world, and things move within it. We note that now things move with respect to this space and not one another. Newton is postulating a new entity, without which it would be impossible to describe the correct physics. He also postulates an entity called "time" (\textit{this is described in lecture \ref{sec:Lecture_10_time}}). An object is described by a function position, parameterized by this time $t$, which establishes its position in the space: $\vec{x}(t)$. The equation of motion are well known, but the important aspect is that position and time are thought of in relational space. 
\begin{figure}[h]                       
\begin{center}                          
\includegraphics[width=8cm]{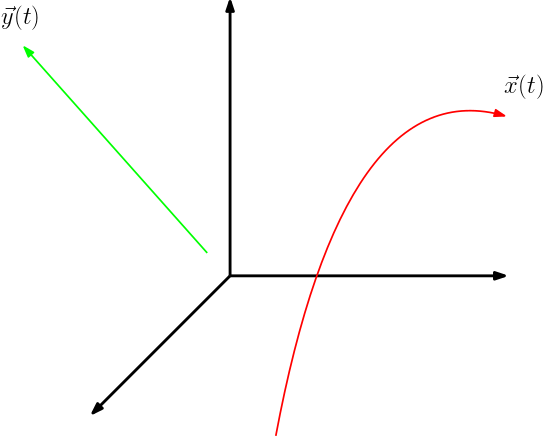}
\caption{The metric space introduced by Newton is a container of the world in which the objects move with respect to the space itself and not one with respect to the other. }
\end{center}
\end{figure}
Acceleration, for Newton, is also described with respect to space and, again, not with respect to other things. \textit{Rovelli briefly describes the bucket experiment at minute 15:21}. Newton says, openly and clearly, that we only have direct access to the common space, while we need to use mathematics and subtle experiment to show the existence of the true one. This second version defines acceleration through the Newtonian equations. Leibniz and others rejected Newton's idea of space. We will use the true space concept since it is used by most people today. There are two more steps to mention. The first one is easy, and the second is crucial. 

\medskip

The first step was made by Einstein in 1905 with the special relativity theory, which still relies on Newton's logic of space (and time). The only difference is that instead of having space and time separated or, in mathematical terms, $\mathbb{E} \times \mathbb{R}$, he unifies the latter in the Minkowski space $\mathcal{M}$. 

\medskip

The second crucial step was also made by Einstein in 1915 with general relativity, which is a discovery that Newton's space and time entities are real and exist. Still, they are not fixed and static: they identify with the gravitational field. So, the \textbf{Einstein's spacetime} is a \underline{dynamical entity}. It is a field very much like $\vec{E}$ and $\vec{B}$: not by chance Einstein was very obsessed with Maxwell's theory, and he strongly believed in it, so these discoveries were computed under the spell of Maxwell and others. This field is what is going to be the quantum gravitational field in QG. It is granular; it interacts in measurements, exhibits superposition, etc. 

\medskip

The important thing to establish is that Newton's notion of space completely disappears in QG, but the old notion of relational space remains! As we underlined, Newton just added a new structure. So we expect the "relational space" concept to remain in QG, not Newtonian. As the last comment to close the story, we mention that Newton took his new concept of time from Democritus. \textit{Rovelli shares other very interesting considerations, see the end of the lecture for details}.
\section{$SU(2)$ group}
\label{Lecture_SU(2)_group}
We now describe the math of $SU(2)$ because it is fundamental in LQG. 
\subsection{Mathematics of $SU(2)$}
$SU(2)$ is the universal covering of $SO(3)$, which is the group of three-dimensional rotations, but now we are interested in its exclusively mathematical description. We mention the fact that it can be identified with quaternions. $SU(2)$ is an abstract group, namely a set with a topology, a product, an inverse element, and an identity. However, it is much easier to think of it as a group of matrices since these completely realize its structure. If we indicate with $ h $ a generic matrix of $ SU(2)$, then we can write:
\begin{equation}
h = \begin{pmatrix}
a & c \\
b & d 
\end{pmatrix} \ ,
\end{equation}
where the matrix elements are complex numbers, $h$ has the following properties:
\begin{equation}
h^{\dagger} = h^{-1} \ ,
\label{Unitary_SU(2)_matrix}
\end{equation}
\begin{equation}
det(h) = 1 \ .
\label{Unitary_det_SU(2)_matrix}
\end{equation}
The symbol $ \dagger $ makes the transpose and then the complex conjugate. Equation \eqref{Unitary_det_SU(2)_matrix} consists of 2 equations because it implies that the real part of the determinant must be equal to one and the imaginary part equal to zero. If we explicitly calculate the equation \eqref{Unitary_SU(2)_matrix}, this results in the following equality:
\begin{equation}
\begin{pmatrix}
\bar{a} & \bar{b} \\ 
\bar{c} & \bar{d}
\end{pmatrix}
= 
\begin{pmatrix}
d & -c \\ 
-b & a
\end{pmatrix}
\end{equation}
So we can express $h$ in the following way:
\begin{equation}
h = \begin{pmatrix}
a & - \bar{b} \\
b & \bar{a}
\end{pmatrix}
\end{equation}
Therefore, there are only two complex numbers that seem to define $h$, but there is also relation \eqref{Unitary_det_SU(2)_matrix}, that applied to the previous expression gives:
\begin{equation}
\left| a \right|^2 + \left| b \right|^2 = 1 \ .
\label{eq:Topology_SU(2)}
\end{equation}
So eventually, there are only three independent numbers out of 4 real numbers (two real numbers specify a complex number). Therefore the group is three-dimensional. If we think of eq. \eqref{eq:Topology_SU(2)} as a real equation and write $ a = x + iy $, $ b = z + ir $, then we get:
\begin{equation}
x^2 + y^2 + z^2 + r^2 = 1 \ ,
\end{equation}
which is the four-dimensional sphere, i.e., in $ \mathbb{R}^4$. Therefore, the $SU(2)$ topology is the simply connected one of $S_{3}$. From this point of view, we immediately see a "natural" measure on $SU(2)$, which corresponds to the Euclidean measure in 4 dimensions, that induces a metric on the sphere itself. It is nothing but the Haar measure in $ SU(2)$. We indicate this measure as $\int dh$. We can therefore integrate group functions with such a $SU(2)$ invariant measure. We can consider functions $\psi(h)$ of $SU(2)$, that is, from $SU(2)$ to complex numbers, that depend on $ h $ and are restricted on the region of the four-dimensional sphere: $ \psi: SU(2) \rightarrow \mathbb{C}$. We can define a scalar product on the functions' space of $ SU (2) $ in the standard way:
\begin{equation}
\left\langle \psi, \phi \right\rangle \equiv \int \bar{\psi}(h)\phi(h) dh \ .
\end{equation}
We can then consider the Hilbert space of functions on $SU(2)$:
\begin{equation}
\mathbb{H} = L_{2}\left[ SU(2) \right] \ .
\end{equation}
In practice, this is the set of functions on $SU(2)$ whose integration does not diverge. That is, which are finite according to the Haar measure. This Hilbert space is the "bread and butter" of the LQG. We will soon write a discrete basis on this Hilbert space. We recall that it is possible to write $\psi (h)$ in Dirac notation as the projection of the abstract element of Hilbert space on the continuous basis given by the elements of the group:
\begin{equation}
\psi(h) = \langle h | \psi \rangle \ .
\end{equation}
As already pointed out, we will soon write other bases of $L_{2}\left[SU (2) \right] $ that are relevant in LQG.

\medskip

Everyone knows that $ SU (2) $ has to do with $ SO (3) $, the three-dimensional group of rotations containing three-dimensional special orthogonal real matrices. A map from $SU(2)$ to $SO(3)$ preserves the group's structure. Namely, the product is sent into the product. But it is not a $1 \rightarrow 1$ isomorphism, actually is $2 \rightarrow 1$, that is only a local isomorphism. We will write this map explicitly in a moment. Let's write these matrices with some indexes:
$
h^A_{\hspace{2mm} B}
$
where $A,B = 0,1$ for the $SU(2)$ elements, while the $SO(3)$ matrices are written as
$
R^i_{\hspace{1mm} j} 
$
and $i,j = 0,1,2$. Therefore we can write the map as follows:
\begin{equation}
h^A_{\hspace{2mm} B} \rightarrow R^i_{\hspace{1mm} j} = h^A_{\hspace{2mm} B}h^C_{\hspace{2mm} D} \sigma^i_{AC}\sigma^{BD}_j \ .
\label{eq:SU(2)_SO(3)_map}
\end{equation}
We used the Pauli matrices:
\begin{equation}
\sigma^i_{AB} = 
\left\lbrace 
\begin{pmatrix}
0 & 1 \\
1 & 0 
\end{pmatrix},
\begin{pmatrix}
0 & -i \\
i & 0 
\end{pmatrix},
\begin{pmatrix}
1 & 0 \\
0 & -1 
\end{pmatrix} 
\right\rbrace \ ,
\end{equation}
where $i=1,2,3$ is the index that indicates the matrix. So one $2 \times 2$ complex matrix is sent into a matrix, which turns out to be real and orthogonal. When discussing the representation, we will specify what is going on from a more abstract point of view. This cannot be a $1 \rightarrow 1$ correspondence since, when we take $-h$ in the right side of \eqref{eq:SU(2)_SO(3)_map}, the $R^i_{\hspace{1mm} j}$ is the same, being the latter a quadratic form. Suppose we imagine $SU(2)$ as a full sphere. In that case, we can think of the whole sphere as the $SU(2)$ elements and of the points on each semi-sphere as $SO(3)$ elements: in this way, it is very intuitive to understand how opposite $SU(2)$ elements have the very same $SO(3)$ projection.  
\begin{figure}[h]                       
\centering                         
\includegraphics[width=6cm]{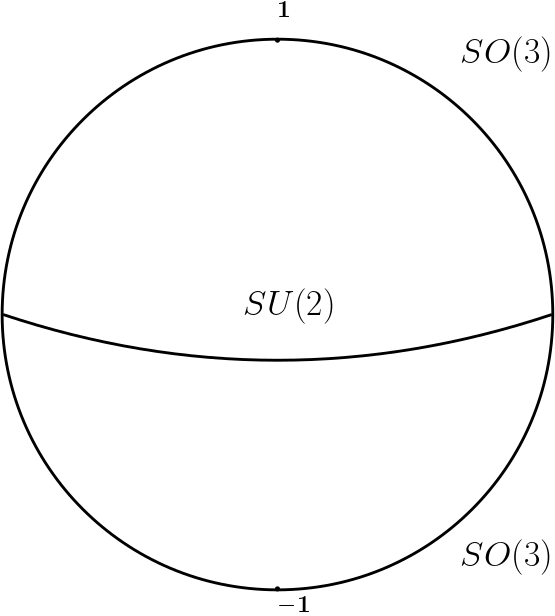}
\caption{3D representation of the $S_3$ sphere, namely the $SU(2)$ topology. Both the identity $\mathbf{1}$ and the anti-identity element $\mathbf{-1}$ of $SU(2)$ are mapped into the identity element of $SO(3)$. The sphere as a hole is $SU(2)$, while the upper and the lower parts are $SO(3)$.}
\label{S33D}
\end{figure}
We can also consider poles as the groups' identity and anti-identity elements. They also coincide from a geometrical point of view, according to the "sphere visualization." \textit{Rovelli explains how to imagine an $S_3$ sphere, see minute $17:40$ of the lecture for details}.
\begin{figure}[h]                       
\begin{center}                      
\includegraphics[width=14cm]{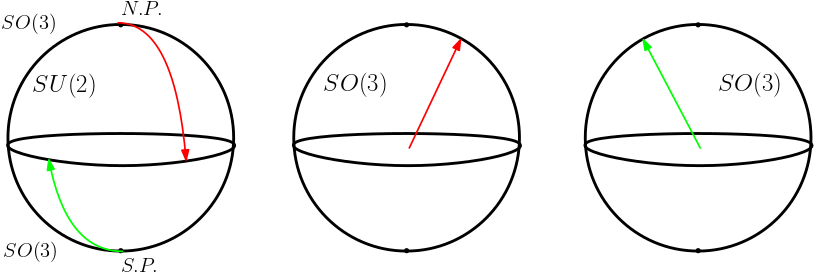}
\caption{On the left is reported the very same 3D representation of the $S_3$ sphere as in Fig. (\ref{S33D}). At the same time, on the right, there are 3D representations of the semi-spheres which constitute its boundaries, namely the $SO(3)$ topology.}
\end{center}
\end{figure}
It is easy to show that every matrix $h$ of $SU(2)$ can be written as: 
\begin{equation}
h = e^{i \alpha^i \sigma_i} \ ,
\label{eq:Exponential_form_group}
\end{equation}
where $\sigma_i$ are the Pauli matrices, while $\alpha^i$ are just three \underline{real} numbers between $0$ and $2 \pi$. If we define a length $|\alpha| = \sqrt{\alpha_1^2 + \alpha_2^2 + \alpha_3^2} \equiv \alpha$ and a direction $\frac{\alpha^i}{|\alpha|} \equiv n^i$, it is an exercise to show that we can write the previous expression as:
\begin{equation}
h = e^{i \alpha^i \sigma_i} = \cos \alpha + i \left( n^i \sigma_i \sin \alpha \right) \ ,
\label{Equazione di Eulero-Pauli Lecture 3}
\end{equation}
where the multiplication by the identity element $\mathbf{1}$ of $SU(2)$ in the right side is implied. Therefore the elements $h$ of the group can be obtained as the exponential of linear combinations of terms $\alpha^i \sigma_i$, which constitute the algebra $\mathbf{su}(2)$ of $SU(2)$.

\medskip

Algebra's group is a linear space, which can be imagined as a tangent plane to the group elements. We see that linear combinations of Pauli matrices form the algebra: $\alpha^1 \sigma_1 + \alpha^2 \sigma_2 + \alpha^3 \sigma_3$, so it consists in $2 \times 2$ matrices, which are hermitian (or anti-hermitian if we multiply the linear combination for $i$, as in the exponential argument in eq. \eqref{eq:Exponential_form_group}). Therefore, the algebra is a vector space with a bracket; we know that $\sigma_1 \sigma_2 = i \sigma_3$, etc.
The usual normalized basis of $SU(2)$ is conventionally not constituted by the $\sigma_i$, but actually $J_i \equiv - \frac{i}{2} \sigma_i$, which are chosen to have the nice property:
\begin{equation}
\left[J_i,J_j\right] = \epsilon_{ij}^{\hspace{2mm}k}J_k \ .
\end{equation}

This is the algebra's structure of $SU(2)$, namely the structure of $\mathbf{su}(2)$. The form \eqref{Equazione di Eulero-Pauli Lecture 3} is a reasonable choice of coordinates of $SU(2)$, but Euler introduced a new choice for rotations (i.e., Euler angles). Mathematica uses these coordinates \textit{(see the minute $25:34$ for the latter or, equivalently, formula (1.42) of the book)}. Of course, this new coordinatization modifies the Haar measure $\int dh$. \textit{See formula (1.44) of the book for details}. 
\subsection{Left Invariant operator}
We have just described the bare Hilbert space of the theory, which is $L_2 \left[ SU(2) \right]$. Let's introduce a key object corresponding to our primary operator since it will play the same role as the momentum operator for one particle's Hilbert space in quantum mechanics. We have functions $\psi(h)$ defined on the $SU(2)$ sphere, and we now want to take derivatives, but these are derivatives on a curved space. We have "natural derivatives," which correspond to the existence of this algebra. Within the algebra, every element corresponds to a direction. Therefore, we can act as follows to define derivatives. Let us define a vector operator (so it has an index) $L^i$, acting on the space $SU(2)$ (i.e., on the functions $\psi(h)$ which are defined in $SU(2)$) as follows:
\begin{equation}
L^i \psi(h) \equiv (-i) \lim_{t \rightarrow 0} \frac{1}{t} \left[\psi(h e^{t J_i}) -  \psi(h)\right] \ .
\label{Left invariant operator Lecture 1}
\end{equation} 
It is called \textbf{left invariant operator} because it is invariant under multiplication for the group elements on the left. It would be possible to define the "right" version of the latter as:
\begin{equation}
R^i \psi(h) \equiv (-i) \lim_{t \rightarrow 0} \frac{1}{t} \left[\psi(e^{t J_i}h) -  \psi(h)\right] \ .
\end{equation}
Anyway, we are interested only in the left version. The three operators \eqref{Left invariant operator Lecture 1} will play a significant role in quantum theory. These are three different operators: $L^i = (L^1, L^2, L^3)\equiv (L^x, L^y, L^z)$, which can be imagined as directions on the sphere defined in every point.
\begin{figure}[h]
\begin{center} 
\includegraphics[width=7cm]{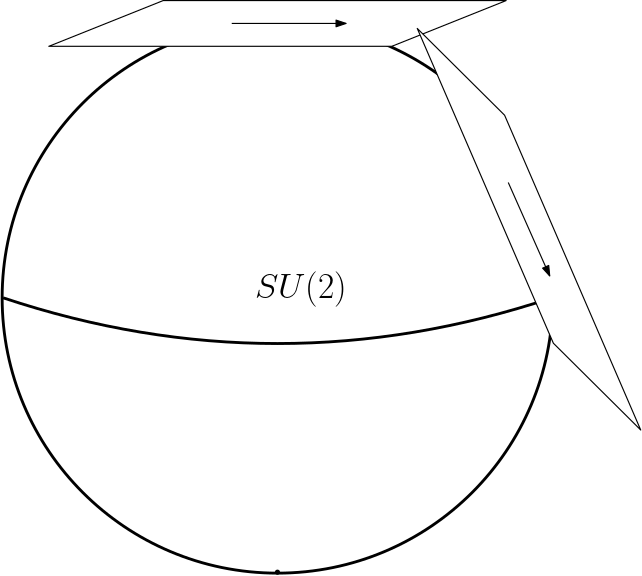}
\caption{An element in the algebra is a direction on $SU(2)$. It can be imagined as an arrow on a tangent plane to the group, defined on every point of the latter (namely, for each element $h$).}
\end{center}
\end{figure}
This is the angular momentum operator. If we take the commutator of two of them, we recover the algebra of $SU(2)$:
\begin{equation}
\left[ L^i, L^j \right] = \epsilon^{ij}_{\hspace{2mm}k}L^k \ .
\end{equation}
We will also use the notation $L^i \equiv \vec{L}$ for the latter.
\subsection{Representation theory}
Representation theory is a fundamental subject since non-relativistic elementary quantum mechanics because it plays a role whenever there is something symmetric in QM. Furthermore, it gives us a natural basis in the Hilbert space.
Representation consists of a vector space $V$ with a set of matrices $D$ acting on the vector space itself, which we indicate as $(V, D)$. Representation is a map from the group to the space of these matrices that preserves the structure of the group itself:
$
G \rightarrow \left\lbrace D \right\rbrace \ .
$
It is unitary if we have a Hilbert structure (basically the scalar product) on $V$ and the matrices are unitary, i.e., $V$ is a Hilbert space, therefore in the following, we always refer to it as $V \equiv \mathcal{H}$. This space is irreducible if there are no invariant sub-spaces when we act with the group. Two representations are inequivalent if there is no unitary mapping from one another. Namely, there is no isomorphism between them. It is possible to list all the unitary inequivalent representations (mathematicians have written beautiful papers). 

\medskip

In particular, for $SU(2)$, these are labeled according to one number $j$ which, for conventions, has values $j = 0,\frac{1}{2},1,\frac{3}{2},...$ and it is called \textbf{spin}. The representation's matrices are functions of the $SU(2)$ elements $h$. There is a family for each $j$ with indices $m,n$: $D^j_{mn}(h)$. What we learn from lists written by mathematicians is that the representation $j$ (thus the Hilbert space $\mathbb{H}^j$, as this is a unitary representation) has dimension $2j+1$: $dim(\mathbb{H}^j) = 2j+1$, so the indices $m,n$ take $(2j+1) \times (2j+1)$ values. They are traditionally taken to be $m,n = -j,-j+1,....j-1,j$. These indices label different bases on $\mathbb{H}^j$, almost always considered the eigenvalues of $L^z$, i.e., the basis in which $L^z$ is diagonal. To compute $D^j_{mn}(h)$ by knowing $h$ explicitly, we can use Mathematica. There is a specific function called "WignerD," which does this. By writing: $WignerD[\{j,m,n \},\psi, \theta, \phi]$ where $\psi, \theta, \phi$ is the Euler's angles, we can visualize the explicit expressions. $D^j_{m,n}(h)$ is a very key object, especially because of a theorem called the "Peter-Weyl theorem," which concerns the harmonic analysis of compact groups such as $SU(2)$. \textit{Rovelli discusses an example not copied here. See minute $38:30$ for details}. 
\subsection{Peter-Weyl theorem}
Peter-Weyl theorem states that $D^j_{mn}(h)$, seen as an element of $L_{2}\left[ SU(2) \right]$, has a finite norm and the set of the latter form an orthogonal basis of $L_{2}\left[ SU(2) \right]$:
\begin{equation}
\int dh \bar{D}^i_{mn}(h) D^j_{m'n'}(h) = \delta^{ij} \delta_{mm'} \delta_{nn'} \frac{1}{2j+1} \ .
\end{equation}
We notice that the norm is not $1$ since the basis is not orthonormal but just orthogonal. We can think at the object $D^j_{mn}(h)$ as:
\begin{equation}
D^j_{mn}(h) = \langle h | j,m,n \rangle \ ,
\end{equation}
i.e., as the projection of the \textbf{discrete} basis elements on the \textbf{continuous} elements of the group. The situation is similar to quantum mechanics, in which $e^{ik\alpha} = \langle \alpha | k \rangle$ is the state $k$ (which are the discrete values of the momentum of a particle in a finite space) in the representation $\alpha$, that represents the change of basis between the continuous basis $\alpha$ and the discrete one $k$.
\section{Intertwiners' space}
\label{Lecture_4_intertwiners_space}
We can think differently about the Peter-Weyl theorem by expressing the same concept but more abstractly, which is the following. Since there is a basis labeled by $j$ of the Hilbert space $L_{2}\left[ SU(2) \right]$, we take all the vectors which are labeled by $j$. This is a subspace with finite dimensions because $m,n$ have a finite range so that we can write:
\begin{equation}
L_{2}\left[ SU(2) \right] = \bigoplus_{j = 0}^{\infty} \left( \mathbb{H}^j \otimes \mathbb{H}^j \right) \ .
\label{eq:Peter_Weyl_decomp_SU(2)} 
\end{equation}
The dimension of the functional Hilbert space $L_{2}\left[ SU(2) \right]$ is infinite and, for each $j$, we have the space $\mathbb{H}^j \otimes \mathbb{H}^j$ of all the matrices $D^j_{mn}(h)$, so the latter has dimensions $(2j+1) \times (2j+1)$. It is constituted by all the linear maps from a finite-dimensional Hilbert space $\mathcal{H}^j$ to itself. Therefore, it is represented by a tensor product. We will have soon several $L_{2}\left[ SU(2) \right]$ attached one another. In this Hilbert space $L_{2}\left[ SU(2) \right]$, we have the left-invariant vector field operator $\vec{L}$ previously introduced. Out of these $L^i$, we can construct the fundamental \textbf{Casimir operator}: 
\begin{equation}
C \equiv \vec{L} \cdot \vec{L} = (L^1)^2 + (L^2)^2 + (L^3)^2 \ .
\end{equation}
It is like the Laplacian but on the $SU(2)$ curved manifold. $C$ turns out to be $SU(2)$ invariant, while $L^i$ has an index, so the latter rotates under $SU(2)$ transformations and, therefore, is not invariant. The eigenspaces of $C$ must transform among themselves under the representations of $SU(2)$, and this is exactly what is going on in \eqref{eq:Peter_Weyl_decomp_SU(2)}: we are breaking up the $L_{2}\left[ SU(2) \right]$ space in eigenspaces, corresponding to different representations of $SU(2)$, which transform among themselves. Because of the Schur Lemma, $C$ must be a multiple of the identity in each of these $j$ subspaces. In fact, when $C$ acts on the subspace labeled by $j$, it sends the latter into itself multiplied by the number $j(j+1)$:
\begin{equation}
C D^j_{mn}(h) =  j(j+1)\, D^j_{mn}(h) \ .
\label{eq:Casimir_operator_action}
\end{equation}   
In other words, the basis $|j,m,n \rangle$ diagonalizes $C$, with eigenvalues $j(j+1)$. 
\subsection{Examples of $SU(2)$ representations}
We now give more pieces of information on $SU(2)$ to do things more explicitly. 
\begin{itemize}
\item $j = 0 \hspace{5mm} $ \textbf{Trivial representation}  \\
This trivial one-dimensional representation is defined on $\mathbb{C}$. All group elements are sent into $1$ by the $C$ operator.
\item $j = \frac{1}{2} \hspace{5mm} $ \textbf{Fundamental representation}  \\
This is the fundamental representation, the dimension of $\mathbb{H}^{j}$ is $2$ and the group is $\mathbb{H}^{\frac{1}{2}} = \mathbb{C}^2$. The elements here are couples of complex numbers which are called \textbf{spinors}:
\begin{equation}
z^A \equiv
\begin{pmatrix}
z^0 \\
z^1
\end{pmatrix}  \in \mathbb{C}^2 \ .
\end{equation}
This is actually what a spinor is, i.e., a vector in the Hilbert space of the fundamental representation of $SU(2)$. It turns out that $\mathbb{C}^2$ is also the fundamental representation of $SL(2,\mathbb{C})$, so spinors are also the mathematical objects that transform under the fundamental representation of $SL(2,\mathbb{C})$. This is a subtle argument because $\mathbb{C}^2$ carries both the fundamental representations of $SU(2)$ and $SL(2,\mathbb{C})$ \textit{(this aspect is treated in lecture \ref{sec:Lecture_15_Unitary_rep_SL(2,C)})}.
\begin{figure}[h]
\begin{center}
\includegraphics[width=3cm]{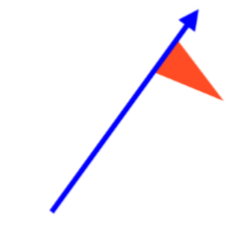}
\caption{Penrose representation of a spinor: a vector with a flag. \textit{Fig. (8.4) of the book}.}
\end{center}
\end{figure}
What is $D^{\frac{1}{2}}(h)$ ? It is very simple because it is just $h$, and the indices $m,n$ are the same things with respect to $A,B$:
\begin{equation}
D^{\frac{1}{2}}_{AB}(h) = h_{AB} \ .
\end{equation}
This is a Hilbert space, and the representation is unitary, so there is a scalar product, which is preserved by the matrix product:
\begin{equation}
\langle z | w \rangle \equiv \bar{z}^A w^B \delta_{AB} \ .
\label{Prodotto scalare in SU(2) Lecture 4}
\end{equation}
We must not confuse it with the invariant product on $\mathbb{C}^2$, given the fact that if we take the tensor product of the Hilbert space with itself, we can decompose the latter according to: 
\begin{equation}
\mathbb{H}^{\frac{1}{2}} \otimes \mathbb{H}^{\frac{1}{2}} = \mathbb{H}^0 \oplus \mathbb{H}^1 \ .
\label{Prodotto tensoriale di C2 per se stesso Lecture 4}
\end{equation}
Namely, decomposing the tensor product's space into orthogonal subspaces: an invariant part and a spin 1 part. So there should be a map from $\mathbb{H}^{\frac{1}{2}} \bigotimes \mathbb{H}^{\frac{1}{2}}$ to $ \mathbb{H}^0$, which is the following:
\begin{equation}
\left( z,w  \right) \equiv z^A w^B \epsilon_{AB} \ ,
\label{Mappa nel prodotto tensoriale Lecture 4}
\end{equation}
where 
$\epsilon_{AB} = \begin{pmatrix}
0 & 1 \\
-1 & 0
\end{pmatrix} $ \ . 
People typically confuse these two things, i.e., the scalar product and the $SU(2)$ invariant map. Note that \eqref{Mappa nel prodotto tensoriale Lecture 4} lives in $\mathbb{H}^0$, so it is a number, while objects living in $\mathbb{H}^1$ will be written within the next example. We will also treat \textit{(this is again done in lecture \ref{sec:Lecture_15_Unitary_rep_SL(2,C)})} the Lorentz group $SL(2,\mathbb{C})$, which is not compact because of boosts. A theorem states that unitary representations of non-compact groups are infinite dimensional, therefore the above representation $\mathbb{C}^2$ for $SL(2,\mathbb{C})$ \underline{is not unitary} and, thus, there is no $SL(2,\mathbb{C})$ invariant scalar product! 
Let's say it more clearly. Let's act with an $SU(2)$ matrix on $z^A$. The result conserves the scalar product \eqref{Prodotto scalare in SU(2) Lecture 4} because the matrix is unitary, but if we act on $z^A$ with an $SL(2,\mathbb{C})$ matrix, the same scalar product is not conserved since the matrix is not unitary (it only has determinant equal to 1). Instead, the map \eqref{Mappa nel prodotto tensoriale Lecture 4} is both $SU(2)$ and $SL(2,\mathbb{C})$ invariant \textit{(see minute 14:45 for a fast proof)}.
\item $j=1  \hspace{5mm} $ \textbf{Adjoint representation} \\
This is a three-dimensional representation. It is known that, for a given element of $SU(2)$, we can write an $SO(3)$ matrix that rotates vectors in the Euclidean $\mathbb{R}^3$ space. These vectors are usually written in the canonical $(i,j,k)$ basis. This creates confusion since this base is not the $m,n$ basis! \textit{Rovelli quickly mentions the correspondence at minute 17:25}. We have seen that it is possible to decompose the tensor product of the Hilbert space with itself as in eq. \eqref{Prodotto tensoriale di C2 per se stesso Lecture 4}, and we are dealing precisely with $\mathbb{H}^1$, so let's split the couples $z^A w^B$ in two parts:
\begin{equation}
z^A w^B \rightarrow 
\begin{cases} z^A w^B \epsilon_{AB}, & \mbox{this is a number (spin 0 part)} \\ 
z^A w^B \sigma_{AB}^i \equiv v^i, & \mbox{this is a vector (spin 1 part)} 
\end{cases}
\end{equation}
In the first line, we used the same expression of \eqref{Mappa nel prodotto tensoriale Lecture 4}, while in the second one, we employed the Pauli matrices to construct a vector from two spinors. This is why sometimes mathematicians say that spinors are the "square root" of vectors.
\end{itemize} 
A general fact is worth mentioning: the spin $j$ representation vectors can always be written as $z^{A_1...A_{2j}}$, a completely symmetric expression. The space of all these symmetric objects transforms into itself, so the latter form an irreducible representation. 

\textit{Historical note: Rovelli mentions that in constructing LQG, at first everybody worked  with the $j=\frac12$ representation only (each $j=\frac12$ giving rise to a ``loop" state), before realizing that working with different types of representations (``spin networks") was more convenient.}
\subsection{Intertwiner}
Let's say two more topics about these tensor representations. If we have two particles with spins $j_1$ and $j_2$, we know that they "live" in two linear spaces $\mathbb{H}^{j_1}$ and $\mathbb{H}^{j_2}$. The tensor product between them can be decomposed according to the following remarkable formula, which establishes the basic angular momentum decomposition:
\begin{equation}
\mathbb{H}^{j_1} \otimes \mathbb{H}^{j_2} = \bigoplus_{j_3 = |j_1 - j_2|}^{|j_1 + j_2|} \mathbb{H}^{j_3} \ .
\end{equation}
Conditions $|j_1 - j_2| \leq j_3 \leq |j_1 + j_2| $ are called the "Clebsch-Gordon conditions" between the three spins. So, every time we have three spins satisfying the Clebsch-Gordon conditions and the request that $j_1 + j_2 + j_3$ must be even, it turns out that every spin is in the  decomposition of the other two or, equivalently, $\mathbb{H}^{j_1} \bigotimes \mathbb{H}^{j_2} \bigotimes \mathbb{H}^{j_3}$ contains $\mathbb{H}^0$. 
\begin{figure}[h]
\begin{center}
\includegraphics[width=9cm]{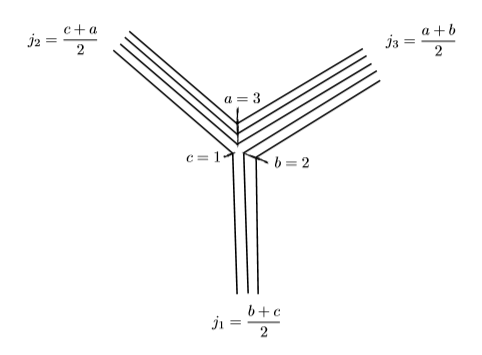}
\caption{Elementary recoupling. \textit{Fig (1.8) of the book}.}
\end{center}
\end{figure}
To say that $\mathbb{H}^{j_1} \bigotimes \mathbb{H}^{j_2} \bigotimes \mathbb{H}^{j_3}$ contains $\mathbb{H}^0$ means that inside this space there is something invariant. Clebsch-Gordon conditions are equivalent to the fact that there exist three non-negative integers $a,b,c$ such that:  
\begin{align*}
2 j_1 & = b+c \\
2 j_2 & = c+a \\
2 j_3 & = a+b \ .
\end{align*}
The generic element of $\mathbb{H}^{j_1} \bigotimes \mathbb{H}^{j_2} \bigotimes \mathbb{H}^{j_3}$ is $z^{A_1...A_{2j_1}, B_1...B_{2j_2}, C_1...C_{2j_3}}$ and the only way to map this (in a number) in a way which is $SU(2)$ and $SL(2,\mathbb{C})$ invariant, is to use a combination of $\epsilon$ since the latter are the only invariant tensors. The number of $\epsilon$ required depends on the spins, and graphically, these invariant tensors are represented by bars connecting the lines, namely the representation's indices. This procedure is unique, up to multiplication by a number. Let's do this explicitly but not in this cumbersome base. Rather, we choose the $m,n$ basis that diagonalizes by definition the left-invariant operator \eqref{Left invariant operator Lecture 1}. We have just seen that in the space $\mathbb{H}^{j_1} \bigotimes \mathbb{H}^{j_2} \bigotimes \mathbb{H}^{j_3}$ if the previous conditions on the spins are satisfied, there is an object $i^{m_1,m_2,m_3}$ which is invariant. Explicitly this means:
\begin{equation}
D^{j_1 m_1}_{\hspace{7mm} n_1}(h)D^{j_2 m_2}_{\hspace{7mm} n_2}(h) D^{j_3 m_3}_{\hspace{7mm} n_3}(h)i^{n_1 n_2 n_3} = i^{m_1m_2m_3} \ .
\label{Invarianza Intertwinters Lecture 4}
\end{equation}
\textit{It corresponds to equation (7.87) of the book in the three-dimensional case}. Note that the argument $h$ is the same in all the Wigner matrices. This element $i^{m_1m_2m_3}$ is called an  \textbf{intertwiner}: it will be fundamental in the following. It is called also \textbf{$3j$ Wigner symbol} and, sometimes, it is written as follows:
\begin{equation} 
\begin{pmatrix}
j_1 & j_2 & j_3 \\
m_1 & m_2 & m_3
\end{pmatrix}
\equiv 
i^{m_1m_2m_3} \ .
\end{equation}
\textit{Please notice that the concept of intertwiner is interchanged here and in the following with the 3j and 4j Wigner symbols.}
Mathematica gives us the explicit forms of the latter (of course, the $j$'s must respect the Clebsch-Gordon conditions). If we take, for example, $j_1 = j_2 = j_3 = 1$ (so we have three vectors) and we use the standard base $(i,j,k)$ instead of $(m_1,m_2,m_3)$, the invariant $\epsilon$ tensor has the usual form $\epsilon_{ijk}$. Therefore:
\begin{equation} 
\begin{pmatrix}
1 & 1 & 1 \\
i & j & k
\end{pmatrix}
\equiv 
\epsilon_{ijk} \ .
\end{equation}
So, if we have three vectors $v^i,w^j,z^k \in \mathbb{R}^3$ the only invariant we can build is the mixed product $v^i w^j z^k \epsilon_{ijk} \equiv (\vec{v} \times \vec{w}) \cdot \vec{z}$. This is the explicit form of the map $\mathbb{H}^{1} \bigotimes \mathbb{H}^{1} \bigotimes \mathbb{H}^{1} \rightarrow \mathbb{H}^0$, namely the expression of the invariant part. 
\subsection{Intertwiners' space}
The last topic for this lecture regards the general notion of intertwiner and, in particular, its use among four representations, which will play an extremely important role in QG. So we consider the space $ \mathbb{H}_{j_1} \bigotimes \mathbb{H}_{j_2} \bigotimes \mathbb{H}_{j_3} \bigotimes \mathbb{H}_{j_4} $ and we define "intertwiners," as before, the elements which live in the invariant part: $ i^{m_1m_2m_3m_4} \in \text{Inv} \left( \mathbb{H}_{j_1} \bigotimes \mathbb{H}_{j_2} \bigotimes \mathbb{H}_{j_3} \bigotimes \mathbb{H}_{j_4} \right)$. They satisfy equation \eqref{Invarianza Intertwinters Lecture 4} but with $4$ matrices instead of $3$. We are talking about intertwiners (instead of a single one) because there is just one possible intertwiner when we have three matrices. When dealing with $4$ matrices, there are generally more choices. There is a finite-dimensional linear space of them, called \textbf{intertwiners space}. The simplest way to "intertwine" four representations is to act as follows since we know how to intertwine three. In order to build an intertwiner $i^{m_1m_2m_3m_4}$ we can use the fact that $i^{m_1m_2m_3}$ is unique:
\begin{equation}
i^{m_1m_2m_3m_4}_k = i^{m_1m_2m} g_{mm'} i^{m' m_2m_3} \ .
\label{Costruzione 4-Intertwiner Lecture 1}
\end{equation}
We glued two indices on the right side by using matrices $g_{mm'}$, where both indices belong to a representation $k$: $g_{mm'} \in \mathcal{H}^k \otimes \mathcal{H}^k$. \textit{For those who are interested, $m' = -m$ and $g_{mm'} = (-1)^{k-m}$ where $k$ is the spin labeling the virtual spin and $i^{m_1m_2m_3m_4}_k$ is a 4jm Wigner symbol}. This representation can be chosen arbitrarily, of course, ascertained that Clebsch-Gordon's conditions are satisfied on both sides. Therefore we have a certain number of representations that we can use to build $i^{m_1m_2m_3m_4}_k$ (this is why we inserted the subscript $k$ explicitly). Each linear combination of these turns out to be an intertwiner. \textit{The previous equation has a very intuitive graphical interpretation, which I do not report here. There's a huge literature on $SU(2)$ graphical calculus for those interested.} 
%
%
%
%
For example, if we have $j_1 = j_2 = j_3 = j_4 = \frac{1}{2}$, the possible values of $k$ are 0 and 1, and the intertwiner is, in general, a linear combination of them. So the intertwiners' space is finite-dimensional, and it is possible to work with it, even if this is laborious. It is denoted with $\mathbb{H}_{j_1j_2j_3j_4}$, and we indicate its basis as $| k \rangle$. The possible values constitute this basis the index $k$ can take, namely the representations in which we split the intertwiners. 

\medskip

We finished with the math, and from now on, we will treat the theory itself. \textit{Rovelli eventually answers some questions}.
\section{LQG kinematics}
\label{Lecture_5_Kinematics}
We will describe the LQG theory as it turns out to be written nowadays, without deriving it from the classical limit, even if there will be the occasion for discussing the connection to the latter. We divide the discussion into two steps: kinematics and dynamics. In this lecture, the treated topic is the first one. 

\medskip

In general, the structure of a quantum theory consists of three elements: $(\mathcal{H},\mathcal{A}, W)$, which are: 
\begin{itemize}
\item a Hilbert space $\mathcal{H}$
\item an algebra of observables $\mathcal{A}$
\item something giving the transition amplitudes $W$
\end{itemize}
We might think of the correspondence with the quantum mechanics of a particle, in which we have $\mathcal{H}$ constituted by the space of the wavefunctions of a particle, $\mathcal{A}$ by the position and momentum operators, and $W$ by the Hamiltonian (that encodes information regarding how things change in time). Instead of the Hamiltonian, we might have the path integral or just the transition amplitude. In QED, we have $\mathcal{H}$ constituted by the Fock space, $\mathcal{A}$ given by the $a$ and $a^{\dagger}$ operators  while $W$ is represented by the Hamiltonian or the action of the Poincarè group on the Fock space, which implements evolution in time. This is the general structure of a quantum theory. 
\subsection{LQG kinematics}
What we call kinematics is constituted by the couple $(\mathcal{H},\mathcal{A})$, while $W$ encodes the dynamics. In kinematics, the important part is the algebra, while the Hilbert space is an auxiliary structure: the core of the thing relies on $\mathcal{A}$. We might think that when Heisenberg, Jordan, and others introduced the matrices' algebra of QM, there was no Hilbert space at all, so the latter is not as important as the algebra itself. Nevertheless, since Schroedinger, Dirac, Von Neumann, etc., it is much easier to work with a concrete algebra realized in a Hilbert space instead of an abstract formulation without it, so we will follow that, even if it would be more intuitive to first start from the abstract algebra. 

\medskip

What is physically the algebra $\mathcal{A}$, i.e., the core of the theory? It is simply the structure that describes variables. The physical system is described by variables (for example, a particle is described by its position, velocity, momentum, energy, angular momentum, etc.), both in classical and quantum mechanics. In QM, these are often called "observables": it is a slightly misleading name since there is no one that "observes," for example, in a distant galaxy. Variables describe how a system interacts with other systems. In quantum theory, it is appropriate to stop talking about "measurements" and start using the term "interactions" because there are no measurements in general (for example, when there are no scientists around), and these interactions are described precisely by the algebra $\mathcal{A}$. For example, if we throw the chalk against the blackboard, where the chalk hits the blackboard is a variable of the chalk, and the effect of the interaction is the sign left by the latter on the blackboard. We are describing in QG the general relativistic gravitational field, which is a piece of spacetime in the Newtonian sense, so it is a region of spacetime. Thus we are describing a four-dimensional region $\mathcal{R}$ of spacetime, and the variables of the latter (its metric, for example) sit on the boundary because it is on the boundary that the region interacts with the rest. Therefore, \underline{variables sit on the boundary} of the four-dimensional compact region $\mathcal{R}$. They are the metric and the derivatives of the latter. Dynamics will describe the differences between points on the boundary of $\mathcal{R}$, such as how metric changes between two different positions on the boundary. As we said, the easiest way to define the variables in $\mathcal{A}$ is to use the Hilbert space $\mathcal{H}$ because the variables of a quantum system form a non-commutative algebra and from the non-commutativity of the latter all the quantum mechanics stuff descends. The easiest way to describe a non-commutative algebra is to think of the latter as operators acting on a Hilbert space. We define $\mathcal{H}$ and then describe the operators acting on it. 
\subsection{Hilbert space of LQG}
The Hilbert space of QG is much more similar to the QED or QCD Hilbert spaces concerning a single particle's space because this is a quantum field theory, so there are infinite degrees of freedom. In QCD or QED, the "clean" way of constructing a Hilbert space (for infinite degrees of freedom's system) is first to define a finite version of the latter and, then, take a sort of limit. The analog way to build a Fock space is to define a space of states up to one particle, then up to two particles, etc., up to $n$ particles and whereupon taking this limit up to infinite. We never take this limit in doing numerical computations. For example, in computing Feynman graphs, we always work with a finite number of particles, so there's always an approximation level. We are always concretely working on a Hilbert space for a finite number of degrees of freedom arbitrarily large: this is precisely what we also do in LQG. QCD on the lattice works more or less in the same way. So let's define the Hilbert space of LQG $\mathcal{H}$ as follows:
\begin{equation}
\mathcal{H} = \lim_{\Gamma \rightarrow \infty} \mathcal{H}_{\Gamma} \ .
\end{equation} 
Where $\mathcal{H}_{\Gamma}$ is the Hilbert space of a truncation of the number of degrees of freedom: we specify what $\Gamma$ is in a moment. We concretely work with $\mathcal{H}_{\Gamma}$ most of the time without worrying about the above limit. $\Gamma$ is a \textbf{graph}, and graphs sit naturally one in another in sub-graph. What is a graph? It is a couple of two things: a set $N$ and a set $\mathcal{L}$, so $\Gamma \equiv (N,\mathcal{L})$. $N$ is just a finite dimensional set with $n$ elements called \textbf{nodes}, while $\mathcal{L}$ is a collection of ordered couples $l$ of elements of $N$, for example $l \equiv (n_1,n_2)$. These are called \textbf{links}. 
\begin{figure}[h]
\begin{center}
\includegraphics[width=8cm]{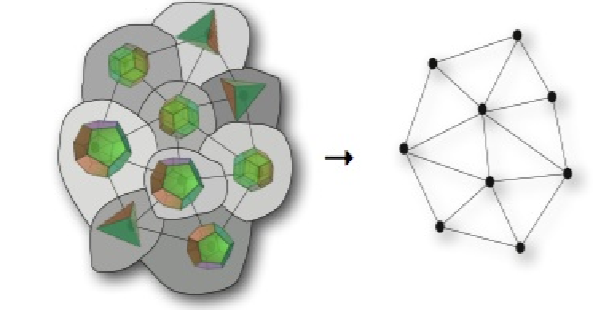}
\caption{A set of adjacent quantum polyhedra and the graph they determine. \textit{Fig. (1.6) of the book}. }
\end{center}
\end{figure}
What matters in graphs is not how we draw them. It is just a combinatorial issue of how to links connect the nodes. Sometimes the above couple is called a "combinatorial graph" to distinguish it from the one embedded in a given manifold, defined as an "embedded graph." Given an oriented link, there is a node from which the link starts, called the \textbf{source} of the link, and the node in which it ends, which is the \textbf{target}. Given a node, its \textbf{valence} is defined as the number of links attached to the link itself. There is a particular sub-class of graphs that is interesting for us.

\medskip

Let's think about a curved 2D surface and imagine triangulating it (i.e., dividing it into triangles). The resulting triangles are attached in a combinatorial way, and we can imagine drawing a point in the center of each triangle and then connecting the latter. By doing this, we obtain a \textbf{dual graph}, describing how the triangles are connected. 
\begin{figure}[h]
\begin{center}
\includegraphics[width=5cm]{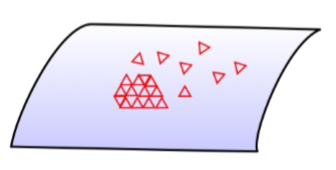}
\caption{Triangulation of a curved 2D manifold. \textit{Taken from page 88 of the book} }
\end{center}
\end{figure}
$\Delta$ usually indicates a given triangulation, while the dual graph obtained by connecting the points (located at the center of triangles) is called "dual triangulation." The latter is usually indicated as $\Delta^*$. If we have a way to fix the orientation of the links, then this is a graph. It is a peculiar graph since all nodes are trivalent.
\begin{figure}[h]
\begin{center}
\includegraphics[width=10cm]{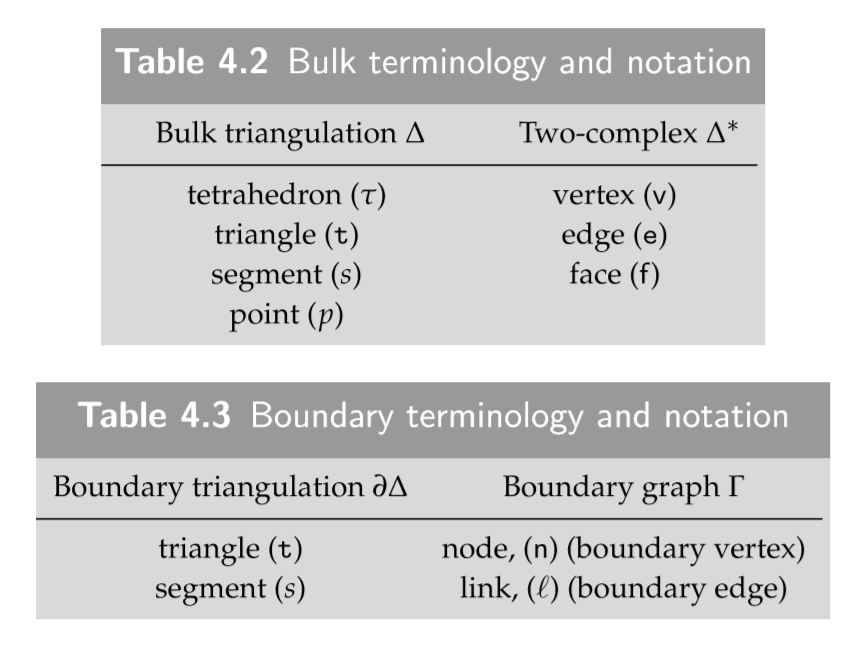}
\caption{Terminology. \textit{Taken from page 93 of the book} }
\end{center}
\end{figure}
We want to repeat this in 3 dimensions, in which we have a piece of space, and we can break it in tetrahedra, imaging to have a point in each center of the latter. So we obtain a graph precisely as before, defined by four-valent nodes forming a triangulation. We are interested in graphs that are dual of triangulation. Most of the time, LQG is defined on these graphs, but not necessarily. This is just one option: we can work with all possible graphs or just with the ones which are dual to triangulation. Does this give different theories? Nobody knows so far. Let's work with graphs that are dual to triangulation, so we consider 4-valent graphs such that there is a fundamental triangulation to which they are dual (but we repeat that the important part is not triangulation). We have not mentioned any metric, curvature, etc.: it is just a question of "pieces of things near to other pieces of things."
In terms of triangulation, a node is dual to a tetrahedron, and a link is dual to a triangle (the common face of two near tetrahedra).
\begin{figure}[h]
\begin{center}
\includegraphics[width=6cm]{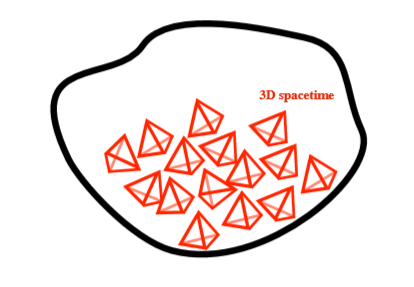}
\label{fig:pag88}
\caption{Triangulation of a 3D region of spacetime. \textit{Taken from page 88 of the book}. }
\end{center}
\end{figure}
If we imagine having one of these graphs, now \underline{we associate an $SU(2)$ element $h_l$ to each oriented link}, supposing that there are $L$ links in total. We consider wavefunctions of these elements, i.e., $\psi(h_l)$.
\begin{figure}[h]
\begin{center}
\includegraphics[width=10cm]{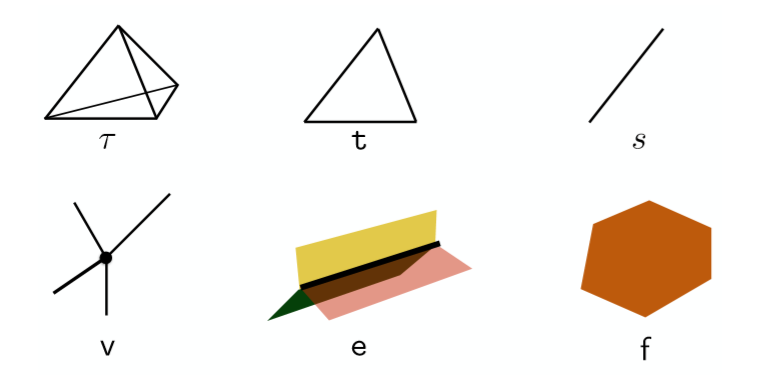}
\caption{Tetrahedra, vertices, edges, faces, links, and their relations. \textit{Fig. (5.1) of the book}. }
\end{center}
\end{figure} 
The Hilbert space is defined as the set of these wavefunctions, namely:
\begin{equation}
\tilde{\mathcal{H}}_{\Gamma} = L_2 \left[ SU(2)^L \right] \ .
\label{Gauge not invariant Hilbert space Lecture 5}
\end{equation}
We have previously defined a larger Hilbert space. The actual one is a subspace of it. This is very similar to Hilbert's space of QCD. If we consider this graph cubic, it is precisely the case. The intuitive meaning will be clear later, but we will try to give some less abstract interpretations for now. \textit{From minute 26:09 to 30:10, Rovelli intuitively explains how to associate (matrices of) rotations to each link of the dual graph, both in 2 and 3 dimensions, describing the space curvature. Essentially, in 3D, we choose an orthonormal frame inside each tetrahedron (centered on its vertex in the dual graph), and we transport it along the links, in the dual graph, by acting with rotation matrices. These matrices are nothing but the $SU(2)$ elements $h_l$, which are associated with each link.} The classical variables are $h_l$, and the quantum state associates an amplitude to a set of the latter. The Hilbert space is just the amplitude for having one set of parallel transport for anyone geometry. The general state is a linear superposition of geometries, which is a state over them. We are going to describe \textbf{quantum geometries}, contained in the Hilbert space $\tilde{\mathcal{H}}_{\Gamma}$. This is the space of quantum geometries, but the actual Hilbert space is smaller. Why? There is a certain amount of arbitrariness (i.e., gauge) in the choice of the orthonormal frame inside every tetrahedron. Precisely like in lattice QCD, there is a gauge that takes $h_l$ (the $SU(2)$ element associated with a link) in the same element multiplied by a group element on the source of the links $\lambda_{s_l}$ and $\lambda^{-1}_{t_l}$, which is a different group element associated with the target link:
\begin{equation}
h_l \rightarrow \lambda_{s_l} h_l \lambda^{-1}_{t_l} \ .
\end{equation}
\begin{figure}[h]
\begin{center} 
\includegraphics[width=5cm]{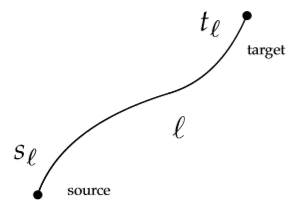}
\caption{\textit{Taken from page 94 of the book}.}
\end{center}
\end{figure}
A gauge transformation, therefore, consists of a group element $\lambda_n \in SU(2)$ for each node. We want these states to be invariant under the previous transformation, so we consider just a subset of wavefunctions satisfying the following property for every choice of $\lambda_n \in SU(2)^N$:
\begin{equation}
\psi(h_l) = \psi(\lambda_{s_l} h_l \lambda^{-1}_{t_l}) \ .
\end{equation}
Thus the set of the wavefunctions which satisfy the previous equation is what we call $\mathcal{H}_{\Gamma} \subset \tilde{\mathcal{H}}_{\Gamma}$ (it is a proper subset since $SU(2)$ is compact and, therefore, there is no "infinity" here). The usual notation is:
\begin{equation}
\mathcal{H}_{\Gamma} \equiv L_2 \left[ SU(2)^L / SU(2)^N \right]_{\Gamma} \subset \tilde{\mathcal{H}}_{\Gamma}  \ .
\label{Gauge invariant Hilbert space Lecture 5}
\end{equation}
We are supposed to know how to factorize this division between groups in concrete applications. The Hilbert space $\tilde{\mathcal{H}}_{\Gamma}$ doesn't know about the graph, but $\mathcal{H}_{\Gamma}$ contains information about it. It is much easier to work with things that are not gauge invariant, i.e., working within $\tilde{\mathcal{H}}_{\Gamma}$, and then pick up the gauge invariant part. 
\subsection{LQG Algebra}
We have just described the Hilbert space. Let's say something about the algebra $\mathcal{A}_{\Gamma}$. We build the latter on $\Gamma$, sitting in $\mathcal{H}_{\Gamma}$. It consists of position and momentum, or $x$ and $p$. These are functions of group elements so that we can multiply any function of two elements, and this is a (diagonal) operator, as in quantum mechanics:
\begin{equation}
\hat{x} \psi(x) = \psi(x) x \ .
\end{equation}
So, any function $f(h_l)$ of $h_l$ turns out to be an operator in $\mathcal{A}_{\Gamma}$, which means that everything that we can measure related to the parallel transport is an operator on the Hilbert space. Within books and papers, it is usually written: 
$$
\hat{h}_l \psi(h_l) = \psi(h_l) h_l \ .
$$ 
Nevertheless, this is confusing since $h_l$ is a group element, not a number. It simply means that the matrix elements $A,B$ of $\left( \hat{h}_l \right)_{AB} \psi(h_l)$ are operators. For example, we can consider the trace operator of $\left( \hat{h}_l \right)_{AB}$, which acts on the wavefunctions by multiplying them for the trace of $ \left( h_l \right)_{AB}$. The interesting part is that the momentum acts as a derivative, similar to what happens in quantum mechanics:
\begin{equation}
\hat{p} \psi(x) = -i \hbar \frac{d}{dx} \psi(x) \ .
\end{equation}
We have already defined derivatives on $SU(2)$ in lecture \ref{Lecture_SU(2)_group}, i.e., the left-invariant operator. So we have a left-invariant vector field $\vec{L}$ for each copy of the group and links label these: $\vec{L}_{l}$. Are these things well defined in $\mathcal{H}_{\Gamma}$ and not only within $\tilde{\mathcal{H}}_{\Gamma}$? No, because $\vec{L}$ is a vector. Therefore, it is not invariant under rotations. We will construct gauge invariant operators within the sub-Hilbert space, but $\vec{L}$ will be our most used operator. We will see soon what this operator means.
\section{Geometrical Operators}
\label{sec:Lecture_6_Geometrical_operators}
As we were saying, $\vec{L}_l$ are the basic operators in the LQG theory. They rotate the variables in the three directions of the algebra. We start reviewing some of the properties they have:
\begin{equation}
\left[ L^i_l, L^j_{l'} \right] = \epsilon^{ij}_{\hspace{2mm}k}L^k \delta_{l,l'} \ .
\end{equation}
\begin{figure}[h]
\begin{center}
\includegraphics[width=6cm]{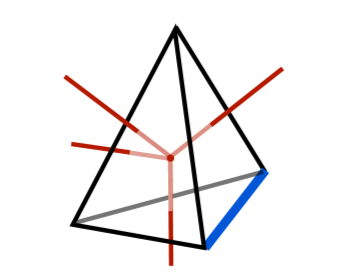}
\caption{Group elements are on the edges (red). Equivalently, algebra elements on the segments (blue) or in their dual faces. \textit{Fig. (4.3) of the book.}}
\end{center}
\end{figure}
So they don't commute if they refer to the same link (otherwise, they do). Of course, they don't commute with $f(h_l)$ since they are derivatives. It is pretty obvious to see that the following equation holds if $\psi$ is a gauge invariant state and if we have a node with four links:
$$
\left( \vec{L}_1 + \vec{L}_2  + \vec{L}_3  +\vec{L}_4  \right) \psi = 0 \ .
$$
So the sum of the link coming out of a node is zero. In general, we write:
\begin{equation}
\sum_{l \in (nth \hspace{1mm} node)} \vec{L}_l \psi(h_l) = 0 \ .
\label{Relazione di chiusura Lecture 6}
\end{equation}
Each operator $\vec{L}$ acts on the left of $h_l$, namely on one of the four links (coming out of the very same node), changing it infinitesimally by turning the latter in one direction of the algebra. The action of $\left( \vec{L}_1 + \vec{L}_2  + \vec{L}_3  +\vec{L}_4  \right)$ on a node does this on all the four links coming out of it. As we said, the operators $\vec{L}$ are defined on $\tilde{\mathcal{H}}_{\Gamma}$, not on $\mathcal{H}_{\Gamma}$, therefore let's see what we can define on the latter.
\subsection{Geometrical operators} 
Since $L^i$ is not gauge invariant (in fact, it has an index), to build a gauge invariant object, we consider the scalar product, defining an operator called $A^2_l$:
\begin{equation}
A^2_l \equiv \vec{L}_l \cdot \vec{L}_l \ .
\end{equation}
This is a self-adjoint operator. Even $\vec{L}$ is, concerning the scalar product defined in $\mathcal{H}_{\Gamma}$. $A^2_l$ is well defined in the gauge invariant Hilbert space since it sends the latter into itself. Another possibility is to consider the same operator but between two links that come out of the same node:
\begin{equation}
G_{l,l'} \equiv \vec{L}_l \cdot \vec{L}_{l'} \ .
\end{equation}
This is also gauge invariant. A third possibility is to define the following operator, in which $l_1$, $l_2$, $l_3$ are three links which, again, come out of the very same node $n$:
\begin{equation}
V^2_n \equiv L^i_{l_1} L^j_{l_2} L^k_{l_3} \epsilon_{ijk} \ .
\end{equation}
If we consider $l_4$ instead of one of the other three links, the operator $V^2_n$ turns out to be the same because of the closure relation \eqref{Relazione di chiusura Lecture 6}, so it depends just on the node. This is the reason why we used the index $n$. 

\medskip

These are all gauge invariant operators. There is a theorem (which we are going to prove) stating that \underline{the set $\left( A_l, V_n \right)$ is a maximal commutative set of operators}, in the sense of Dirac. So it is like the set $\left( H, L^2, L_z \right)$ for the hydrogen atom. In other words, we can diagonalize the operators $\left( A_l, V_n \right)$ concurrently since there is a basis that does it, and there is no degeneracy. We will write this basis explicitly, solving the spectrum problem for these operators, but not in this lecture. The geometrical interpretation of the operators just introduced will be clearer when we briefly discuss the classical limit.
\subsection{Triads}
If we imagine having a tetrahedra triangulation on a three-dimensional Riemannian curved space with a metric $q_{ab}(x)$ (namely a metric space), we can approximate this metric with the triangulation by assuming that the tetrahedra themselves are flat. The curvature is concretely computed when we start from a vertex and move to the near vertices through the links. When we return to the vertex from which we started, we have a measure of the curvature around the segment of the tetrahedron around which we have moved, so the curvature is physically measured relative to the segments. If we know the geometry of each of these tetrahedra (so we know their volumes, angles, lengths, etc.), then we construct the geometry of the metric space itself. For example, if the sum of all the angles around a segment is $2 \pi$, the spacetime region is flat, while if it is more (less), there turns out to be negative (positive) curvature along the segments itself. If we know the geometry, we can compute the area associated with each triangle and the volume associated with each tetrahedron. It will turn out that the operator $A_l$ associated with the links (which are dual to triangles) corresponds to this area. In contrast, $V_n$ associated with the nodes (dual to tetrahedra) corresponds to the volume. When we go back to classical GR, writing the equations explicitly, we will argue that it is much better to think of GR in terms of tetrads instead of using metric or, in three dimensions, in terms of \textbf{triads} because it is a much more general formulation. For example, the metric is not a good tool for fermions. Therefore in three dimensions, instead of using $q_{ab}(x)$ to describe the geometry, we use the triad fields $e^i_a(x)$, which have two indices: $a, i = 1,2,3$. $a$ is a general relativistic index while $i$ is an internal flat index. The relation between the two formalisms is:
\begin{equation}
q_{ab}(x) = e^i_a(x) e^j_b(x) \delta_{ij} \ .
\label{Triadi e metrica Lecture 6}
\end{equation}
The inverse relation turns out to be:
\begin{equation}
q_{ab}(x)e_i^a(x) e_j^b(x) = \delta_{ij} \ ,
\end{equation}
where $e^a_i(x)$ is the inverse of the $3 \times 3$ matrix $e_a^i(x)$. Sometimes called "triad fields" and "co-triad fields," respectively. We can think at $e^a_i(x)$ as a three-vector field:
\begin{equation}
e^a_i(x) = (\vec{e}_1(x), \vec{e}_2(x), \vec{e}_3(x)) \ .
\end{equation}
The expression $q_{ab}(x)e_i^a(x) e_j^b(x)$ is nothing but the physical scalar product $\vec{e}_i \cdot \vec{e}_j$. So these vectors are orthogonal, and they have lengths equal to one. Instead of giving the geometry with the metric, we can specify the latter by introducing three orthogonal vectors (basically a reference frame) in each point. If we do this, it is equivalent to knowing the metric because of relation \eqref{Triadi e metrica Lecture 6}. This formulation is due to Weyl and Cartan, who developed the latter independently. We call it \underline{triad formulation} in three dimensions. Thus we have these triads, and in the triangulation case, we can put a triad inside each tetrahedron, constant on the tetrahedron itself. Within each tetrahedron, we can choose a Cartesian reference frame, and when we go to the next one, we rotate it. When we go around a segment, the amount of rotation gives the curvature. Suppose we have a tetrahedron and the quantities $e^i_a(x)$. In that case, we can do something with the triangles (the faces of the tetrahedrons): it is possible to associate a vector with them. Let's first use the (easier) language of forms and then use coordinates. $e^i_a(x)$ defines a 1-form $e^i$ in the following way:
\begin{equation}
e^i_a(x) \rightarrow e^i_a dx^a = e^i \ .
\end{equation}
From a 1-form, we can build a 2-form by considering $e^i \wedge e^j$, an antisymmetric expression in both indices. Being the previous expression antisymmetric, we can put another tensor and write it as $e^i \wedge e^j \epsilon_{ij}^{\hspace{2mm} k}$: this is 2-form, and it is known that the latter can be integrated on surfaces. 
\subsection{Meaning of geometrical operators}
If we integrate the 2-from $e^i \wedge e^j \epsilon_{ij}^{\hspace{2mm} k}$ on a triangular face of a tetrahedron, we obtain a quantity called $E^k$:
\begin{equation}
E^k \equiv \frac{1}{2} \epsilon_{ij}^{\hspace{2mm} k} \int_{\Delta} e^i \wedge e^j \ .
\label{Definizione dei vettori E Lecture 6}
\end{equation}
\textit{I copy here the expression (1.11) or (3.92) of the book, in which there is a factor $\frac{1}{2}$ }. So to each triangle, we can associate a quantity $E^k$, i.e., a vector. The space spanned by the triads is a three-dimensional Cartesian space. We could put coordinates to the triangles and drop the forms notation. What is this vector $E^i$? If we choose Cartesian coordinates on the tetrahedron such that $e^i_a =  \delta^i_a$ (we can effectively choose a triad that is both constant and diagonal since the information is only when we go around segments), we would get a vector $E^i$ normal to the face on which we are integrating. The triangles' area would come up from integrating the previous expression. Therefore the norm of $E^i$ is equal to the area of the face times the normal: 
\begin{equation}
\vec{E} = A \hat{n} \ .
\end{equation}
This implies that $\vec{E} \cdot \vec{E} = A^2$. 
\begin{figure}[h]
\begin{center} 
\includegraphics[width=4cm]{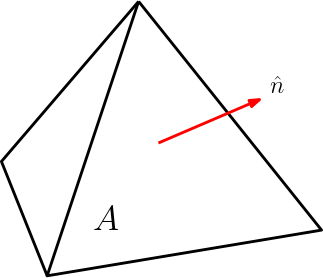}
\caption{\textit{Normal to the triangular face of a tetrahedron, where $A$ is the area of the face itself.}}
\end{center}
\end{figure}
What can we say about the volume of the tetrahedron? If we know the four vectors $E^i$ coming out of the same tetrahedron, we can compute its volume. These four vectors have a property. What happens if we sum them? Of course, we obtain zero \textit{(Rovelli mentions a brilliant demonstration given by Penrose)}:
\begin{equation}
\vec{E}_1 + \vec{E}_2 +  \vec{E}_3 +  \vec{E}_4 = 0	 \ .
\end{equation}
The volume must be scalar, and we should be able to compute it by knowing at least three vectors $E^i$ since the previous relation gives the last one. It should also be invariant under rotations. We can construct just one number out of three vectors that satisfy the above properties, namely the mixed product. So we write:
\begin{equation}
V^2 = \alpha \epsilon_{ijk} E^i_1 E^j_2 E^k_3 \ ,
\end{equation}
where $\alpha$ is a number we could easily determine because it must be the same for all tetrahedrons. \textit{Rovelli starts computing it explicitly at minute 31:10. The final correct result is, according to formula (1.10) of the book, $\alpha = \frac{2}{9}$}. Summarizing, we can describe the geometry of the tetrahedrons by using the vectors $E^i$, satisfying the above properties. It turns out that, as we shall see, the operators $\vec{L}_l$ associated with links and dual to triangles are exactly the quantization of $\vec{E}$. This also descends from the canonical quantization approach, which we will discuss in future lectures. Once written the action, the conjugate momentum to the connections is $E^i$. More precisely, since the GR action $S$ consists of an integral multiplied by $\frac{1}{16 \pi G}$, the conjugate momentum to the connection will be $E^i$ divided by $G$. So the operator corresponding to the momentum will be proportional to $E^i/G$. Since this operator acts as a derivative, $E^i$ will be proportional to $\hbar G$ times a derivative. Eventually, the relationship between these quantities is:
\begin{equation}
E^i_{\Delta} \varpropto G \hbar L^i_l \ ,
\end{equation}
where $l$ is the link dual to the triangular face $\Delta$. This is the meaning of the quantum geometrical operator $E^i$. From a dimensional point of view, it is easy to be convinced that the above relation is effectively correct. \textit{At the end of the lecture, Rovelli answers some questions}.
\section{Quanta of Space, spin networks and discreteness}
\label{sec:Lecture_7_Quanta_space}
Now we treat real physics. Let's start by wondering about a question: what was the main physical discovery of quantum mechanics? Of course, there were several new concepts: probability, quantum interactions, etc., but the most remarkable was the peculiar discreteness concept. In phase space of quantum theory, we cannot localize the system better with respect to within a cell with the Planck length dimension. \textit{Rovelli shares an interesting anecdote on Schroedinger teaching in Dublin in the fifties}. The reason which atoms are stable is that there is a minimum energy level for the electron, and the reason why we do not observe the ultraviolet divergence is that when frequencies are very high since there is an energy level that separates the vacuum and the first excited state, it requires a lot of energy to pass from the one to the other, so we cannot fill it with the available energy. How do we concretely compute discreteness in QM? We know a not commutative algebra exists, and the latter's observables have a spectrum. As already pointed out, we prefer to work explicitly with it in the Hilbert space. The best way to define it is by examining the eigenvalue equation for an observable $A$ in the algebra $\mathcal{A}$:
\begin{equation}
A | \psi_n \rangle = | \psi_n \rangle a_n \ .
\end{equation}
Somehow, the key interpretation concerning (one of) the QM's postulate(s) is that only the values $a_n$, which the observable $A$ can take, are within the spectrum of $A$. We have seen that in LQG, the Hilbert space \eqref{Gauge not invariant Hilbert space Lecture 5} is not gauge-invariant and the invariant one \eqref{Gauge invariant Hilbert space Lecture 5}. We have operators within them: the left-invariant vector fields associated with the links (the latter being defined only in the not gauge invariant Hilbert space) and the others defined in the previous lecture, namely $A^2_l$, $V^2_n$ and $G_l,l$. These are all well-defined in the gauge invariant Hilbert space. We have briefly mentioned their geometrical interpretation, i.e., $ \hbar G \vec{L}_l$ is a vector normal to a face of a tetrahedron in a triangulation of the spacetime. The correspondence between classical and quantum theory will be something like this:
\begin{equation}
\vec{E}_l \longrightarrow \hbar G \vec{L}_l \ .
\end{equation}
As we have seen, $A_l$ is the area of the triangle, while $V_l$ is the volume. It is easy to see that the operator $G_{l,l'}$ gives essentially the angle between two faces of a tetrahedron, to which $l$ and $l'$ are the dual links. If we can compute these three geometrical operators, we reconstruct, in principle, the geometry of all the tetrahedrons and, within this approximation, of the spacetime region that we are describing. The more (smaller and smaller) tetrahedra we use, the more precisely we can reconstruct spacetime geometry. If we count the degrees of freedom, we see that for each node, we have three vectors, each with three components, but the invariant part is up to rotation (i.e., up to $SO(3)$). So, there are $9-3 = 6$ degrees of freedom. The space of the possible shapes of the flat tetrahedron is six-dimensional: it is captured either by the six lengths of its segments or by three vectors (each one having three components) up to three-dimensional rotations. Of course, four vectors are coming out from the very same node. Still, because of the closure relation \eqref{Relazione di chiusura Lecture 6}, they are not independent, so the full exact counting of degrees of freedom would be eventually written as $12 - 3 - 3 = 6$.
\subsection{Gauge invariant states}
We aim to solve the eigenvalue equations for the area and volume operators. In a certain sense, we have already done it within the mathematical description of $SU(2)$. We said that in a single Hilbert space $L_2 \left[ SU(2) \right]$ there is a basis made up of orthogonal bases of $SU(2)$ consisting in Wigner matrices: $ D^j_{mn}(h) = \langle h | j,m,n \rangle $. This notation means that, for a fixed $j$, $D^j_{mn}(h)$ is the projection of the ket $| j,m,n \rangle$ (which turns out to be the basis of the Hilbert space $\mathbb{H}^j$, chosen to diagonalize the operators $L^i$) on the abstract group elements of $SU(2)$. For what concerns the space $\tilde{\mathcal{H}}_{\Gamma} = L_2 \left[ SU(2)^L \right]$, we have $L$ copies of $SU(2)$, so a possible basis for $\tilde{\mathcal{H}}_{\Gamma}$ is given by the following Wigner matrices' product:
$$
D^{j_{1}}_{m_1 n_1}(h_l)...D^{j_{L}}_{m_L n_L}(h_l) = \langle h_l | j_l ,m_l ,n_l \rangle \ .
$$
Abstractly we could write:
\begin{equation}
L_2 \left[ SU(2)^L \right] = \bigotimes_l \left( \bigoplus_{j_l} \left( \mathcal{H}_{j_l} \otimes \mathcal{H}_{j_l} \right) \right) \ .
\end{equation}
\textit{If compared with formula (5.18) of the book, the above version seems more precise, since each $j$ depends on the link $l$ which we are considering in the product}.
That is, a tensor product of all the $L_2 \left[ SU(2) \right]$ copies and on each link $l$ we have a basis given by $|j_l,m_l,n_l \rangle$, namely (a part from the representation spin index $j_l$) an $m_l$ index in the Hilbert space $\mathcal{H}_{j_l}$ and an $n_l$ index in the other one: $\bigoplus_{j_l} \left( \mathcal{H}_{j_l} \otimes \mathcal{H}_{j_l} \right)$. Now we can rewrite the previous expression interchanging the sum with the product:
$$
\bigotimes_l \left( \bigoplus_{j_l} \left( \mathcal{H}_{j_l} \otimes \mathcal{H}_{j_l} \right) \right) 
= 
\bigoplus_{j_l} \left( \bigotimes_{l} \left( \mathcal{H}_{j_l} \otimes \mathcal{H}_{j_l} \right) \right) \ .
$$
Now comes the interesting part. 
\begin{figure}[h]
\begin{center}
\includegraphics[width=5cm]{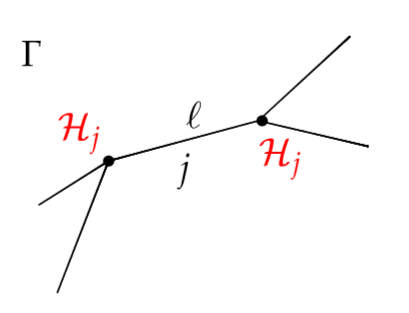}
\caption{\textit{Taken from page 107 of the book}.}
\end{center}
\end{figure}
Considering a link $l$ attached between two nodes, we know that the matrix $D^j_{mn}(h_l)$ is associated with the link itself, where the indices $m$ and $n$ live in different spaces. When we act with gauge transformations, the gauge transformation on one node acts on one index while the one acting on the other node acts on the other index. This is true because when we act on $D^j_{mn}(h_l)$ from the left (right), we act on the $m$ ($n$) index. In other words, the gauge invariance of the source node acts on $m$, whereas the gauge invariance of the target node acts on $n$. So we can imagine the two indices directly "attached" to the two nodes we are considering. Therefore each link has a couple of Hilbert spaces $\mathcal{H}_l$ to its edges, so the idea is to re-group them by nodes. Thus we can write:
$$
\bigoplus_{j_l} \left( \bigotimes_{l} \left( \mathcal{H}_{j_l} \otimes \mathcal{H}_{j_l} \right) \right)
=
\bigoplus_{j_l} \left( \bigotimes_{n} \left( \mathcal{H}_{j_{l_1}} \otimes \mathcal{H}_{j_{l_2}} \otimes  \mathcal{H}_{j_{l_3}} \otimes  \mathcal{H}_{j_{l_4}} \right) \right) \ ,
$$
where $\mathcal{H}{j_{l_i}}$ is the $i$-th Hilbert space coming out of the $n$-th node (we, of course, are imaging a tetrahedra triangulation in which four links are coming out of each node).
\begin{figure}[h]
\begin{center}
\includegraphics[width=4cm]{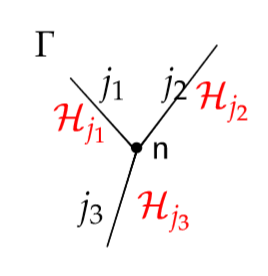}
\caption{\textit{Taken from page 108 of the book}.}
\end{center}
\end{figure}
We do this passage because actually, we want a basis in $\mathcal{H}_{\Gamma}$, not in $\tilde{\mathcal{H}}_{\Gamma}$, i.e., a basis made up of gauge invariant objects. A gauge transformation on a given node $n$ only acts on the Hilbert spaces $ \mathcal{H}_{j_{l_1}} \otimes \mathcal{H}_{j_{l_2}} \otimes  \mathcal{H}_{j_{l_3}} \otimes  \mathcal{H}_{j_{l_4}}$. So if $\bigoplus_{j_l} \left( \bigotimes_{n} \left( \mathcal{H}_{j_{l_1}} \otimes \mathcal{H}_{j_{l_2}} \otimes  \mathcal{H}_{j_{l_3}} \otimes  \mathcal{H}_{j_{l_4}} \right) \right)$ is (spanned by) a basis of $\tilde{\mathcal{H}}_{\Gamma}$, then a basis of $\mathcal{H}_{\Gamma}$ must span $\bigoplus_{j_l} \left( \bigotimes_{n} \hspace{1mm} Inv \left( \mathcal{H}_{j_{l_1}} \otimes \mathcal{H}_{j_{l_2}} \otimes  \mathcal{H}_{j_{l_3}} \otimes  \mathcal{H}_{j_{l_4}} \right) \right)$, i.e., the invariant part of the spaces' tensor product. $Inv \left( \mathcal{H}_{j_{l_1}} \otimes \mathcal{H}_{j_{l_2}} \otimes  \mathcal{H}_{j_{l_3}} \otimes  \mathcal{H}_{j_{l_4}} \right)$ is the object we call \textbf{intertwiner space} among these four Hilbert spaces. So, from this point of view, it is much easier to understand the intertwiner space. 

\medskip

We can indeed write the states of $\tilde{\mathcal{H}}_{\Gamma}$ as $| j_l, (m_1,m_2,m_3,m_4)_n  \rangle$, by considering for each node $n$ the four outgoing links. The states $| j_l, (m_1,m_2,m_3,m_4)_n  \rangle$ and $| j_l,m_l,n_l \rangle$ are different possible bases of $\tilde{\mathcal{H}}_{\Gamma}$, the only difference is that with the first one, we are considering the four indices associated with the links coming out of each node, while with the second one, we are considering the two indices attached to the source and the target of every link, each one living in different Hilbert spaces.  How can we write gauge invariant states out of these?
The gauge invariant states form just a subspace of $\tilde{\mathcal{H}}_{\Gamma}$, that is, they span $\mathcal{H}_{\Gamma} \subset \tilde{\mathcal{H}}_{\Gamma}$. Therefore we can use linear combination of $| j_l, (m_1,m_2,m_3,m_4)_n  \rangle$ in order to construct them. Let's write these states as \textit{(see minute 28:27 for comments on the formula, since it is easy to understand but difficult to write)}:
\begin{equation}
| \psi \rangle = \sum_{m_1...m_4} i^{m_1^{n_1}m_2^{n_1}m_3^{n_1}m^{n_1}_4}_{n_{1}}... i^{m_1^{n_n}m_2^{n_n}m_3^{n_n}m^{n_n}_4}_{n_{n}} | j_l, (m_1^n,m_2^n,m_3^n,m_4^n)_n  \rangle  \ .
\end{equation}
This state is gauge invariant if, when we act with rotations on $(m_1^n,m_2^n,m_3^n,m_4^n)_n$, this is gauge invariant, so also the above coefficients $i^{m_1^{n_1}m_2^{n_1}m_3^{n_1}m^{n_1}_4}_{n_{1}}... i^{m_1^{n_n}m_2^{n_n}m_3^{n_n}m^{n_n}_4}_{n_{n}}$ of the linear combination must be gauge invariant. Therefore they must satisfy the following:
\begin{equation}
D^{j_1 m_1}_{\hspace{7mm} n_1}(h)D^{j_2 m_2}_{\hspace{7mm} n_2}(h) D^{j_3 m_3}_{\hspace{7mm} n_3}(h) D^{j_4 m_4}_{\hspace{7mm} n_4}(h) i^{n_1 n_2 n_3 n_4} = i^{m_1 m_2 m_3 m_4} \ ,
\end{equation}
where each index of the Wigner rotation matrices lives in a different representation. \textit{For a more formal discussion there is, for example, section 5.2.2 of the book}. The above condition on the coefficients $i^{n_1...n_4}$ ensures the gauge invariance of the state $| \psi \rangle$. If we want a basis of gauge invariant things, then we need a basis of the intertwiner space $Inv \left( \mathcal{H}_{j_{l_1}} \otimes \mathcal{H}_{j_{l_2}} \otimes  \mathcal{H}_{j_{l_3}} \otimes  \mathcal{H}_{j_{l_4}} \right)$. Namely, a set of independent elements $i^{m_1...m_4}$ for each node such that any other element can be written as a linear combination of them. At the end of lecture 4, we have seen that the simplest possibility to construct $i^{m_1...m_4}$ is to use (unique) three-dimensional intertwiners. 
%
%
We can label a basis of this intertwiner space by $| k \rangle$. 
Thus we can write a gauge invariant state as $| \psi \rangle = | j_l, k_n \rangle  $, having constructed a basis in the physical gauge invariant Hilbert space $\mathcal{H}_{\Gamma}$ by choosing two quantum numbers: the spin associated with every link $j_l$ and an element of a basis in the intertwiner space $k_n$ \underline{associated with each node}. This is, in the end, just one of the possible choices. 
\subsection{Eigenvalue equation for the area operator}
To solve the eigenvalue problem, we can now use this new basis. The first operator $ A^2_{l} = \vec{L}_l \cdot \vec{L}_l = C $ is the Casimir operator, and it is diagonal in the Peter-Weyl decomposition. It has the value $j(j+1)$ in each representation $j$. So the previous operator acts on the gauge invariant states constructed above as:
\begin{equation}
A_l^2 | j_l, k_n \rangle = | j_l, k_n \rangle j_l (j_l + 1) \ .
\end{equation}
The area operator in the geometrical interpretation is this multiplied by $(\alpha \hbar G)^2$, where $\alpha$ is a (not still specified) real number. Therefore the eigenvalue equation for the area operator turns out to be: 
\begin{equation}
A_{\Delta} | j_l, k_n \rangle = | j_l, k_n \rangle \alpha \hbar G \sqrt{j_l(j_l + 1)} \ .
\end{equation} 
\textit{This last passage should be formally taken as a redefinition of the area operator: $A_l^2 \equiv \vec{L}_l \cdot \vec{L}_l \rightarrow A^2_{\Delta} \equiv (\alpha \hbar G)^2 \vec{L}_l \cdot \vec{L}_l$}.
We postpone the discussion of the volume operator to the following lecture.
The area operator is completely blind to intertwiners: it transforms the intertwiners' space into itself. It just sees the  spin associated with the links.
The point is that $j_l = 0, \frac{1}{2}, 1 ...$ are half-integers, so the area operator turns out to have a discrete spectrum, which means that \underline{we cannot have an arbitrarily small area}. But we said that we could take a portion of three-dimensional space and describe its geometry better as we take smaller and smaller tetrahedra, namely refining it better and better. Thus quantum theory states a bottom level of the order of $\hbar G$, at which we have to stop! This is the first quantum effect in LQG theory, which comes from an eigenvalue calculation (i.e., not a postulate or an assumption). If something is cutting off the continuity of geometry, we have a hint of why UV divergence comes out in QFT.
\section{Volume operator and Quantum space}
\label{sec:Lecture_8_volume_operator}
We have written the eigenvalue equation for the area, and now we do the same with the volume, also examining the physical consequences that turn out to be remarkable.
\subsection{Eigenvalue equation for the volume operator}
Remember that the volume is (up to a number, i.e., a power of $\hbar G$) $V^2 = \epsilon_{ijk} L^i_{l_1} L^j_{l_2} L^k_{l_3}$, where $l_1,l_2$ and $l_3$ are three links coming out of the very same node (the fourth is not independent because of the closure relation) that is dual to the tetrahedron of which we are calculating the volume. $\vec{L}_{l_i}$ are the generators of rotations on the $i$-th link and they act on the indices $(m_1^n,m_2^n,m_3^n,m_4^n)_n$ of the states $| j_l, (m_1^n,m_2^n,m_3^n,m_4^n)_n  \rangle$. This can be written explicitly:
\begin{equation}
L^i D^{j}_{mn}(h) =  D^j_{m n'}(h) L^{i}_{n'n} \ ,
\end{equation}
where we see exactly on which index the operator $\vec{L}$ acts. Namely, it is just the generator $L^{i}_{n'n}$ of $SU(2)$ in the representation $j$ acting on the right index of the Wigner matrix, and it physically represents a gauge transformation on $h$. The point is that the three operators $L^i_{l_1} L^j_{l_2} L^k_{l_3}$ act on the indices $(m_1^n,m_2^n,m_3^n,m_4^n)_n$ \underline{on the same node}, so they transform three indices of the coefficients $i_n^{m_1...m_4}$. The volume operator, therefore, is a linear operator in the finite-dimensional Hilbert space $\mathcal{H}_{j_1} \otimes \mathcal{H}_{j_2} \otimes \mathcal{H}_{j_3} \otimes \mathcal{H}_{j_4}$. This space contains the (linear) intertwiners subspace, and since the volume is gauge invariant, it transforms the intertwiners space into itself. 
%
%
Some programs diagonalize the matrix $v_k^{\hspace{1mm} k'}$, whose elements are precisely the eigenvalues of the volume. Of course, for large $j$, this is a very large matrix. If we now define as $v$ the eigenvalues of the volume, then we can define states which are just a linear combination of states $| k \rangle$ with some coefficients:
\begin{equation}
| j_l, v_n \rangle = \sum_k  c_{v, k} |  j_l, k_n \rangle \ .
\end{equation}
Therefore, instead of using a basis of elements $ | j_l, k_n \rangle $, we can define a basis $ | j_l, v_n \rangle $ where the $v_n$ are the volume's eigenvalues. These states are called \textbf{spin network states}. It is boring to calculate the volume's eigenvalues explicitly. Still, the point is that, since the space $\mathcal{H}_{j_1} \otimes \mathcal{H}_{j_2} \otimes \mathcal{H}_{j_3} \otimes \mathcal{H}_{j_4}$ is finite-dimensional, the eigenvalues $v_n$ are discrete. Therefore it turns out that both the quantum numbers $j_l$ and $v_n$ are discrete.
Thus we arrive at the important conclusion that \underline{both area and volume's eigenvalues are discrete}. This means that the tetrahedra's area and volume must have a minimum size. But there is something which is even more interesting. The geometry of the tetrahedron consists of $6$ degrees of freedom. We have just written explicit a basis $| j_{l_n},v_n \rangle$, in which quantum numbers depend on the links on the node, so for each tetrahedron, there is one volume quantum number $v_n$ and four quantum numbers related to the area of the faces $j_{l_n}$. Therefore, in this representation, each tetrahedron has a fixed volume and fixed area of the faces. Thus, we find five numbers in total... what is missing? 
\subsection{Quantum features of spacetime}
It is well known that, in QM, the classical angular momentum variables $\left( L_x, L_y, L_z \right)$ can be quantized, but we can concurrently diagonalize only two of them, typically $\left( L^2, L_z \right)$. So a little object (for example, an electron) has a fixed angular momentum in one direction and a spread angular momentum in the other two directions. The same argument is true here: out of 6 classical variables that describe the tetrahedron, we can sharpen at least five variables simultaneously. There are other states (obtained as linear combinations of these states) that spread these variables and sharp other ones: it is just a different choice of basis. For example, we could write a basis in which $G_{l,l'}$ is sharp (so the angle between two faces is determined), and these five variables are not sharp anymore. So these are not simply tetrahedra. They are "fuzzing" tetrahedra because they are quantum without a clean Euclidean geometry.
\begin{figure}
\centering 
\includegraphics[width=6cm]{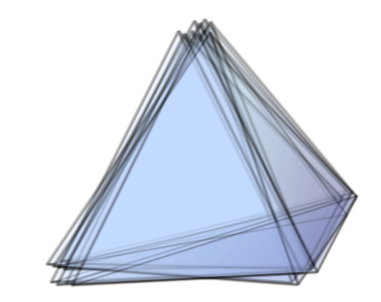}
\caption{\textit{Taken from pag. 17 of the book}.}
\end{figure}
So, in addition to \textbf{discreteness} (which results from the fact that the eigenvalues of area and volume are discrete), the second crucial aspect of LQG geometry is \textbf{fuzziness}, which comes out of the fact that all the variables describing the geometry of tetrahedra don't commute between them. There is a third aspect that is implicit here: the following. As a particle can go through two holes at the same time (in the famous double slit experiment) since the position is a quantum variable and the latter can be spread, in LQG, we find that the \underline{possible states of geometry are arbitrary linear combinations} of the states $|j_l,v_n \rangle$. Consequently, a tetrahedron can have a geometry that turns out to be a linear combination, for example, of different volume eigenvalues. This is just the standard linearity of QM. All this fuzzy quantum geometry is clearly defined with the mathematical apparatus of the theory, which we have built so far.

\medskip

This theory predicts that measuring something (for example, a cross-section) smaller than the length of the Planck scale order will never be possible. \textit{For other interesting considerations concerning measurements, experiments, etc., see the lecture from minute 16:23 to 20:09}. If we have a graph, a spin network state $| j_{l_n}, v_n \rangle$ describes the latter by using the quantum numbers $j_{l_n}$, i.e., the spins (which sit on the links) and the invariant intertwiners $v_n$ (sitting on the nodes, that can be chosen as eigenvalues of the volume). The graph labeled by these numbers is called \textbf{spin network}. The basis of eigenvalues of the volume is a "natural" basis since it diagonalizes a macroscopic observable, but it is not easy to work with. Computing the eigenvalues of the volume is time-consuming. The easy basis to work with is $|j_l,k_n \rangle$, which can be constructed by contracting trivalent intertwiners and computed using Mathematica. Of course, it is not unique because four intertwiners can be split differently. A simple relation exists between the states $|j_l,k_n \rangle$ and the other possible states, which can be obtained by connecting the intertwiners differently. 
%
%

\medskip

The coefficients in the linear combination specifying the change of basis are called $6j$ Wigner symbols. They are also given explicitly by Mathematica. They are variables that depend on six numbers that can be constructed from three-dimensional intertwiners. They allow us to realize the change of basis from the bases where we use different intermediate representations to pair the intertwiners themselves. The state $|j_l,k_n \rangle$ has a transparent physical meaning since it is the eigenstate of the left-invariant operator acting on links, which are dual to faces of the tetrahedron, so $k$ is the quantum number associated with the "dihedral" angle between the faces. Thus, we can construct states diagonalizing four faces and one angle between two while the remaining angles are completely spread. But we know, by using the $6j$ Wigner symbols, how to change the basis from one angle to another. To have something similar to the wave packets of QM, namely, not having sharped variables, whereas others are completely spread. Simone Speziale and others have enormously contributed and developed a sort of semi-classical coherent states. \textit{Rovelli tells the class how Speziale and Livine used the theory of $SU(2)$ coherent states to introduce these new states, which are treated in lectures \ref{sec:Lecture_18_Regge_calculus}-\ref{sec:Lecture_19_coherent_states} }.
\begin{figure}[h]
\begin{center}
\includegraphics[width=7cm]{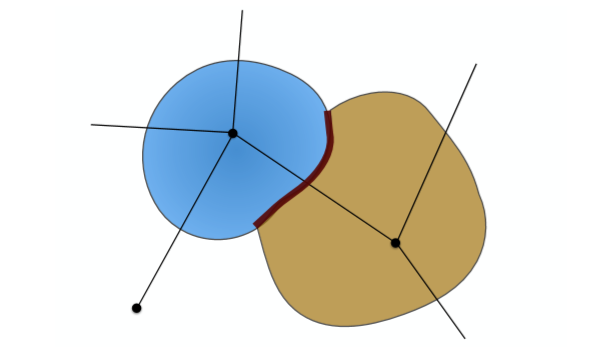}
\caption{Chunks of space with quantized volume are associated with nodes. The area of the surface shared between two cells is quantized as well. \textit{Fig. (7.3) of the book.}}
\end{center}
\end{figure}
Intuitively, it is appropriate to think about nodes not as sharp geometrical points but as a minimum "quantum blob" of spacetime, which is all spatially connected. Therefore, the "spacetime entity" in its wholeness on which we live in the sense of Newton turns out to be not a continuous thing: it is a structure made up of quanta, and it exhibits all properties of the quantum stuff. That is, it can be in superposition, there is discreteness, we find minimum "chunks" of spacetime, etc. Of course, matter lives on these quantum portions of spacetime, which can only be upon them, not in between. The adjacency among these chunks is exactly what the links describe, while quantum numbers tell us how "big" they are (in the sense of geometry). In relational space, we can answer "Where are we?" by specifying which chunks of spacetime we are located. We have finished the kinematical description of LQG.
\section{Penrose spin networks, Clebsch-Gordan, ...}  
\textit{This lesson is dedicated to answering some questions, so I don't write down the details here since no new topics are introduced.}
\section{Time}
\label{sec:Lecture_10_time}
We first discuss the time, emphasizing some crucial conceptual points not to be confused about LQG. Then we talk about the Hamiltonian, the fundamental object that describes the system's evolution. In the first part, we will follow the same approach we used throughout the discussion of the space in the second lecture \ref{Lecture_2_space}.
\subsection{Concepts of Time}
A lot of confusion comes from the fact that with "time," as it happens for space, generally, we refer to distinct concepts. The first distinction that needs to be clarified is the concept of time that arises when referring to "When?". We answer by saying "tomorrow, yesterday, etc." i.e., we locate events one respect to the others temporally. The physics world is a world of events. Aristotle also describes the time concept in detail: "Time is the number of changes with respect to the before and the after." So it is a way of numbering changes, which means that \underline{if nothing changes, then there is no time}. Or, equivalently, if nothing happens, there is no time. If we were in a dark room with all objects at rest, then the "changing events" would be the processes in our minds. We can summarize by saying that, according to Aristotle, time is "numbering" processes events. 

\medskip

After that, Newton came up with his new concept of \textbf{Newtonian time} and, once again, he states that the "common time" effectively exists, but there is \underline{also} the Newtonian time $t$, which is a variable that passes by itself, \underline{whether or not nothing else changes}. We again emphasize that, according to Newton, the two concepts coexist. Quite importantly, Newton strongly pointed out that the $t$ variable is not directly observable: we tend to think that Newtonian time coincides with our direct perception of time, while Newton explicitly says that this time cannot be computed, perceived, or detected out of the movement. When we compute the movement of objects with respect to this time, we can write beautiful (Newtonian) equations of motion. This means the very important fact that the clocks are just "processes" designed to be as close as possible to the Newtonian idea of time. Summarizing, \underline{Newtonian time is not directly observable}, but turns out to be approximated by real clocks. 

\medskip

Once again, when Einstein introduced special relativity in 1905, he merged the Newtonian concepts of space and time in the Minkowski space. 

\medskip

In 1915, with GR, we already pointed out that space and time become a field $g_{\mu \nu}(x)$, and a clock becomes an object whose dynamic is such that it carries out, along its worldline $\gamma$, a feature of the gravitational field. A clock in spacetime moves with a little pendulum, ticking at units of its own time, which is just a measure of an aspect of the gravitational field $g_{\mu \nu}(x)$. Namely, the time measured by the clock during the motion along its worldline $\gamma$:
\begin{equation}
T_{\gamma} = \int_{\gamma} \sqrt{g_{\mu \nu} dx^{\mu} dx^{\nu}} \ .
\end{equation}
The clock, therefore, measures an entity, and once again, Newton correctly intuited the latter. But this entity is not "above" all the other things of the world since it is just "a thing" of the world, i.e., the gravitational field. Newtonian time is, therefore, just reduced to a variable of the ingredients of the Universe. 
\subsection{Peculiarity of time}
So far, the time concept is completely analogous to the space concept, but something is still missing. If $t$ is just a variable, why does it behave so differently with respect to the other world's variables? Why, for example, can we go back in space but not in time? There are several aspects for which the time variable behaves differently. The first one is \underline{irreversibility}, and the other one is that time \underline{flows}: we cannot "remember" the future, as we do not compute probabilities in the past. This topic would require a very long discussion, so we sketched it. The key observation is that "irreversibility" is not in mechanics but is present in thermodynamics. Thus, all phenomena in which there is no heat (or something analogous) are reversible. We understand quite well the second principle of thermodynamics up to an important point. The mystery to clarify is why entropy was low in the past since nobody has satisfactorily explained this. \textit{See lecture from minute 16:41 to 21:50 for interesting considerations regarding this aspect}. The only possible answer to the question "Why do not we remember the future?" is that \underline{entropy was low in the past}. This is a big mystery concerning the second law of thermodynamics. Nothing else allows us to explain why we do not have future memories. Also, in QM, a lot of confusion arises regarding this topic. The truth is that also this theory is "blind" to the direction of time precisely as classical mechanics. As Einstein wrote in a paper, we remember the past just because we have traces of it and continuously use it. Still, if we used the QM to calculate probabilities in the past, we would obtain probabilities for it and the future. Traces in the past, which make us perceive the past as "fixed" and the future as "open," exist because entropy was low in the past, i.e., there is a gradient of entropy in time. The brain uses memories of the past to  predict the future, and this is precisely the "flow" of time that we perceive. This has nothing to do with quantum gravity, classical mechanics, or quantum mechanics: it concerns our brain, statistical mechanics, and thermodynamics. Disentangling these concepts is crucial to understanding that asking for a "flow" of time in quantum gravity is incorrect since this is just a brain's perception of the world. For example, we tend to think a red T-shirt is more vivid than a grey one. Still, from a physical point of view, they have two different frequencies: the reason for which we perceive them differently in terms of showiness relies upon our brain (of course, also, this aspect could maybe be treated with physics, but we are still not smart enough). In quantum gravity, we have events, we can number them, and the convenient way to do that is to use proper time. The proper time is computed along a world line, i.e., is not global: general relativity teaches us that there is no "common time" in the Universe, and the notion of "simultaneity" is just an approximation. What we can compute in classical theory is the evolution of a state not in time but of the variables described in one respect to the others. In quantum  theory, given some values, we can compute probabilities of other values determining how they are related to one another, one of which could be proper time (if we have a clock measuring it). 

\medskip

Summarizing, this is the structure that we have briefly discussed: no preferred time variable or direction, no flow but yes events, yes local clock times, and yes gravitational field.
In GR, given initial data on an initial three-dimensional surface at a fixed time, classically, we can predict the evolution of the data on a final three-dimensional surface (which lies within the light cone of the initial surface in spacetime). From a quantum point of view, we can do the same by computing the probabilities of the data on the final surface. This is precisely how we will work in QG, namely, computing amplitudes $| W \rangle $ that tell us probabilities of obtaining a final quantum state $| fin \rangle $ given an initial quantum state $| in \rangle $: $ \langle W | \left( | in \rangle \otimes  | fin \rangle \right)$.
\begin{figure}[h]
\begin{center}
\includegraphics[width=4cm]{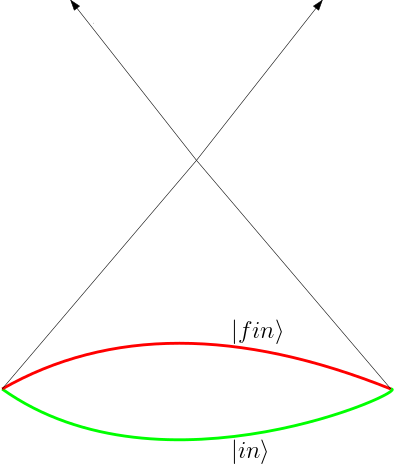}
\caption{\textit{Three-dimensional space-like surfaces: initial (green) and final (red), on which initial and final states are defined. The final surface lies within the light cone of the initial one.}}
\end{center}
\end{figure}
The spin networks sit on the initial state and the final one, so we will concretely compute probabilities associated with them. The tensor product of the initial and the final state $| in \rangle \otimes  | fin \rangle$ is called the \textbf{boundary Hilbert space} $\mathcal{H}_{boundary}$ and the dynamics will tell us how to compute amplitude on every boundary state. So, in other words, the dynamics is just a bra of the Hilbert space we have already described. We do not need time since the states include a gravitational field, and any clock is a function of the gravitational field. \textit{From minute 34:30 to 37:30, Rovelli illustrates how to decline these concepts in a concrete scattering experiment}.
\subsection{Hamilton function}
Let's talk about the Hamilton function, which plays a role in formalizing these concepts. The Hamilton function is a function of some variables which can be defined, in classical mechanics, as follows: imagine we have a system described by a variable $q(t)$ and equations of motion given in terms of a Lagrangian $\mathcal{L}(q, \dot{q})$. Therefore we have an action:
\begin{equation}
S[q] = \int dt \mathcal{L} \left( q(t), \dot{q}(t) \right) \ .
\end{equation}
If we vary the action and put it to zero, we obtain the equations of motion, which have some solutions $q(t)$ that satisfy the Euler-Lagrangian equations. If this is a "standard situation," the solutions are uniquely determined by initial positions $q_0$ and velocity $\dot{q}_0$, or, up to degeneracy, they can be determined by initial positions and final positions $q_f$. 
\begin{figure}[h]
\begin{center}
\includegraphics[width=5cm]{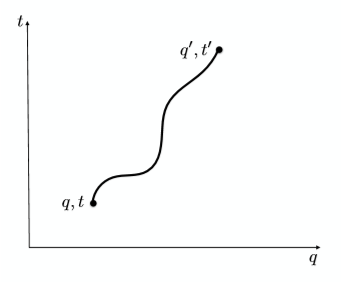}
\caption{\textit{Taken from page 32 of the book. In this picture of course $q' \equiv q_f$, $t' \equiv t_f$}.}
\end{center}
\end{figure}
Let us choose an initial position at an initial time $(q_0,t_0)$ and a final position at a final time $(q_f, t_f)$. Generically, let's say there is only one solution of the equations of motion going from the initial to the final state (for a free particle, this is precisely the case). So, given four numbers, there is just one solution of the equations of motion $q_{q_0,t_0,q_f,t_f}(t)$ and if we insert it in the action of the system, we obtain the \textbf{Hamilton function}:
\begin{equation}
S(q_0, t_0, q_f, t_f) = \int dt \mathcal{L} \left[ q_{q_0,t_0,q_f,t_f}(t), \dot{q}_{q_0,t_0,q_f,t_f}(t) \right] \ .
\end{equation}
Contrary to the action, the Hamilton function is not a functional of the full trajectory (giving a number for any function $q(t)$) but is just a function of four variables. Therefore the two objects are related: the Hamilton function is the action of a particular trajectory determined by its boundaries. Mr. Hamilton said that if we know this function, we can generally compute the equations of motion. Knowing the Hamilton function of a system is, therefore, equivalent to knowing the solution of its equations of motion. This happens because the Hamilton function has some beautiful properties, and we spend some time on them since they are essential. 
\begin{itemize}
\item The derivative of the Hamilton function with respect to the initial position gives the initial momentum, whereas the same relation holds between the final position and the final momentum (except for a sign):
\begin{align}
\frac{\partial S(q_0, t_0, q_f, t_f)}{\partial q_0} & = - p_0(q_0, t_0, q_f, t_f) \ ,
\label{Momenti coniugati ottenuti con la funzione di Hamilton Lecture 10} \\
\frac{\partial S(q_0, t_0, q_f, t_f)}{\partial q_f} & = p_f(q_0, t_0, q_f, t_f) \ .
\end{align}
\item The derivatives with respect to time turn out to be related to the energy of the system:
\begin{align}
\frac{\partial S(q_0, t_0, q_f, t_f)}{\partial t_0} & = E_0 \ ,
\\
\frac{\partial S(q_0, t_0, q_f, t_f)}{\partial t_f} & = - E_f \ .
\end{align}
\textit{I took these equations from formulas (2.10) and (2.14) of the book, in which it is explained how to derive them}.
\item Just by looking at \eqref{Momenti coniugati ottenuti con la funzione di Hamilton Lecture 10}, it emerges that knowing the Hamilton function is equivalent to knowing the equations of motion, since if $p_0(q_0, t_0, q_f, t_f)$ is known then we can invert this equation to obtain $q_f(q_0,p_0,t_0,t_f)$ that is the final position as a function of initial data and time, i.e., the solution. So, Mr. Hamilton was right. This peculiar formulation treats time and space $(q,t)$ on the same ground. At the same time, in the Lagrangian formalism, $t$ is the evolution parameter (namely the parameter in which we are evolving our system), while $q$ is variable, so they have a different status. In Hamiltonian formulation, it is unnecessary to specify the time and the position. This formulation, therefore, captures everything we need. 
\item How can we find the Hamilton function without going through the Lagrangian? Another property of the Hamilton function is being a solution of the Hamilton-Jacobi equation, so we can directly characterize a system through the latter:
\begin{equation}
\frac{\partial S}{\partial t} = - H \left( q, \frac{\partial S}{\partial q} \right) \ .
\end{equation}
\textit{I wrote $(q,t)$ to state that both $(q_0,t_0)$ and $(q_f,t_f)$ can be replaced in the Hamilton-Jacobi equation since the latter is a solution in both the set of variables}.
We can rewrite the latter in a slightly different way, treating $q$ and $t$ on the same ground by writing:
\begin{equation}
C \left( q, t, \frac{\partial S}{\partial q}, \frac{\partial S}{\partial t} \right) = 0 \ ,
\end{equation}
where $C \left( q,t,S_q,S_t \right) \equiv S_t + H(q,S_q)$. The previous equation is how the Wheeler-DeWitt equations are usually written, so it is not necessarily at all to treat time and space on different grounds. This is exactly the case of GR, in which time (the proper time of objects moving in spacetime) is "hidden" in the metric, so we don't have it separated from the other variables. In QM, we can do the same by writing transition amplitudes $W$ that, in the semi-classical limit, are given by the exponential of the Hamilton function:
\begin{equation}
W(q_0, t_0, q_f, t_f) \sim e^{\frac{i}{\hbar}S(q_0, t_0, q_f, t_f)} \ .
\label{Ampiezza di transizione nel caso semiclassico Lecture 10}
\end{equation}
The quantity $W$ can be seen as a function of all variables without distinguishing the dependent/independent variables. So this is how we can do both classical and quantum mechanics, i.e., simply describing variables evolving with respect to each other without referring to a "privileged" time. This is precisely the way QG work. We will construct these variables explicitly for the spin network. \textit{At the end of the lecture, Rovelli answers some questions}.
\end{itemize}
\section{Structure of timeless mechanics}
\label{sec:Lecture_11_structure_timeless_mechanics}
By definition, all system variables $q_i$ belong to the configuration space $\mathcal{C}$. Dynamics aim to understand how they evolve in time, i.e., to determine $q(t)$. Evolution is determined by the position and the momenta $(q_i p^i) \in \Gamma$ at some time, where $\Gamma$ is the phase space. The dynamic motion is described by an \textbf{Hamiltonian}, which is a function of the phase space $H(q_i p^i)$ that defines evolution through Hamilton's equations. Let's say that in quantum theory, where $(q_i,p^i)$ become operators, exists a maximal set of not commutative eigenstates $| q_i \rangle$ labeled by quantum numbers $q_i$ (as it happens for a particle, we say that this set is "complete" in the sense of Dirac). Then, quantum dynamics is captured by a transition amplitude from an initial to a final state: 
\begin{equation}
W(q^{in}_i, t_{in}, q^{fin}_i, t_{fin}) = \langle q_i^f | e^{-\frac{i}{\hbar}(t_f - t_i)H} | q_i^{in} \rangle \ .
\end{equation}
\textit{It is obvious that the upper suffix $in$ in $q^{in}_i$ means "initial"}. The latter can be written as a path integral over history. In a limit where we can take $\hbar$ small (i.e., the system's action is small compared to $\hbar$), the functional integral is dominated by its classical trajectory. In this case, the amplitude is proportional to the following:
\begin{equation}
W \sim e^{\frac{i}{\hbar}S[q_i^i, t_i, q_i^f, t_f]} \ ,
\end{equation}
where $S[q_i^i, t_i, q_i^f, t_f]$ is the classical action of the physical trajectory, and we defined it as the Hamilton function during the last lecture. This is what we already pointed out in eq. \eqref{Ampiezza di transizione nel caso semiclassico Lecture 10}, so the Hamilton function also gives (at a first approximation) the transition amplitudes. 

\medskip

We spent some time discussing classical and quantum mechanics with time, but we know that in GR, if we move two clocks, afterward, they do not sign the same time, so which one is the "correct" time? It is well known that in GR, there is no preferred time. Therefore, we need a formulation of physics in which we do not start from a privileged (Newtonian) time variable $t$: we need to change the above structure by taking $t$ and treating it on the same ground with respect to the $q_i$ variables.
\subsection{Extended configuration and phase space} 
We introduce a new "extended" configuration space $\mathcal{C}_{ext}$, having a new set of variables $q_a$, that in the Newtonian case would be the set $(q_i,t)$. Here we do not distinguish them. It turns out that $i=1...N$ where $N$ is the number of degrees of freedom, while $a=1...N+1$. To extend the previous formalism, including also conjugated momenta to the variables, it is necessary to introduce an extended phase space $\Gamma_{ext}$ with elements $(q_a,p^a) = (q_i,t,p^i,p_t)$, where $p_t$ is the extra variable conjugated to time $t$, that turns out to be related to $-E$, i.e., minus the energy of the system. The Hamiltonian does not give the dynamics in this language, but by a single function $C(q_a,p^a)$ which has different names: \underline{relativistic Hamiltonian} or \textbf{Hamiltonian constraint}. The \textbf{relativistic Hamilton-Jacobi equation} is written as:
\begin{equation}
C \left( q_a, \frac{\partial S}{\partial q_a} \right) = 0 \ .
\label{Equazione di Hamilton Jacobi relativistica Lecture 11}
\end{equation}
If we know the solution, we can write the solution of the equations of motion. In particular, a family of solutions can be written from the Hamilton function $S(q_a^i, q_a^f)$ by requiring that the latter is a solution of eq. \eqref{Equazione di Hamilton Jacobi relativistica Lecture 11} in both the initial and final sets of variables. This determines the evolution of the system according to the following:
\begin{equation}
\frac{\partial S(q_a^i, q_a^t)}{\partial q_a^i} = - p^a_i(q_a^i, q_a^f) \ .
\end{equation}
We thus get a relation between initial positions, initial momenta, and final positions. So, we need a relation between all the final variables for every set of initial positions and momenta in $\Gamma_{ext}$. We can rewrite eq. \eqref{Equazione di Hamilton Jacobi relativistica Lecture 11} as:
\begin{equation}
C \left( q_a,p^a \right) = 0 \ .
\label{Hamilton Jacobi relativistica con momenti Lecture 11}
\end{equation}
So all motions we consider are described by points in the extended phase space, which are in the subset of it determined by the Hamiltonian constraint \eqref{Hamilton Jacobi relativistica con momenti Lecture 11}. Another way of viewing this is that the symplectic space $\Gamma_{ext}$ (i.e., there are Poisson brackets defined on it between the conjugated variables $q_a$ and $p^a$) when we write eq. \eqref{Hamilton Jacobi relativistica con momenti Lecture 11}, we are defining a surface within $\Gamma_{ext}$ called \textbf{constrained surface}. In a symplectic space, when we have a constrained surface, the symplectic form reduced on the surface forms "flows directions" that define some lines on it, which in turn define the system's motions. 
\begin{figure}[h] 
\centering 
\includegraphics[width=12cm]{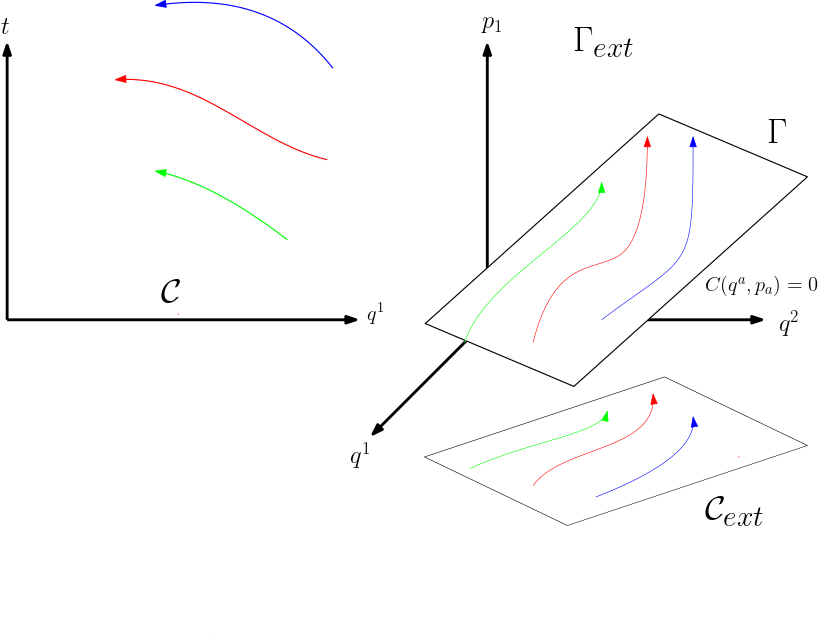}
\caption{\textit{Relation between $\Gamma_{ext}$, $\mathcal{C}_{ext}$, $\Gamma$ and $\mathcal{C}$ for a system with two coordinates $q^1,q^2$ and two momenta $p_1,p_2$ with three different motions explicitly represented. $\Gamma_{ext}$ should be 4D, so I have suppressed $p_2$. The standard configuration space $\mathcal{C}$ is one-dimensional and constituted only by the $q^1$ axis.}}
\label{Giotto}
\end{figure}
If we project it down to $\mathcal{C}_{ext}$ (which is just half of the phase space), we get motions in the extended configuration space. These motions are very similar to what we treated before in the standard configuration space $\mathcal{C}$, except that in the latter, we supposed there was a given time $t$ labeling the points. At the same time, there is no preferred (time) variable here. Trajectories in $\mathcal{C}$ were embedded in an $N$ dimensional space, while those in $\mathcal{C}_{ext}$ are in an $N+1$ dimensional space. Therefore, the relation between this configuration space and the "usual one" is very easy.  \textit{I tried to represent all these spaces graphically in Fig. (\ref{Giotto})}. The Newtonian case, in which $q_a = (q_i,t)$, occurs when the Hamiltonian constraint has the form $C(q_a,p^a) = p_t + H(q_i,p^i)$, where $H(q_i,p^i)$ is the usual Hamiltonian. In this case, $p_t$ turns out to be (minus) the energy, and the relativistic Hamilton-Jacobi equation reduces to the standard Hamilton-Jacobi equation:
\begin{equation}
\frac{\partial S}{\partial t} + H \left( q^i, \frac{\partial S}{\partial q^i} \right) = 0 \ .
\end{equation}
In particular, the Hamilton function solution of that is $S(q^{in}_i, t_{in}, q^{fin}_i, t_{fin})$, which satisfies it in both sets of variables. From this solution, we can calculate the following:
\begin{equation}
\frac{\partial S(q^{in}_i, t_{in}, q^{fin}_i, t_{fin})}{\partial q_i^{in}} = - p^i_{in}(q^{in}_i, t_{in}, q^{fin}_i, t_{fin}) \ .
\end{equation}
As discussed before, by inverting the $p_{in}^i$ function, we can extract a function $q_i^{fin}$ for each point in phase space $(q^i_{in},p_i^{in})$. There is no preferred time in the Universe, so we have to use this configuration formalism and throw away the old $t$ variable. We examine a few examples to render this formalism as concrete as possible.
\subsection{Examples of systems treated with totally covariant formulation}
\begin{itemize}
\item Example 1 (with Newtonian time): \textbf{One free particle} \\
This is the simplest example we can imagine. In the extended configuration space, it is comfortable to call the variables "space" and "time": $\mathcal{C}_{ext} = \left( q, t \right)$. The relativistic Hamiltonian is:
\begin{equation}
C = p_t + \frac{p^2}{2m} \ .
\end{equation}
The Hamilton-Jacobi equation is the following:
\begin{equation}
\frac{\partial S}{\partial t} + \frac{1}{2m} \left( \frac{ \partial S}{\partial q} \right)^2 = 0 \ .
\end{equation}
The Hamilton function, which is a solution of this equation (in both sets of initial and final variables), turns out to be:
\begin{equation}
S = \frac{m}{2}\frac{\left(q_f - q_i \right)^2}{\left(t_f - t_i \right)} \ .
\end{equation}
From the latter, we immediately get the equations of motion because if we write $\frac{\partial S}{\partial q_i} = - p_i$, then we get $m \left(q_f - q_i \right) = p_i \left(t_f - t_i \right)$ which implies:
\begin{equation}
q_f = q_f(t_f) = \frac{p_i}{m} \left( t_f - t_i \right) + q_i \ .
\end{equation}
So the Hamilton function, given an initial position and momentum, provides us a relation between $q$ and $t$ at every time. This is just a \underline{relation} between these variables: nothing forces us to think this is the evolution of $q$ in $t$! \textit{Except for the fact that I explicitly underlined it by writing $q_f(t_f)$ in the second passage, as written in eq. (2.75) of the book.}
\item Example 2 (without \underline{preferred} Newtonian time): \textbf{One free special relativistic particle} \\
We have $\mathcal{C}_{ext} \equiv \mathcal{M} $ (Minkowski spacetime), while the relativistic Hamiltonian/Hamiltonian constraint reads:
\begin{equation}
C = p^2 - m^2 
\end{equation}
It is then easy to write all the other things but we are not interested in them. Nothing prevents us from choosing $x^0 \equiv t $ (Newtonian time), but this is true only in a particular Lorentz frame. 
\begin{figure}[h] 
\begin{center}
\includegraphics[width=6cm]{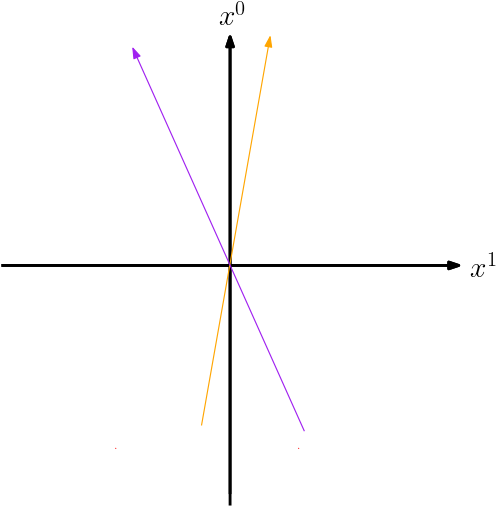}
\caption{\textit{Standard linear motions in Minkowski spacetime along the $x^1$ direction. The motion represented by the purple line is moving faster (and in the opposite direction) with respect to the orange motion.}}
\end{center}
\end{figure}
In another frame, we could choose as the time the variable (supposing we only have a motion along the $x^1$ direction) $\tilde{x}^0 = \gamma \left( x^0 + \beta x^1 \right)$. Which one is the correct time? As we have pointed out several times, there's no preferred time.
\item Example 2 (without Newtonian time at all): \textit{Example 2.5.1 of the book}  \\
The system we will study can be formulated \underline{only} in this total covariant manner, i.e., without the classical time description. Let us set $\left( a,b \right) \in \mathcal{C}_{ext}$. The Hamiltonian constraint of that is the following:
\begin{equation}
C \left( a,b,p_a,p_b \right) = \frac{1}{2}\left( p_a^2 + p_b^2 + a^2 + b^2 \right) - E \ ,
\end{equation}
where $E$ is a constant. The calculations, which can be found in the book \textit{(in section 2.5.1)}, lead to an ellipse in the $a,b$ graph.
\begin{figure}[h]   
\begin{center}   
\includegraphics[width=6cm]{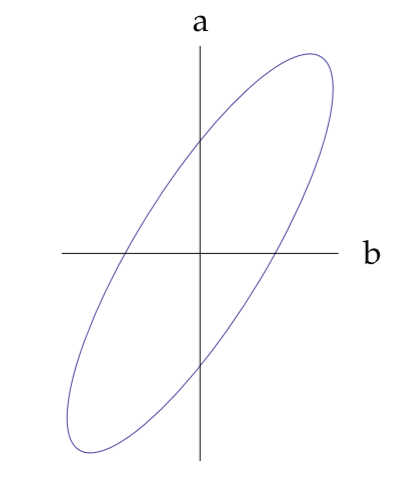}
\caption{The motion in the space (a,b). Which of the two variables is the time variable? \textit{ Fig. (2.3) of the book. } }
\end{center}
\end{figure}
In its wholeness, this motion cannot be thought of as $a(b)$ of $b(a)$ since the ellipse is not a function. Therefore this formalism is an \underline{extension} of the Newtonian one, which also includes (general) relativistic cases.
\end{itemize}
\subsection{Time parameter in timeless mechanics}
For many people, it is very hard to handle the formalism that we have just described, i.e., doing physics without referring to a "global" time. Still, unless we understand this, then QG becomes incomprehensible. We emphasize some more points about timeless formalism. If the motion in $\mathcal{C}_{ext}$ is just a "curve" which gives us a relation between the variables as $q_i(q_j)$, that states how one changes with respect to others, such as the semi-ellipses in the last example, no one can prevent us from choosing an arbitrary parameter $\tau$ and use the latter to parametrize the curve. This is what in literature is usually called \underline{parameter time}: it is not a variable we can measure. It has no particular physical meaning (if we arbitrarily choose the latter), but sometimes it is convenient. For example, in special relativity, we know that the motion of a particle is sometimes described in terms of $\vec{x}(x^0)$, where $x^0$ is the time of a clock in a particular reference frame. But we know that the more comfortable way to describe the motion is in terms of $x^{\mu}(\tau)$ where $\tau$ is an arbitrary parameter. 
\begin{figure}[h]
\begin{center} 
\includegraphics[width=6cm]{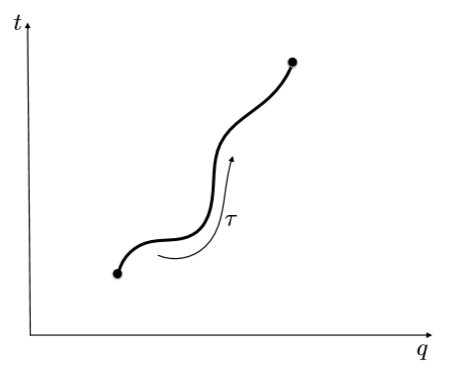}
\caption{\textit{ Taken from page 42 of the book.} }
\end{center}
\end{figure}
We can write the evolution of the action in the parameter $\tau$: 
\begin{equation}
S = m \int d \tau \sqrt{ \frac{d x^{\mu}}{d \tau} \frac{d x^{\nu}}{d \tau}\eta_{\mu \nu}} \ .
\label{Azione relativistica particella Lecture 11}
\end{equation}
\textit{At this moment, there was a technical problem, and the remaining part of the lecture was not taped. Rovelli restarts the discussion}.
We were saying that the Euler Lagrange equations, extracted by the action \eqref{Azione relativistica particella Lecture 11}, don't fix the parametrization. Actually, they are invariant under reparametrization of $\tau$ according to $\tau \rightarrow \tau'(\tau) $. If we start from this action by doing the Hamiltonian treatment, we find that the Hamiltonian is:
\begin{equation}
H = p_{\mu} \dot{x}^{\mu} - \mathcal{L} = 0 \ ,
\end{equation}
where $p_{\mu} \equiv \frac{\partial \mathcal{L}}{\partial \dot{x}^{\mu}}$. The Hamiltonian is zero, and it follows from the definition of $p_{\mu}$ that $p^2 = m^2$, so in there's the constraint $C = p^2 - m^2 = 0$ in the extended phase space $\Gamma_{ext} \equiv (p_{\mu}, x^{\mu})$. We have four variables, the $x^{\mu}$, for three degrees of freedom (one particle moving in space) $\tau$ is not a real physical parameter. According to the previous treatment, we call $C$ the relativistic Hamiltonian/Hamiltonian constraint. We must not confuse $\tau$ with the "time" parameter. When we do the GR version of the Hamilton-Jacobi equation, we will describe the metric $g_{\mu \nu}(\vec{x},t)$. Still, this $t$ is analogous to the $\tau$ parameter in this case, namely just an arbitrary parametrization of spacetime and not one of the theory's variables, which has nothing to do with the clock. We already pointed out that what a clock measures along its worldline (there is only one clock for every worldline and the other way around) is just:
\begin{equation}
T_{\gamma} = \int_{\gamma} \sqrt{g_{\mu \nu}dx_{\mu}dx^{\mu}} \ .
\label{Tempo misurato da un orologio Lecture 11}
\end{equation}
The quantity \eqref{Tempo misurato da un orologio Lecture 11} is an observable in the extended phase space. The metric components are the variables that we find in $\mathcal{C}_{ext}$, not $(\vec{x},t)$. Einstein was struggling to understand the physical meaning of $(\vec{x},t)$ since they are not (unlike special relativity) clocks and rulers. 
\section{Hamilton-Jacobi General Relativity}
\label{sec:Lecture_12_Hamilton_Jacobi_GR}
In GR, the coordinates $(\vec{x},t)$ of $g_{\mu \nu}(\vec{x},t)$ are arbitrary parameters of the motions exactly like the parameter time $\tau$ at the end of the previous lecture. The physical variables are "hidden" in the metric itself. We have seen, for example, that the value $T_{\gamma}$ measured by a clock along a worldline $\gamma$ is given by \eqref{Tempo misurato da un orologio Lecture 11}. So let's do this concretely, examining the Hamilton-Jacobi formulation for General Relativity. 
\subsection{Hamilton-Jacobi formulation for General Relativity}
This was originally done in 1962 by Perès, whose work was based on the ADM formulation of GR and others. Dirac was the first who understood that Hamiltonian mechanics needed to be generalized.
We need an extended configuration space and the general relativistic Hamiltonian (or the general relativistic Hamilton-Jacobi equation). The extended configuration space $\mathcal{C}_{ext}$ can be taken to be the space of all the three-dimensional Euclidean Riemannian metrics $q_{ab}(\vec{x})$, where $a,b = 1,2,3$. The latter is a field in three dimensions, and its physical interpretation relies upon the fact that it is a metric (we will be more precise in a moment). Out of this, the extended phase space is $\Gamma_{ext}$, with elements $\left(q_{ab}(\vec{x}), p^{ab}(\vec{x})  \right)$, of course both $q_{ab}$ and $p^{ab}$ are symmetric fields. The Hamilton-Jacobi equation is the following:
\begin{equation}
G^2 \left(q_{ab}q_{cd} - \frac{1}{2}q_{ac}q_{bd}  \right) \frac{\delta S[q_{ab}]}{\delta q_{ac}(\vec{x})}\frac{\delta S[q_{ab}]}{\delta q_{bd}(\vec{x})} - det \left( q_{ab}(\vec{x}) \right) \mathcal{R}[q_{ab}](\vec{x}) = 0 \ ,
\label{Hamilton-Jacobi relativistica della GR Lecture 12}
\end{equation}
where $\mathcal{R}[q_{ab}](\vec{x})$ is the three-dimensional Ricci scalar of the metric $q_{ab}$, while $G$ is the Newton's constant. This equation should be supplemented with another equation:
\begin{equation}
S[q_{ab}] = S[\tilde{q}_{ab}] \ ,
\label{Classe di diffeomorfismi della GR Lecture 12}
\end{equation}
where $q_{ab}$ and $\tilde{q}_{ab}$ are related by a diffeomorphism:
\begin{equation}
q_{ab}(\vec{x}) \rightarrow \tilde{q}_{ab}(\tilde{x}) = \frac{\partial x^c}{\partial \tilde{x}^a} \frac{\partial x^d}{\partial \tilde{x}^b} q_{cd}(\vec{x}(\tilde{{x}})) \ .
\end{equation}
\textit{In which $\tilde{x}$ is supposed to be a vector, so it would have been more appropriate to write $\vec{\tilde{x}}$, but I didn't want to weigh down the notation}. Eq. \eqref{Classe di diffeomorfismi della GR Lecture 12} states that $S[q_{ab}]$ is actually a function not simply of $q_{ab}$, but of the equivalence class of these fields, which turn out to be related by diffeomorphisms, so $S[q_{ab}]$ is the same independently of the way in which we write coordinates: it is a function of the geometry of the surfaces. Note that giving \eqref{Hamilton-Jacobi relativistica della GR Lecture 12} is equivalent to giving the Einstein equations (we are treating the version without matter). Again, we stress that there is no "time coordinate" in this equation since it is not needed in GR. It turns out to be only an auxiliary variable to simplify the calculation, which contains no real information. We will see in a minute that this is actually 4D GR. We note that \eqref{Hamilton-Jacobi relativistica della GR Lecture 12} is not, in general, an easy equation to solve. We now examine a particular class of solutions.
\subsection{Hamilton function of General Relativity}
We repeat that the Hamilton function is a function of the initial metric $q_{ab}^i(\vec{x})$ and the final one $q_{ab}^f(\vec{x})$: $S[q^i_{ab}(\vec{x}), q^f_{ab}(\vec{x})]$. It is a solution of eq. \eqref{Hamilton-Jacobi relativistica della GR Lecture 12} in both sets of variables, so we can determine $q^f_{ab}(\vec{x})$ as a function of $q^i_{ab}(\vec{x})$ and $p^i_{ab}(\vec{x})$. We do this in the standard way:
\begin{equation}
\frac{\delta S}{\delta q_{ab}^i} = p^{ab}_i(\vec{x}) = p^{ab}_i \left[ q_{ab}^i, q_{ab}^f \right] \ .
\label{Momenti della GR derivati dalla Hamilton function Lecture 12}
\end{equation}
Namely, we do not obtain one final metric by inverting the last equality. We know that the explicit form of $S$ is given by:
\begin{equation}
S[q^i_{ab}(\vec{x}), q^f_{ab}(\vec{x})] = \int d^4x \sqrt{det(g_{\mu \nu}(\vec{x}))} \mathcal{R} \ ,
\label{Hamilton function della GR in funzione della metrica Lecture 12}
\end{equation}
where $g_{\mu \nu}(\vec{x})$ solves the Einstein equations in the 4D space bounded by $q^i_{ab}(\vec{x})$ and $q^f_{ab}(\vec{x})$.
\begin{figure}[h]
\begin{center}
\includegraphics[width=6cm]{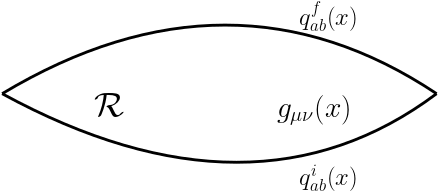}
\caption{\textit{Geometrical interpretation of the four-dimensional region $\mathcal{R}$ of spacetime, described by the metric $g_{\mu \nu}(x)$. The three-dimensional boundaries of the latter are constituted by the space-like surfaces on which the initial $q^f_{ab}(x)$ and final $q^I_{ab}(x)$ Riemannian metrics are defined.}}
\end{center}
\end{figure}
So, we can essentially face two different situations:
\begin{itemize}
\item \underline{If we know the solution $g_{\mu \nu}$ of the Einstein equations}, we can easily calculate the Hamilton-Jacobi function (given an initial and a final three-dimensional metric) and compute whatever is in between, integrating eq. \eqref{Hamilton function della GR in funzione della metrica Lecture 12} on the 4D space bounded by the metrics
\item The other way around, \underline{if we have the Hamilton-Jacobi function}, we can compute the evolution of $q^f_{ab}(\vec{x})$ given the initial metric and momenta from the relation \eqref{Momenti della GR derivati dalla Hamilton function Lecture 12}.
\end{itemize}
We note that everything is \textbf{local} in spacetime: we don't need an asymptotic limit since GR is a local theory that tells us what happens in a region of spacetime. The momentum has a geometrical interpretation related to the time derivative of the metric, which in turn is related to the extrinsic curvature of the metric itself. The latter contains information on how the hypersurfaces are embedded in the 4D spacetime. For example, if the surface is curved and it moves parallel to itself, its time derivative, of course, cannot be zero, while if it is flat, then this is precisely the case. 
\subsection{Approaches to quantization of GR}
Formally, which means with ill-defined mathematics, in quantum theory, the coordinates (namely, the three metric) $q_{ab}(\vec{x})$ and their conjugated momenta $p^{ab}(\vec{x})$ should go become not-commutative operators, realizing an algebra. So, up to ill-defined mathematics (the following lectures are devoted to defining this formal structure correctly), let's say that $q_{ab}(\vec{x})$ forms a maximal commutative set. So we expect to have something like the eigenstates of a three-geometry $| q_{ab} \rangle$, that is not simply a metric because of the relation \eqref{Classe di diffeomorfismi della GR Lecture 12}. What we want to compute effectively is the quantum amplitude related to the probability of going from an initial three-geometry to a final one:
\begin{equation}
W = \langle q^f_{ab} | q^i_{ab}  \rangle = \int \mathcal{D}g e^{\frac{i}{\hbar}\int d^4x \sqrt{g}\mathcal{R}} \ .
\label{Ampiezza quantistica come integrale funzionale Lecture 12}
\end{equation} 
We have written the amplitude formally as a functional integral over a 4D geometry. The quantity in the exponential turns out to be the Einstein-Hilbert action, which is dominated by the classical solution. Assuming that there is just one classical solution going from $q^i_{ab}$ to $q^f_{ab}$, the amplitude is well approximated by the Hamilton-Jacobi function:
\begin{equation}
\langle q^f_{ab} | q^i_{ab}  \rangle \sim e^{\frac{i}{\hbar}S\left[ q^i_{ab}, q^f_{ab} \right]}  \ .
\label{Limite semi-classico dell'Ampiezza Lecture 12}
\end{equation}
We expect this quantity to solve the quantum version of eq. \eqref{Hamilton-Jacobi relativistica della GR Lecture 12}. The relativistic Hamilton-Jacobi equation should be the iconal approximation of the quantum equation satisfied by the amplitude $\langle q^f_{ab} | q^i_{ab}  \rangle$.

\medskip

We now spend some time treating this quantum timeless version in the finite-dimensional case. We remember that in finite dimensions we had $C \left(q^a,  p_a \right) = 0$ and the Hamilton-Jacobi equation $C \left( q^a, \frac{\partial S}{\partial q^a} \right) = 0$ with solutions $S \left[q_a^i, q_a^f \right]$. The quantum transition amplitudes are given as a functional integral of trajectories over boundary conditions, indicated by ${b.\hspace{1mm} c.}$, specified by $q^i_a$ and $q^f_a$:
\begin{equation}
W \left( q^i_a, q^f_a \right) = \int_{b.\hspace{1mm} c.} d q(\tau ) e^{\frac{i}{\hbar}S[q(\tau)]} \ .
\end{equation}
This transition amplitude satisfies, in both the sets of variables, the \textbf{Wheeler-DeWitt equation}:
\begin{equation}
C \left( q_a, -i \hbar \frac{\delta }{\delta q_a} \right) W \left( q^i_a, q^f_a  \right) = 0 \ .
\label{Wheeler-DeWitt equation Lecture 12}
\end{equation}
In the special case in which the (quantum) Hamiltonian constraint has the form $C = p_{t} - H \left( p_i,q^i \right)$, the Wheeler-DeWitt equation becomes the Schroedinger equation:
\begin{equation}
\left[ i \hbar \frac{\partial}{\partial t} - H \left( q_i, i \hbar \frac{\partial}{\partial q_i} \right) \right] W = 0 \ .
\end{equation}
So the Schroedinger equation is nothing but a special case of the Wheeler-DeWitt equation, satisfied by the transition amplitudes in the general covariant formalism, in which the relativistic Hamiltonian has the "Newtonian" form. Therefore the Schroedinger equation is valid for quantum mechanics when there is a Newtonian time, but now we know its generalization, namely eq. \eqref{Wheeler-DeWitt equation Lecture 12}. So, does QM need to be changed in QG? Yes, a bit, since it has to be generalized to consider that there is no Newtonian time because the time variable is mixed with the others. 
The quantum algebra of operators that we need is the general relativistic one, with $q_a$ and $p_a$. The dynamical equation that the amplitudes satisfy is the Wheeler-DeWitt equation \eqref{Wheeler-DeWitt equation Lecture 12}, which reduces to the Schroedinger equation in the special case in which we can choose one particular variable, in which the Hamiltonian has the form $C = p_{t} - H \left( p_i,q^i \right)$. So the QM structure remains the same in the not-Newtonian case but with a more general form of the Schroedinger equation. We don't have to panic about the lack of time since we have seen that the above structure is a generalization. 

\medskip

When we formally go to quantum gravity, the transition amplitudes $\langle q^f_{ab} | q^i_{ab}  \rangle $ satisfy eq. \eqref{Wheeler-DeWitt equation Lecture 12} in which the Hamiltonian constraint is \eqref{Hamilton-Jacobi relativistica della GR Lecture 12}, where we are forced to use functional derivatives instead of ordinary ones. This is how, in 1963, DeWitt wrote the Wheeler-DeWitt equation in its first original formulation. If we define: 
\begin{equation}
G_{abcd} \equiv G^2 \left(q_{ab}q_{cd} - \frac{1}{2}q_{ac}q_{bd} \right) \ .
\end{equation}
As well as $\langle q^f_{ab} | q^i_{ab}  \rangle \equiv W \left[ q^i_{ab}, q^f_{ab} \right]$, then we can rewrite the Wheeler-DeWitt equation as:
\begin{equation}
\left[ G_{abcd} \frac{\delta}{\delta q_{ab}}\frac{\delta}{\delta q_{cd}} - g \mathcal{R} \right] W \left[ q^i_{ab}, q^f_{ab} \right] = 0 \ .
\label{Wheeler-DeWitt versione 1964 Lecture 12}
\end{equation}
\textit{From minute 27:50 to 32:17, Rovelli discusses the Copenaghen interpretation of QM}. In QG, we have values on the initial and final 3D space-like boundary surfaces: given some data on the first, we obtain a probability distribution on the second. This is spectacularly nice because QM comprises relations between systems that interact, while GR deals with interactions between contiguous spacetime regions. So, the boundaries are, of course, artificial, but they are needed to do QM. The difference is that in QM, we talk about systems interacting, whereas in QM, we deal with spacetime regions. The theory itself is "smart" enough not to depend on where we put the boundaries themselves. In other words, there is nothing physical about the boundaries since the information relies only on how the system we study affects its outside. In contrast, the latter, in turn, affects its outside (in a manner that the QM of it determines), and so on and so forth. Therefore, in QG, the \underline{system is identified with the spacetime region}, and it turns out that the interaction happens where the spacetime is contiguous (namely, at the boundaries). This is one of the most significant discoveries of the 20th century, i.e., the \textbf{locality}. To describe this quantum interaction, we need to compute the quantities $\langle q^f_{ab} | q^i_{ab}  \rangle $ and the operators $q_{ab},p^{ab}$. We have already studied these operators, but we still need to see concretely how to compute the transition amplitudes. So we are going to write these quantities $W \left[ q^i_{ab}, q^f_{ab} \right]$ explicitly, at least at some order of approximation. The condition on the quantities of the amplitudes is that, in the classical limit, they should reproduce GR, which means that "when $\hbar$ is small," we should have (at the first order):
\begin{equation}
W \left[ q^i_{ab}, q^f_{ab} \right] \sim e^{\frac{i}{\hbar}S \left[ q_{ab}^i,q_{ab}^f \right]} \ ,
\label{Ampiezza di transizione quantistica limite semi-classico Lecture 12}
\end{equation}
where $S \left[ q_{ab}^i,q_{ab}^f \right]$ is the Hamilton-Jacobi equation of GR. This is what we want. Can we define such transition amplitudes? This is what we are going to do during the next lecture. To do that, we should formally treat the functional integral \eqref{Ampiezza quantistica come integrale funzionale Lecture 12} over a 4D geometry bounded by regions with given three metrics. This is not well defined for several reasons:
\begin{itemize}
\item The measure $\mathcal{D}g$ is ill-defined, so we must write it more precisely.
\item The Wheeler-DeWitt equation \eqref{Wheeler-DeWitt versione 1964 Lecture 12} is not well defined since the product $\frac{\delta}{\delta q_{ab}}\frac{\delta}{\delta q_{cd}}$ has no mathematical meaning because $\frac{\delta}{\delta q_{ab}(\vec{x})}$ is a distribution and product of distribution has no meaning at all. We need to get around this problem in some way.
\item We have to find a way to define the operators $q_{ab},p^{ab}$. In a sort of way, we already did since we have seen that they define a "quantum geometry" of a three-dimensional region (that is, the boundary of the 4D region on which we are integrating). But this is what we did within the kinematics part when we described the Hilbert space of operators. But we must remember that the quantum amplitudes must be the quantum numbers of the operators. So, we need amplitudes not in terms of "initial classical three-geometry" and "final classical three-geometry" but from eigenstates of initial and final geometries. So the amplitudes will be functions of spin networks: 
\begin{equation}
W \left[ q^i_{ab}, q^f_{ab} \right] \longrightarrow W \left[ \Gamma^i, j^i_l,v^i_n; \Gamma^f, j^f_l,v^f_n \right] \ .
\end{equation}
In the end, given some initial and final spin networks, we want something that gives us a complex number such that in the limit $\hbar \rightarrow 0$, the latter reduces to eq. \eqref{Ampiezza di transizione quantistica limite semi-classico Lecture 12}, where $S\left[ q^i_{ab}, q^f_{ab} \right]$ is the Hamilton function of the 4D geometry. The latter is determined by the geometry approximated by the quantum discrete geometries, which bound the 4D region. We will build explicitly the quantities $W \left[ \Gamma^i, j^i_l,v^i_n; \Gamma^f, j^f_l,v^f_n \right]$ and we will see that they are finite, namely without UV divergences.
\end{itemize}
\section{K = yL, part 1}
\label{sec:Lecture_13_K_gamma_L}
Now we talk about classical GR, building the main ingredients to write the dynamical transition amplitudes. We put some order in the different formalisms we have used so far, also writing the GR action, the canonical formalism, and, most importantly, discussing a fundamental equation:
\begin{equation}
\vec{K} = \gamma \vec{L} \ ,
\end{equation}
whose meaning is going to be examined at the end of this brief review. This equation seems to play a key role in GR, coming from various angles, and we have not completely clarified its implications. 
\subsection{Classical General Relativity}
Let's discuss the main formulations of GR developed and used throughout time:
\begin{itemize}
\item \textbf{Einstein} \\
Einstein was the first one to give a formulation of GR in terms of a field $g_{\mu \nu}(\vec{x})$, which is the Riemann's metric field that Einstein turned around into pseudo-Riemannian manifolds (so, essentially it is Riemannian geometry). The main point about GR is identifying the gravitational field with this metric. Still, there is a crucial point to underline to understand quantum gravity: Einstein discovered that the spacetime metric is the gravitational field, not the other way around. If we invert the previous statement, reducing the gravitational field to be the spacetime metric, we will be confused about its quantum interpretation. If we consider the relational spacetime notion, viewing the metric as something "extra" makes it easier to interpret QG. From $g_{\mu \nu}(\vec{x})$ Einstein obtained the definition of Levi-Civita connection $\Gamma^{\mu}_{\nu \rho}(x)$ and of the Riemann tensor $R^{\mu}_{\hspace{1mm} \nu \rho \sigma}(x)$. He used these mathematical structures to write his famous equations, which he got step by step:
\begin{equation}
R_{\mu \nu} - \frac{1}{2}g_{\mu \nu}R + \lambda g_{\mu \nu} = 8 \pi G T_{\mu \nu} \ .
\end{equation}
The "cosmological constant term" $\lambda g_{\mu \nu}$ was obtained in 1917, while the rest was definitely written in 1915 (there were several versions and attempts during previous years, as well known). The term $\lambda g_{\mu \nu}$ is a fundamental theory aspect. In fact, the "biggest mistake of Einstein's life" was thinking of being able to obtain a static solution by using this term without realizing that the theory is unstable. But there is no reason to drop it, which must be inserted in the field equations. Lemaitre repeatedly told Einstein that this term was essential and that the Universe was not static. In fact, he was the first to see the data of nebulae. The GR theory itself depends on two constants: $G$ and $\lambda$. $G$ has been measured since the times of Newton, while $\lambda$ was measured only a few years ago through the expansion of the Universe. This is the reason why the term $\lambda g_{\mu \nu}$ has a not negligible effect only on large distances. As a first approximation, we will neglect it because it is much easier to work without the latter (but it also turns out to have an effect in QG). We don't consider the matter's contribution, but the book discusses how to couple this theory to matter \textit{(see chapter 3)}. The kinematics and dynamics in QG can be extended to the matter treatment, thinking that it "sits" on spin networks, as it happens for lattice QCD.
\item \textbf{Hilbert} \\
Immediately after Einstein, Hilbert wrote the field equations in terms of action, from which they can be derived, namely the Einstein-Hilbert action. We employ the geometric units system $8 \pi G = 1$, so we write:
\begin{equation}
S \left[ g \right] = \frac{1}{2} \int d^4x \sqrt{g} R \left[ g \right] \ .
\label{Einstein-Hilbert action Lecture 13}
\end{equation}
Sometimes this is called "second-order" formalism.
\item \textbf{Palatini} \\
Some years later, Palatini realized there was another way this action could be written. The action \eqref{Einstein-Hilbert action Lecture 13} is a functional of the metric field, $S \left[ g \right]$, but we can astutely consider the fields $g_{\mu \nu}$ and $\Gamma^{\mu}_{\nu \rho}$ to be independent variables in the action:
\begin{equation}
S \left[ g, \Gamma \right] = \frac{1}{2} \int d^4x \sqrt{g} g^{\mu \nu} R^{\sigma}_{ \hspace{1mm} \mu \sigma \nu} \left[ \Gamma \right] \ .
\label{Palatini action Lecture 13}
\end{equation}
By varying eq. \eqref{Palatini action Lecture 13} with respect to $\Gamma^{\mu}_{\nu \rho}$ we obtain the relation between $g_{\mu \nu}$ and $\Gamma^{\mu}_{\nu \rho}$, while varying it with respect to $g_{\mu \nu}$ gives the field equations. This is an alternative formulation of the theory, which is sometimes called "first-order" formalism. The same arguments can be applied with the Lagrangian formalism when we have $q$ and $\dot{q}$: if we consider the latter to be independent when we vary the action with respect to $\dot{q}$ we get the relation between $q$ and $\dot{q}$.
\item \textbf{Cartan and Weyl} \\
The previous formulations have several problems. For example, it is not possible to couple fermions. In fact, for this step, we actually need to change variables. This was first realized by Weyl and Cartan, who introduced tetrad formalism (independently and in different manners with respect to each other). The key idea is to use tetrads instead of $g_{\mu \nu}$ by introducing a field $e^{I}_{\mu}(x)$, where both the indices run over the same numbers $\mu, I = 0,1,2,3,$, but $I$ is a flat index, i.e., raised and lowered by Minkowski metric $\eta = (1,-1,-1,-1)$. The relation between $g_{\mu \nu}$ and $e^{I}_{\mu}(x)$ can be written as:
\begin{equation}
g_{\mu \nu} (x) = e^{I}_{\mu} (x) e^{J}_{\nu} (x) \eta_{I J} \ .
\label{Relazione fra tetradi e metrica Lecture 13}
\end{equation}
If we consider the action \eqref{Einstein-Hilbert action Lecture 13} and use eq. \eqref{Relazione fra tetradi e metrica Lecture 13} we get:
\begin{equation}
S \left[ e \right] = \frac{1}{2} \int det(e) R \left[ e \right] d^4x \ .
\end{equation}
By varying this action, we discover that tetrad fields satisfy Einstein equations, as the metric field does. Nevertheless, we have introduced some redundancies. In fact, by making a Lorentz transformation on the flat indices in eq. \eqref{Relazione fra tetradi e metrica Lecture 13}:
\begin{equation}
e_{\mu}^I(x) \rightarrow \Lambda^I_{\hspace{1mm} J} (x)e^J_{\mu}(x) \ ,
\end{equation}
where $\Lambda^I_{\hspace{1mm} J}(x) \in SO(3,1)$ is a matrix of the Lorentz group, it turns out that $g_{\mu \nu}$ remains the same, so there are more fields within the tetrad formulation. 
\begin{figure}[h] 
\begin{center}
\includegraphics[width=5cm]{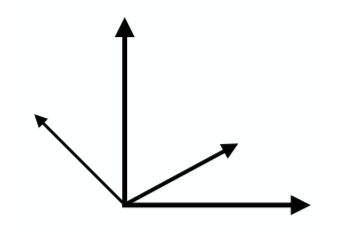}
\caption{\textit{ Taken from page 61 of the book. } }
\end{center}
\end{figure}
In fact, this formalism actually introduces a \underline{local Lorentz invariance}. We can thus "rotate" the 4D frame at every point of spacetime. This is a much more fundamental formulation of GR with respect to previous ones since, as we pointed out, it allows coupling to fermionic matter. Technically speaking, the formulation with the metric $g_{\mu \nu}$ turns out to be not completely correct because fermions effectively exist in nature, therefore the Weyl-Cartan description is definitely more general.
\item \textbf{Tetrad first order} \\
Of course, we can repeat the "Palatini trick" by considering the tetrad field and the associated connection as independent field variables. We don't write explicitly the equations for its connection $\omega_{\mu}^{IJ}(x)$, which turns out to be directly related to the Levi-Civita one. Since the connection is in the algebra, we can say that the latter is actually an $SL(2,\mathbb{C})$ or a $SO(1,3)$ connection (the algebra of these groups is the same). The variables are the tetrad fields $e^I_{\mu}(x)$ and its connection $\omega_{\mu}^{IJ}(x) = -\omega_{\mu}^{JI}(x)$, which is a 1-form with a lower spacetime index and two upper flat Lorentz antisymmetric indices in the $SL(2,\mathbb{C})$ algebra. We write the GR action as
\begin{equation}
S \left[ e, \omega \right] = \frac{1}{2} \int det(e) e^{\mu}_I e^{\nu}_{J} F^{IJ}_{\mu \nu} \left[ \omega \right] d^4x \ ,
\end{equation}
where $F^{IJ}_{\mu \nu}$ is the standard curvature (the same object we use in linear theory), which turns out to be a function of the connection:
\begin{equation}
F^{IJ} \left[ \omega \right] = d \omega^{IJ} + \omega^I_{\hspace{1mm} K} \wedge \omega^{KJ} \ .
\end{equation}  
This is nothing but GR with explicit $SL(2,\mathbb{C})$ or $SO(1,3)$ gauge invariance. \textit{In order to rewrite the action in a previous way, it is useful to use equation (3.15) of the book}. It is very much a Lorentz connection with an additional connection, similar to a Yang-Mills theory. The action can be rewritten by considering that $e^{\mu}_I(x)$ and $e_{\mu}^I(x)$ are opposite $4 \times 4$ matrices, therefore:
\begin{equation}
S \left[ e, \omega \right] = \frac{1}{2} \int  
\epsilon_{IJKL} \epsilon^{\mu \nu \rho \sigma} e_{\rho}^K e^L_{\sigma} F^{IJ}_{\mu \nu} d^4x \ .
\end{equation}
This is useful since it allows to rewritten $e^I$ and $\omega^{IJ} = \omega^{IJ}_{\mu}dx^{\mu}$ as a 1-form, namely $e^{I} = e^{I}_{\mu}dx^{\mu}$ and $F^{IJ}$ as a two form: $F^{IJ} = F^{IJ}_{\mu \nu}dx^{\mu} dx^{\nu}$. We see therefore that the action can be rewritten nicely and cleanly using the wedge product:
\begin{equation}
S \left[ e, \omega \right] = \frac{1}{2} \int e^{K} \wedge e^{L} \wedge F^{IJ} \epsilon_{IJKL}  \ .
\end{equation}
We now define, as usually the Hodge dual tensor $\frac{1}{2} F^{IJ} \epsilon_{IJKL} \equiv  \left( F^{*} \right)_{KL}$. \textit{I follow the convention of the book, see pages 62-63}. When we have tensors contracted within the very same algebra, indices are typically omitted:
\begin{equation}
S \left[ e, \omega \right] = \int e \wedge e \wedge F^{*} \ . 
\label{Azione GR con F star Lecture 13}
\end{equation}
Now we have a beautiful GR action \eqref{Azione GR con F star Lecture 13}, which is quite general. This will not be, however, the action used in QG since there is a final step to do.
\item \textbf{Ashtekar Barbero Holst} \\
Ashtekar was the first one who found a very interesting and intriguing canonical transformation in the phase space of GR. Since canonical transformations are equivalent to adding terms to the action, people now think of its transformation by adding a term in  eq. \eqref{Azione GR con F star Lecture 13}. The piece he added leads to new total action:
\begin{equation}
S \left[ e, \omega \right] = \int \left( e \wedge e \wedge F^{*} + \frac{1}{\gamma} e \wedge e \wedge F \right) \ ,
\label{Azione QG con pezzo di Ashtekar Lecture 13}
\end{equation}
where $\gamma$ is a constant. This does not change the theory since \underline{on the solution} of equations of motion, the term $ \frac{1}{\gamma} e \wedge e \wedge F $ turns out to be just a boundary term. So, varying the action with or without it gives exactly the same classical equations of motion and $\gamma$ drops out of the latter. \textit{From minute 25:33 to 27:38 Rovelli explains how to visualize this feature directly just by looking at the action}. Therefore, classically this is still GR, but from a quantum point of view we are changing the theory: there is a one-parameter $\gamma$ family of possible actions, giving the same classical equations of motion. It will turn out that in the limit $\gamma \rightarrow \infty$ we get trouble with quantum theory, while the action \eqref{Azione QG con pezzo di Ashtekar Lecture 13} is one that we will use in LQG. \textit{At minute 28:38 Rovelli explains how to think to the $\gamma$ term with respect to lattice QCD. At 31:25, he talks about interesting discoveries made by Perez regarding the possible terms that one can add to this action, which could actually explain why GR is not renormalizable. Finally, from 331:52 to 35:37 Rovelli does some excursus on $f(R)$ theories}. Let us rewrite eq. \eqref{Azione QG con pezzo di Ashtekar Lecture 13} in a slightly different form, by shifting the star on the $e$:
\begin{equation}
S \left[ e, \omega \right] = \int \left[ \left( e \wedge e \right)^{*} + \frac{1}{\gamma} \left(e \wedge e \right) \right] \wedge F \ .
\end{equation}
This can also be written as:
\begin{equation}
S \left[ e, \omega \right] = \int B \wedge F \ ,
\label{Azione LQG Lecture 13 con B}
\end{equation}
where we introduced the 2-form $B_{\mu \nu}^{IJ}$ with values in the Lorentz algebra $SL(2,\mathbb{C})$, while $F$ is still the curvature for the Lorentz connection. $B$ is constrained to have the form: 
\begin{equation}
B =  \left( e \wedge e \right)^{*} + \frac{1}{\gamma} \left(e \wedge e \right) \ .
\label{Condizione su B Lecture 13}
\end{equation}
An action written in  form \eqref{Azione LQG Lecture 13 con B} is called a \textbf{BF action} and plenty of properties are known about these theories, including their quantization procedure (which consists, essentially, in quantizing the two fields separately). Of course, for arbitrary $B$, this does not coincide with GR. Therefore we discover that GR can be recast as a BF theory plus the condition \eqref{Condizione su B Lecture 13} on $B$. \textit{From minute 39:50 to 41:13 Rovelli briefly describes BF theories in the general case}. 
\end{itemize}
\subsection{Time gauge}
Some interesting relations emerge from the condition \eqref{Condizione su B Lecture 13}. We want to study the theory on a compact 4D spacetime region $\mathcal{R}$, bounded by a three-dimensional surface $\Sigma$ \textit{(see next lecture for a graphical representation)}. In the Einstein formulation with the metric $g_{\mu \nu}(\vec{x},t)$, when we are on the upper surface we can always choose coordinates such that, locally, the latter is described by $t = const.$ and the $\vec{x}$ coordinates describe it. So every surface is labeled by a different value of $t$, and locally it is always possible to choose a gauge such that:
\begin{equation}
g_{\mu \nu} =
\begin{pmatrix}
1 & 0  \\
0 & q_{ab}
\end{pmatrix} \ . 
\end{equation}
where $g_{ab}$ is the three-metric induced on the boundary surface by the 4D metric. This is called \textbf{time gauge} (in the metric formalism). We have the initial three-metric $q_{ab}^{in}$ on the lower boundary surface and the final one $q_{ab}^{fin}$ on the upper. The rest is fixed. 
\section{K = yL, part 2}
\label{sec:Lecture_14_again_K_gamma_L}
In the tetrad formalism, we have $e_{a}^{I in}(x)$ and $e_{a}^{I fin}(x)$ defined on the lower and the upper boundary surfaces of the compact spacetime region we are considering. We can also choose the very same time gauge in the tetrad formalism:
\begin{equation}
e_{a}^{I} = 
\begin{pmatrix}
1 & 0  \\
0 & e_{a}^i
\end{pmatrix} \ ,
\label{Time gauge tetrad Lecture 14}
\end{equation}  
where $a,i = 1,2,3$ and $e_{a}^i$ are the triad fields, discussed in lecture 6. Since tetrads are generally not symmetric, we must impose two different conditions in \eqref{Time gauge tetrad Lecture 14}. The time gauge in tetrad formalism is actually stronger since here we don't have just the lapse function equal to one and the shift vector equal to zero, but we are \underline{"orienting" the Lorentz frame}. In fact, if we sit on a boundary three-surface, the latter defines a frame for us: the condition of being on such a surface picks up a preferred sub-group of $SO(3,1)$, where $t$ is fixed by the three-surface we are sitting on. Therefore we have the breaking of Lorentz invariance:
\begin{equation}
SO(3.1) \longrightarrow SO(3) \ .
\end{equation}
Equivalently, in terms of algebra:
\begin{equation}
\mathbf{sl}(2,\mathbb{C}) \longrightarrow \mathbf{su}(2) \ .
\end{equation}
This symmetry breaking is a characteristic of the fact that we are sitting on the boundary. Now we can take the 2-form $B$ \eqref{Condizione su B Lecture 13} and restrict it on $\Sigma$: 
\begin{equation}
B \longrightarrow B|_{\Sigma} \equiv B_{\Sigma} \ .
\end{equation}
In terms of coordinates, we are just looking at the "spatial" $B$ components sitting on the boundary surfaces:
\begin{equation}
B_{\mu \nu}^{IJ} \longrightarrow B_{ab}^{IJ} \ .
\end{equation}
If we have, in a 4D spacetime, a three-dimensional surface on which are defined coordinates $(\sigma^1, \sigma^2, \sigma^3)$, we always have a 1-form defined on the surface itself that can be written abstractly written as:
\begin{equation}
n_{\mu} = \epsilon_{\mu \nu \rho \sigma} \frac{dx^{\nu}}{d \sigma^1} \frac{dx^{\rho}}{d \sigma^2} \frac{dx^{\sigma}}{d \sigma^3} \ ,
\end{equation}
where $x^{\mu}(\sigma^1, \sigma^2, \sigma^3)$ is the embedding map of the surface within the spacetime. If we use the particular coordinates in which the surface is identified by the condition $t = const$, then $(x^1,x^2,x^3) = (\sigma^1, \sigma^2, \sigma^3)$, therefore the 1-form in coordinate form turns out to be:
\begin{equation}
n_{\mu} = (1,0,0,0) \ .
\label{1-forma ortogonale alle ipersuperfici Lecture 14}
\end{equation}
But, since we have a tetrad, from this 1-form, we can define an object with Minkowski indices:
\begin{equation}
n_{\mu} \longrightarrow n_I = e^{\mu}_I n_{\mu} \ .
\end{equation}
While $n_{\mu}$ is a geometrical object, $e^{\mu}_I$ is a field, so the product of the two gives us an object in the Minkowski space defined in every point. 
\begin{figure}[h] 
\begin{center}  
\includegraphics[width=9cm]{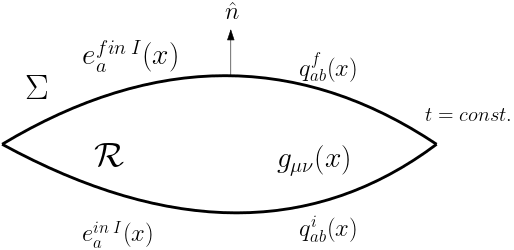}
\caption{\textit{Geometrical interpretation} }
\end{center}
\end{figure}
With this, we can break $B$ into two separate parts:
\begin{equation}
B_{\Sigma}^{IJ} = 
\bigg \{
\begin{array}{rl}
B_{\Sigma}^{IJ} n_J \equiv K^I \\
\left( B_{\Sigma}^* \right)^{IJ}n_J \equiv L^I \\
\end{array} \ .
\label{L e K Lecture 14}
\end{equation}
\textit{I report directly the definitions taken from eq. (3.42) of the book}. In coordinate notation and in the time gauge, with coordinates $(x^1,x^2,x^3) = (\sigma^1, \sigma^2, \sigma^3)$, we can write: 
\begin{align}
n_I & = (1,0,0,0) \ ,
\\
B_{\Sigma}^{IJ} & = 
\bigg \{
\begin{array}{rl}
 K^I & = B^{I0} = (0, \vec{K}) \\
 L^I & = \frac{1}{2} B^{KL}\epsilon_{KL0}^{\hspace{6mm}I} = (0, \vec{L}) \\
\end{array} \ .
\end{align}
The $4 \times 4$ matrix $B^{IJ} $ is written in terms of $\vec{L}$ and $\vec{K}$ in the same way in which the E.M. tensor $F^{\mu \nu}$ is written in terms of the magnetic $\vec{E}$ and electric $\vec{B}$ fields, respectively. So, since the Lorentz symmetry is broken, $B$ is shifted in an electric and a magnetic part. Still, the two components are separated only in the specific reference frame determined by our being on the three-dimensional boundary surface. In other frames, the two components are mixed in general. This will play a great role in the future. By inspecting the expression $B \wedge F$, remembering that $F = d \omega + \omega \wedge \omega$, we see that the conjugate variable to $\omega$ turns out to be precisely $B$. In Yang-Mills canonical theories, the conjugate variable to the connection sits in the algebra. Its Poisson bracket gives the algebra, so $B$ generates $SL(2,\mathbb{C})$ transformations. Therefore in canonical theory, $B$ is gonna be the generator of $SL(2,\mathbb{C})$ transformations, while $\vec{L}$ and $\vec{K}$ play the role of generators of rotations and boosts. We again emphasize that, in an arbitrary frame, it is generally impossible to separate rotations from boosts since the latter are mixed. We have defined $L^I$ and $K^I$ as \eqref{L e K Lecture 14}, but we still have to consider the constraint \eqref{Condizione su B Lecture 13} on $B$. \textit{Rovelli starts doing an explicit calculation of $L^I$ and $B^I$ at minute 13:18. It is easy using coordinates, remembering expressions \eqref{Time gauge tetrad Lecture 14} and \eqref{1-forma ortogonale alle ipersuperfici Lecture 14}. I only report the results here:}
\begin{align}
K^I & = 
\bigg \{
\begin{array}{rl}
 K^0 & = 0 \\
 K^i & = \epsilon^i_{\hspace{1mm}jk} e^j \wedge e^k \\
\end{array} \\ 
L^I & = 
\bigg \{
\begin{array}{rl}
 L^0 & = 0 \\
 L^i & = \frac{1}{\gamma} \epsilon^i_{\hspace{1mm}jk} e^j \wedge e^k \\
\end{array} \ .
\end{align}
From the previous equations, we can extract the following fundamental relation:
\begin{equation}
\vec{K} = \gamma \vec{L} \ ,
\label{Equazione cardine Lecture 14}
\end{equation}
where $\vec{K}$ are the boosts and $\vec{L}$ are the rotations. Summarizing, we can say that on the boundary, $B$ has an electric and a magnetic part. The constraint of $B$ implies that the latter is proportional, with a constant $\gamma$. So, GR is an $SL(2,\mathbb{C})$-BF theory where, on the boundary, the generator of $SL(2,\mathbb{C})$ (i.e., the generators of boosts and rotations) are proportional to each other. This \underline{entirely defines General Relativity}: the latter is the unique theory with the action that exhibits the above properties. The entire construction of quantum theory will be based on the key equation \eqref{Equazione cardine Lecture 14}, which is very interesting. We now describe its properties a little bit more in detail. Given a spacetime region satisfying Einstein equations, if we choose an arbitrary three-dimensional space-like surface and consider the above fields on the latter, \eqref{Equazione cardine Lecture 14} is always satisfied. This is trivial. The opposite is remarkable: these conditions are necessary and sufficient to define GR. Let's comment on \eqref{Equazione cardine Lecture 14}.
\subsection{Comments on $\vec{K} = \gamma \vec{L}$}
\textit{I am unsure about the formulas in this subsection, so please don't take the equations too seriously.}
\begin{itemize}
\item It is an equation involving two 2-forms, and, as we pointed out, the latter can be integrated on surfaces. For example if we integrate $\vec{L}$ on a triangle $\Delta$ we have:
\begin{equation}
\int_{\Delta} L^i d^2x = \frac{1}{2} \int_{\Delta} \epsilon^i_{\hspace{1mm}jk} e^j \wedge e^k  \ .
\end{equation}
The operator we called $L^i$, defined directly on the Hilbert space of LQG, was geometrically interpreted as the integral on a triangle of the triangulation of the quantity $\epsilon^i_{\hspace{1mm}jk} e^j \wedge e^k $, up to a constant.
\begin{figure}[h] 
\begin{center} 
\includegraphics[width=5cm]{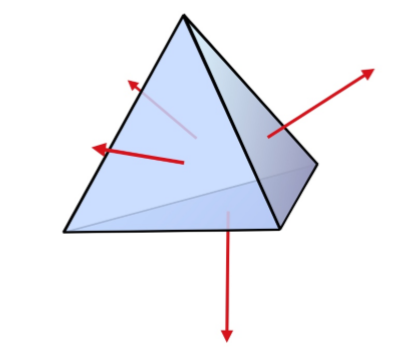}
\caption{The four vectors $\vec{L}_a$, normal to the faces. \textit{Fig (1.4) of the book.} }
\end{center}
\end{figure}
So $L^i$ has to be identified with the operators sitting on the links (dual to faces) of spin networks, up to a factor $8 \pi G$. More precisely, if we write the "old" operators as $\hat{L}^i$ in order not to confuse, the correspondence turns out to be:
\begin{equation}
L^i = \frac{1}{8 \pi G \gamma} \hat{L}^i \ .
\end{equation}
This is the oriented vector orthogonal to the face and whose length is proportional to the area, whereas the area itself is given by:
\begin{equation}
A^2 = \hat{\vec{L}} \cdot \hat{\vec{L}} = (8 \pi G \gamma )^2  \vec{L} \cdot \vec{L} \ .
\end{equation}
Therefore $\gamma$ enters the eigenvalues of the area. Putting back all the constants, the latter can be written as:
\begin{equation}
A = 8 \pi G \hbar \gamma \sqrt{j(j+1)} \ .
\end{equation}
So this fixes the argument we left open before by using precisely the Poisson bracket given by the action \eqref{Azione QG con pezzo di Ashtekar Lecture 13} and reinserting all the constants. This also tells us that $\vec{L}$ is related to the area. \textit{For further details, see section 5.2 of the book}.
\item $\vec{K}$ is the generator of boosts and $\vec{L}$ turns out to be related to the area. From eq. \eqref{Equazione cardine Lecture 14} we get:
\begin{equation}
|\vec{K}|^2 = \gamma^2 |\vec{L}|^2 = \gamma^2 \frac{A^2}{\left(8 \pi G \gamma  \right)^2} \ .
\end{equation}
Thus the absolute value of the boost generator turns out to be:
\begin{equation}
|K| = \frac{A}{8 \pi G} \ .
\end{equation}
Frodden, Ghosh, and Perez found a remarkable equation called \textbf{FGP equation} that we see in a moment. If we are sitting at a distance $d$ from a black hole, to stay put, the acceleration required is proportional to $1/d$. Therefore, locally we are accelerating in a Minkowski spacetime, and the generator of the corresponding trajectory is a boost. 
\begin{figure}[h] 
\begin{center}   
\includegraphics[width=8cm]{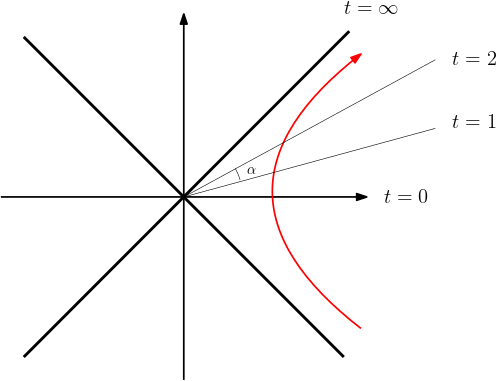}
\caption{ \textit{Hyperbolic structure of an accelerated Minkowski reference frame. The red trajectory corresponds to a static observer at a distance $d$ from the black hole, whereas the angle $\alpha$ is the boost parameter.}}
\end{center}
\end{figure} 
The Hamiltonian, which generates the evolution on this trajectory with respect to the proper time along it, is a scaling of the boost parameter: $H \sim a K$, so $K \sim \frac{H}{a}$. This means that the Hamiltonian generating this motion is:
\begin{equation}
H = \frac{a A}{8 \pi G} \ .
\end{equation}
If we have something falling in the black hole and we measure the energy of the latter $dE$ in the falling frame, the area of the black hole increases by an amount $dA$, and the FGP relation turns out to be:
\begin{equation}
dE = \frac{a}{8 \pi G}dA \ .
\end{equation}
This is another way of rewriting the relation \eqref{Equazione cardine Lecture 14}. If we compute the ADM energy of this black hole, the energy measured at a distance $d$ is precisely given by the previous formula. In an accelerated frame, Unruh was able to show that there is a significant temperature $T_{U} = \frac{a \hbar}{2 \pi }$ (where $a$ is the acceleration) and, if we rewrite $a$ as a function of the temperature, we get a relation between $E$, $A$ and $T$:
\begin{equation}
E = \frac{A k_B T}{4 \hbar G} \ .
\end{equation}
Remembering that the entropy is given by $S = E/T$, we get the \textbf{Hawking entropy}:
\begin{equation}
S = \frac{k_B A}{4 \hbar G} \ .
\end{equation}
Therefore, the relation \eqref{Equazione cardine Lecture 14} plus the Unruh temperature gives the Hawking entropy. \textit{For further considerations, consult section 10.2 of the book}. 
\item Jacobson wrote a remarkable paper in which, starting from the considerations of the previous point, he derived the Einstein equations:
\begin{equation}
\bigg \{
\begin{array}{rl}
S & = \frac{k_B A}{4 \pi G} \\
T_U & = \frac{a \hbar}{2 \pi} \\
\end{array} 
\longrightarrow \text{GR field equations} \ .
\end{equation}
If we forget about thermodynamics, replacing $S$ with $A$ and $T$ with $a$ using the above Jackobson results means that from the relation \eqref{Equazione cardine Lecture 14}, we can derive the Einstein equations! Therefore, it is possible to derive Einstein equations by just considering that there is a \underline{local} relation in the geometry between an area element and an acceleration. We still do not understand what is precisely going on here. \textit{Rovelli thinks that there is something crucial concerning this relation. He keeps sharing further exciting considerations. See the end of the lecture for details}.
\end{itemize}
\section{Unitary representations of SL(2,$\mathbb{C}$) and the Y map}
\label{sec:Lecture_15_Unitary_rep_SL(2,C)}
This will be a little bit more technical lecture. We have seen that action \eqref{Azione LQG Lecture 13 con B} encodes GR, where $\omega$ is an $SL(2,\mathbb{C})$ connection, $B$ is an $SL(2,\mathbb{C})$ 2-form and $F$ corresponds to the curvature of $\omega$. This is a sort of "bulk" of the theory. It turns out that $B$ restricted on the boundary has an electric $\vec{L}$ and a magnetic $\vec{K}$ part related by the remarkable relation \eqref{Equazione cardine Lecture 14}. We have seen that this fact "breaks" the Lorentz group into rotations: this defines the structure on the boundary.

\medskip

To treat the quantum theory, we need to write the Hilbert space. Unsurprisingly, this space carries representations of the groups, so we must deal with unitary representations for the latter. Therefore, we now discuss the mathematics of $SL(2,\mathbb{C})$ and its representations.
\subsection{Structure and representations of $SL(2,\mathbb{C})$}
We could think we know everything about the Lorentz group and its representations. Physicists typically use the finite-dimensional representations of $SL(2,\mathbb{C})$, which are not unitary since the Lorentz group is not compact. So, physicists do not know the unitary representations of $SL(2,\mathbb{C})$. If we are interested in this topic, there is a math book by Ruhl in 1974 describing all this mathematics. So we now start slowly by studying the representations we need to describe quantum gravity, following the same steps we did for $SU(2)$. We indicate the group elements with $g \in SL(2,\mathbb{C})$. We can think of these elements as matrices:
\begin{equation}
g = \begin{pmatrix}
a & c \\
b & d
\end{pmatrix} \ ,
\label{Matrice di SL(2,C) generica Lecture 15}
\end{equation}
which, however, are not unitary. The only condition on $g$ is the (complex) equation $det(g) = 1$, which encodes two real equations up to 4 complex numbers. Therefore, we have six real numbers specifying the dimensions of this group. This is not surprising since we know that the Lorentz group is a six-dimensional group defined by three rotations and three boost parameters. Being a Lie group, we can always write its elements as the exponential of a linear combination (of Pauli matrices) with three \underline{complex} parameters: 
\begin{equation}
g = e^{i \alpha^i \sigma_i}  \ .
\end{equation}
If we take $\alpha_1, \alpha_2, \alpha_3$ to be real, we return to the generic exponential form of an $SU(2)$ element, as in eq. \eqref{eq:Exponential_form_group}. $SU(2)$ is a sub-group of $SL(2,\mathbb{C})$ and the three imaginary parts in the complex numbers $\alpha_1, \alpha_2, \alpha_3$ specify precisely the boosts parameters. The generators are usually written in the adjoint representation, and the covariant way of writing the algebra is:
\begin{equation}
\left[ J^{IJ}, J^{KL}  \right] = -\eta^{IK}J^{JL} + \eta^{IL}J^{JK} - \eta^{JK}J^{IL} + \eta^{JL}J^{IK} \ ,
\end{equation}
where we used Lorentz antisymmetric indices for the generator $J$. There is a more elegant way to write it; we will do it soon. In fact, instead of the above covariant form, we can pick an $SU(2)$ subgroup (breaking $SL(2,\mathbb{C})$ down to $SU(2)$), which in this language means to choose a "time" direction $t_I$, i.e., selecting a Lorentz frame. We can choose the coordinates to have $t^I = (1,0,0,0)$, breaking the generator $J^{IJ}$ in the electric $K^{I}$ and the magnetic $L^J$ part. So we have:
\begin{align}
L^I & \equiv \frac{1}{2} \epsilon^I_{\hspace{1mm}JKL}J^{JK}t^L = \left(0, \vec{L} \right) \ , \\
K^I & \equiv J^{IL}t_L = \left(0, \vec{K} \right) \ .
\end{align}
We can define these objects for every form of $t_I$, but adapting the coordinates to the form $t_I = (1,0,0,0)$ allows us to have only the spatial parts of $K^I$ and $L^I$ different from zero. The $L^i$ are the generators of rotations (which generate $SU(2)$, that lies inside the Lorentz group):
\begin{equation}
\left[ L^i, L^j \right] = \epsilon^{ij}_{\hspace{2mm}k}L^k \ .
\end{equation}
The mixed commutator between $L^i$ and $K^j$ tells us that the latter behave like vectors under rotations:
\begin{equation}
\left[ L^i, K^j \right] = \epsilon^{ij}_{\hspace{2mm}k}K^k \ .
\end{equation}
Finally, the commutator of two boosts turns out to be a rotation:
\begin{equation}
\left[ K^i, K^j \right] = \epsilon^{ij}_{\hspace{2mm}k}L^k \ .
\end{equation}
Not by chance, we used the same notations and names employed in the previous lecture, in which we described GR. Two Casimir operators are in $SL(2,\mathbb{C})$. In $SU(2)$, we had just one Casimir, the square of $L^i$. Therefore, we must have two invariant scalars under $SL(2,\mathbb{C})$ transformations. These are $J^{IJ}J_{IJ}$ and the "star" version $ \epsilon_{IJKL}J^{IJ}jJ^{KL}$. If we write them in terms of $\vec{K}$ and $\vec{L}$ we have:
\begin{align}
C_1 & \equiv |\vec{K}|^2 - |\vec{L}|^2 \\
C_2 & \equiv \vec{K} \cdot \vec{L} \ .
\end{align}
\textit{These operators are defined by formulas (7.129) and (7.130) of the book}.  

\medskip

Let's talk about representations of $SL(2,\mathbb{C})$. We all know the finite-dimensional representations of the Lorentz group, and the simplest one is the fundamental representation, in which the group element $g$ acts on the vector space of spinors $\mathbb{C}^2$, with elements $
\begin{pmatrix}
z^0 \\
z^1
\end{pmatrix}  $,
precisely with the matrix \eqref{Matrice di SL(2,C) generica Lecture 15}. So, it is the same vector space that we have discussed for the fundamental representation of $SU(2)$, but we must be careful since \underline{it is not the same Hilbert space}. In fact, for $SU(2)$, this was also the Hilbert space since the scalar product \eqref{Prodotto scalare in SU(2) Lecture 4} turns out to be an $SU(2)$ invariant, but we already pointed out that it is not also $SL(2,\mathbb{C})$ invariant. Actually, in $\mathbb{C}^2$, it is impossible to find such a scalar product. Therefore this is not a unitary representation (in fact, it is a finite-dimensional representation for a not-compact group). We said that there was an $SL(2,\mathbb{C})$ invariant antisymmetric quadratic expression, which was \eqref{Mappa nel prodotto tensoriale Lecture 4}, defined to be the invariant part of the tensor product of the space with itself. Using the scalar product \eqref{Prodotto scalare in SU(2) Lecture 4} means breaking Lorentz down to $SU(2)$, and, physically, it is equivalent to choosing a Lorentz frame. We will examine this in more detail. 

\medskip

We also know the Lorentz representation acting on four-vectors $v^I$ in Minkowski space $\mathcal{M}$. There is a map from $SL(2,\mathbb{C})$ to $SO(3,1)$, exactly as seen for the $SU(2) \rightarrow SO(3)$ case \eqref{eq:SU(2)_SO(3)_map}. \textit{See from minute 19:21 to 20:12 for a brief oral discussion about the properties concerning this map}.
There is also the finite adjoint representation, namely the one in which the algebra lives, used for the E.M. acting on tensors $F^{IJ}$. 

\medskip

All the above are finite-dimensional representations, but we need the infinite-dimensional ones for QG. We now list some facts regarding them, remanding to the following lectures for more detailed discussions. The key point is that since there are two Casimir operators, these representations will be labeled by two quantum numbers. Remember that the $SU(2)$ representations were labeled by $j$, and in the $j$ representation, we had a basis with a magnetic quantum number $m$: $|j,m \rangle$. The representation was concretely specified in terms of Wigner $D_{mn}^j(h)$ matrices, depending on the functions of the group elements. For $SL(2,\mathbb{C})$, we have two numbers labeling the representations, and since the group is not compact, one of these has to be continuous. This continuous quantum number is denoted with $p$, while the other discrete one is called $k$. It turns out that $p$ is a real number, whereas $k$ is a not negative half-integer. Finding a basis in the $(p,k)$ representation is very easy because there is a theorem that states that the unitary representation $(p,k)$ is defined on a Hilbert space $\mathcal{H}^{(p,k)}$ and, since $SU(2)$ lives in $SO(1,3)$, the latter are also representations of $SU(2)$. So, the representation $(p,k)$ can be decomposed into irreducible representations of $SU(2)$:
\begin{equation}
\mathcal{H}^{(p,k)} = \bigoplus_{j=k}^{\infty} \mathcal{H}^j \ .
\label{Decomposizione rappresentazioni di Lorentz unitarie Lecture 15}
\end{equation} 
This immediately allows us to write a basis in each $j$ space. We can diagonalize $L_z$. The \underline{canonical basis} will be denoted as $|p,k;j,m \rangle$. The books give us the following form for the representation matrices: $D^{PK}_{jmj'm}(g)$. \textit{Rovelli does not think Mathematica has implemented the explicit form of these, which are given by complicated integral expressions. So, they cannot be written in terms of elementary functions as it happens for the Wigner matrices}. Remember that for the Casimir operator of $SU(2)$, we had the relation \eqref{eq:Casimir_operator_action}, which told us that it is a multiple of the identity in each representation. For $SL(2,\mathbb{C})$, this is gonna be the same for the 2 Casimir operators (mathematicians can derive all the eigenvalues):
\begin{align}
C_1 |p,k;,j,m \rangle & = |p,k;j,m \rangle (p^2 - k^2 - 1) \ , \\
C_2 |p,k;,j,m \rangle & = |p,k;j,m \rangle  pk \ .
\end{align}
\textit{These expressions correspond to eq (7.11)-(7.12) of the book}. Since we have just stated mathematical notions, we examine an interesting result that will be derived. 
\subsection{The $Y_{\gamma}$ map}
First, we take one of the $\mathcal{H}^{(p,k)}$ spaces and choose a subset of the representation with a particular relation between $p$ and $k$:
\begin{equation}
p = \gamma k \ , 
\end{equation}
where $\gamma$ is a real number. Within this special subset of representations, we take the lowest component in the sum \eqref{Decomposizione rappresentazioni di Lorentz unitarie Lecture 15}:
\begin{equation}
k = j \ .
\end{equation}
Therefore we look only for the spaces with states $|\gamma j, j; j, m \rangle$. In this state, one has:
\begin{equation}
\langle \vec{K} \rangle = \gamma \langle \vec{L} \rangle \ .
\label{Relazione fra stati Lecture 15}
\end{equation}
We can map a state $| j;m \rangle \in \mathcal{H}^j$ into a state $|p,k;j,m \rangle \in \mathcal{H}^{(p,k)}$, where $k=j$ and $p=\gamma j$. This is the \textbf{$Y_{\gamma}$ map}, which remarkably depends on $\gamma$:
\begin{align}
Y_{\gamma} : & SU(2) \longrightarrow SL(2,\mathbb{C}) \ , \\
Y_{\gamma} & | j;m \rangle =  |\gamma j,j;j,m \rangle \ .
\end{align}
Equation \eqref{Relazione fra stati Lecture 15} is true up to terms that, in quantum theory, will be neglectable. Furthermore, $\gamma$ has not to be necessarily real, but people typically assume this condition. This $Y_{\gamma}$ map will play a key role in the quantum theory. In the next lecture, we will explicitly write the transition amplitude and examine this map in more detail.
\section{LQG dynamics}
\label{sec:Lecture_16_LQG_dynamics}
Now we see how the $Y_{\gamma}$ map implements the relation \eqref{Equazione cardine Lecture 14} into the quantum theory.
\subsection{Transition amplitudes of LQG}
Let's start thinking about B-F theories, so we consider an action such as \eqref{Azione LQG Lecture 13 con B}:
\begin{equation}
S \left[ e, \omega \right] = \frac{1}{2} \int B \wedge F \ .
\end{equation}
The quantum theory of this action turns out to be very simple to derive. We want to define, formally, a path integral:
\begin{equation}
Z = \int \mathcal{D}B \mathcal{D}\omega e^{\frac{i}{\hbar}\int B \wedge F} \ .
\end{equation}
We want to introduce it in the same way that Feynman first defined path integrals in his Ph.D. thesis, which consists of (in one dimension) discretizing the time on a line, by writing transition amplitude step by step and inserting the resolution of the identity in every single integral. These multiple integrals formally become, in the limit, a path integral, which Feynman takes as a definition. Therefore we want to discretize and write multiple integrals for $Z$, even if this is not well defined from a mathematical point of view. To discretize this quantity, we need to discretize a 4D region of spacetime geometrically. We choose a triangulation, writing the fields on it. In lattice CQD, we have a background metric; here, we don't. Therefore, we are doing the most "covariant" things as possible. In 4D, instead of tetrahedra, we have the so-called \textbf{4-simplex}.
\begin{figure}[h] 
\begin{center}  
\includegraphics[width=4cm]{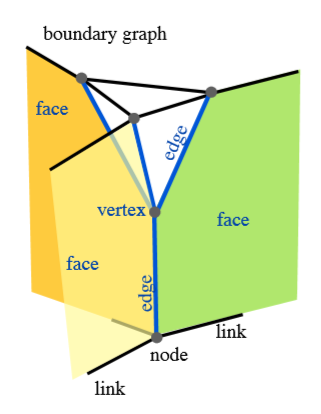}
\caption{Triangulation and two-complex terminology in 4D. \textit{Fig (7.1) of the book.}}
\end{center}
\end{figure} 
Five tetrahedra bound a 4-simplex, and it can be imagined as a tetrahedron with an extra point where the spacetime "shrinks". As we did in three dimensions, it is much easier to think in terms of dual triangulation:
\begin{itemize}
\item \textbf{4 simplex $\leftrightarrow$ vertex} \\
To each 4-simplex in the triangulation, we associate a vertex in the center of it in the corresponding dual graph.
\item \textbf{tetrahedron $\leftrightarrow$ edge} \\
To each three-dimensional tetrahedron in the 4D triangulation, we associate a one-dimensional edge in the dual graph. This \underline{five} edges come out from each vertex described above.
\item \textbf{triangle $\leftrightarrow$ face} \\
Finally, the triangle faces bounding the tetrahedra correspond to faces in the dual graph.
\end{itemize}
\begin{figure}[h] 
\begin{center} 
\includegraphics[width=10cm]{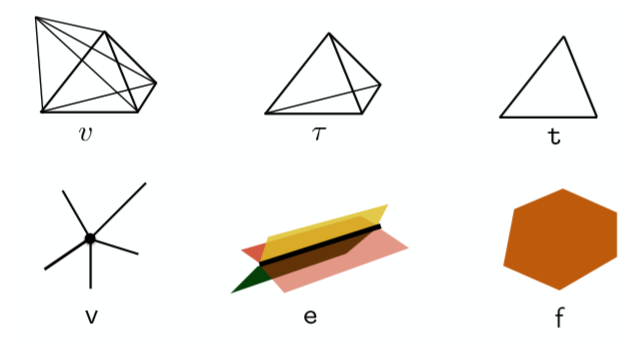}
\caption{A vertex is a point inside the 4-simplex. It has five edges corresponding to the five tetrahedra bounding the 4-simplex. A triangle is dual to a face that “wraps around it” because, in 4D, we can go around a triangle in the $(x,y)$ plane, moving in the $(t,z)$ plane. \textit{Fig. (7.2) of the book.} }
\end{center}
\end{figure}
\begin{figure}[h]
\begin{center} 
\includegraphics[width=10cm]{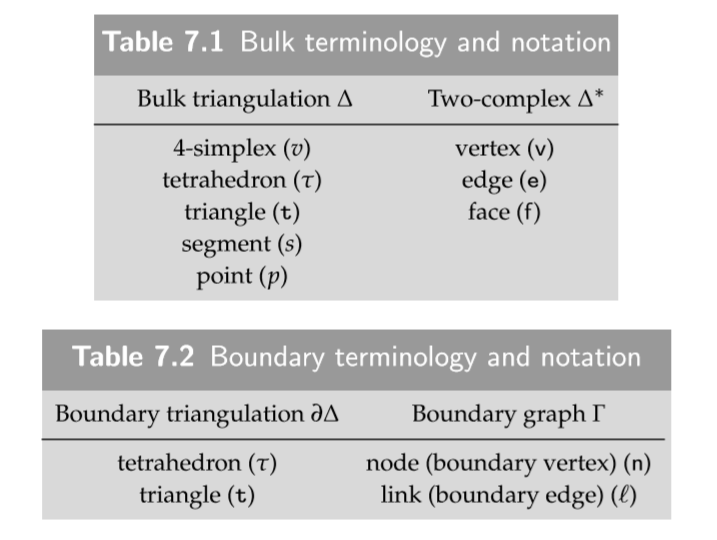}
\caption{\textit{Taken from page 138 of the book.} }
\end{center}
\end{figure}
The structure created within the dual triangulation by these faces, edges, and vertices is called a \textbf{2-complex}. 
\begin{figure}[h] 
\begin{center} 
\includegraphics[width=4cm]{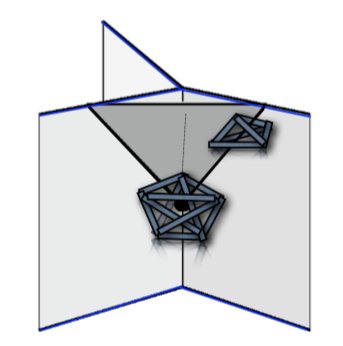}
\caption{\textit{Taken from page 138 of the book.} }
\end{center}
\end{figure}
The $B$ field is a 2-form; we can integrate the latter on a triangle face to discretize it. So, to each triangle, which means to each face in the dual graph, we can associate a discrete version of $B$. The connection is a 1-form, and we can integrate it along the lines. The most natural thing to do is to integrate it on the lines connecting two vertices associated with two 4-simplices (or, in terms of gravity, two different reference frames). 
So, the $B$ field transforms into something integrated on faces, as connection transforms into something integrated over edges. When we integrate the connection, we don't want to integrate just the 1-form but its exponentiation. Therefore we define a generic element of a generic group $g$, which is defined as its integral along the edge:
\begin{align}
B & \longrightarrow B_{faces} = \int_{faces} B \\
\omega & \longrightarrow \omega_{edge} = g \sim e^{\int_{edge}\omega} \ .
\end{align}
This way, $B$ has a value in the algebra, while $\omega$ is in the group. By discretizing spacetime, the continuous field theory turns out to be truncated. Still, the price to do this in such a covariant manner (compared to lattice QCD) has a dramatic consequence, which will be examined during the following lecture. Once we have done this, we can integrate (whatever it means) over all the $B_{face} \equiv B_f$ and $g_{edges} \equiv g_e$, rewriting the path integral as:
\begin{equation}
Z = \int \mathcal{D} B_{f} \int_G d g_{e} e^{\frac{i}{\hbar} \sum_{f} B_f \prod_{e \in f} g_e} \ .
\end{equation}
We have discretized the curvature by using the connection defined on the edges, while $\prod_{e \in f} g_e$ is the product of group elements sitting on the edges around each (dual) face, which gives a measure of the curvature. Then we can formally perform the integral over $B$, ending up with something well-defined by using the fact that, up to some factors, $\int dp e^{ipx} \propto \delta(x) $. Therefore:
\begin{equation}
Z = \int_G d g_{e} \prod_{f} \delta (g_{e_1}...g_{e_n}) \ ,
\end{equation}
where $g_{e_1}...g_{e_n}$ are the group elements on the edges around the face $f$, and the integral is performed over a copy of $G$ for each edge. This is a well-defined expression if the group is compact. Otherwise, we might have infinite quantities inside the integral, but we will treat this case briefly. This $Z$ is a number, which depends only on the topology of the manifold $\mathcal{M}$ (\textit{the fact that the partition function does not depend on the chosen triangulation is briefly discussed later}), so we can write $Z \equiv Z_{\mathcal{M}}$. We have taken this as the definition of path integral, but we want to rewrite it slightly differently.
\begin{figure}[h] 
\begin{center}      
\includegraphics[width=6cm]{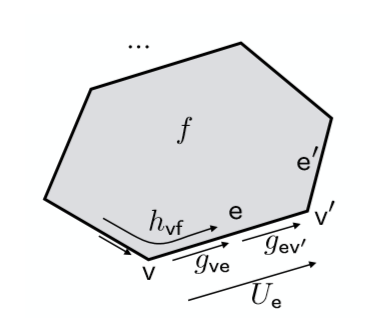}
\label{Splitting cosmetico delle facce}                            
\caption{Splitting the group elements on the edges in two. \textit{Fig. (5.5) of the book. } }
\end{center}
\end{figure}
First, we double the variables by introducing the quantities vertex-edge $g_{ve}$, rewriting the integral as:
\begin{equation}
Z = \int dg_{ve} \prod_f \delta \left(  \prod_f g_{ve} \right) \ .
\end{equation} 
Then, we introduce the variables vertex-face $h_{vf}$, rewriting the above expression by trading the $g_{ve}$ variables for these new ones:
\begin{equation}
Z = \int dg_{ve} \int dh_{vf} \prod_{f} \delta(h_{f v_1}...h_{f v_n}) \prod_{vf} \delta \left( h_{vf} g_e g_{e'} \right) \ ,
\end{equation}
This allows us to get the form we need because now we can collect all the vertices:
\begin{equation}
Z = \int dh_{vf} \prod_{f} \delta(h_{f v_1}...h_{f v_n}) \prod_{v} A_v \left( h_{vf} \right) \ ,
\label{Funzione di partizione LQG Lecture 16}
\end{equation}
where the vertex amplitudes $A_v$ are given by:
\begin{equation}
A_v = \int dg_{ve} \prod_{f \in V} \delta(h_{vf} g_e g_{e'}) \ .
\end{equation}
If the group is compact, we can view this expression as the path integral of a BF theory. We have put $Z$ in this form because there is a general formula valid for every compact group, which establishes that a delta function in the group can be written as a sum of representations as:
\begin{equation}
\delta(g) = \sum_{j} (2j+1)Tr_j(g) \ ,
\end{equation}
where $2j+1$ is the dimension and $Tr_j$ is the trace in the representation $j$. Therefore we can rewrite $A_v$ as:
\begin{equation}
A_v = \int dg_{ve} \prod_{f \in V} \sum_{j} (2j+1)Tr_j(h_{vf} g_e g_{e'}) \ .
\label{Ampiezza vertex con elementi di SL nella traccia Lecture 16}
\end{equation}
\textit{This can be compared with eq.(7.47) of the book, although the latter is in 3D}. We want to interpret this for a theory where the group is $SL(2,\mathbb{C})$, but on the boundaries, we have $SU(2)$, imposed by the remarkable relation $\vec{K} = \gamma \vec{L}$. How can we do it? Remembering that each vertex corresponds to a 4-simplex, the amplitudes $A_v$ is associated with the 4-simplex, which depend on the group elements $h_{vf}$ sitting on its boundary, a three-dimensional triangulated surface. We have a graph on the boundary with nodes and links, namely a discretized spin network. \textit{From minute 24:10 to 25:29, Rovelli starts illustrating how to imagine the 3D boundary and discussing the interpretation}. We want $SU(2)$ quantum states of gravity on the boundary. We want the theory's full $SL(2,\mathbb{C})$ invariant states within the bulk. The amplitude $A_v(h)$ is just a function of the ten $h$ sitting on each link of the boundary, so we have to interpret this $ h$ as the $SU(2)$ elements defining the quantum geometry on the boundary, keeping the $SL(2,\mathbb{C})$ within the bulk: $g \in SL(2, \mathbb{C})$, $h \in SU(2)$. In other words, we want an $SL(2,\mathbb{C})$ theory, but on the boundary, we want to represent our states in the time gauge. Therefore the integral $\int dh_{vf}$ in $Z$ is actually over $SU(2)$, while the integral $\int dg_{ve}$ in $A_v$ urns out to be performed over $SL(2,\mathbb{C})$. The last thing to fix is that in the amplitude \eqref{Ampiezza vertex con elementi di SL nella traccia Lecture 16}, inside the trace over $SU(2)$ spins, we have the product of $g_e g_{e'}$, which are elements of the Lorentz group! We need to map these objects into $SU(2)$ elements to compute the trace, so we use the $Y_{\gamma}$ map. The \textbf{LQG amplitude} can be finally written as:
\begin{equation}
A_v(h) = \int_{SL(2,\mathbb{C})} dg_{ve} \prod_{f \in V} \sum_{j} (2j+1)Tr_j(h_{vf}Y^{\dagger}_{\gamma} g_e g_{e'} Y_{\gamma}) \ ,
\label{Ampiezza della LQG Lecture 16} 
\end{equation}
where
\begin{equation}
Tr_j(h_{vf}Y^{\dagger}_{f}g_e g_{e'} Y_{f}) := D^j_{m'm}(h_{vf}) D^{jj}_{jm jm'}(g_e g_{e'}) \ .
\end{equation}
\textit{These expressions can be compared to formulas (7.49) and (7.50) of the book.}
But, instead of the partition function, we want the transition amplitudes, so how can we compute them? If we consider the 4D region of the spacetime and the boundary associated with the latter, the 4-simplices are bounded by tetrahedra, creating a spin network that constitutes the boundary itself. 
\begin{figure}[h] 
\centering   
\includegraphics[width=6cm]{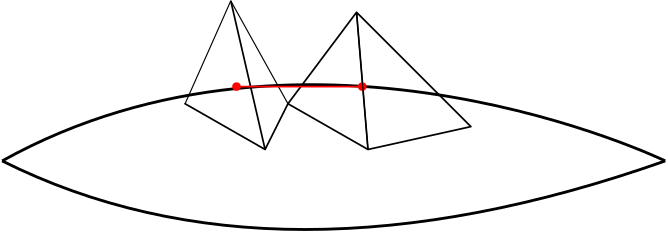}
\caption{\textit{Representation of the 3D spin network boundary of a two-complex. The red link connects two nodes drawn in the center of the tetrahedra that realize the boundary itself.} }
\label{3Dboundary}
\end{figure}
When we compute the expression $\prod_{f} \delta(h_{f v_1}...h_{f v_n})$ inside the integral defining $Z$ in eq. \eqref{Funzione di partizione LQG Lecture 16}, we are implicitly including in the integration over $SU(2)$ also some of the $h$ sitting on the boundary, but these are not integrated over. Therefore $Z$ is a function only of the $h$ sitting on the boundary since the other ones are integrated. This means that $Z$ is a function in the Hilbert space $L_2 \left[ SU(2)^L \right]$, where $L$ is the total number of links on the boundary. This space can also be intuitively rewritten as:
\begin{equation}
L_2 \left[ SU(2)^L \right] = \mathcal{H}^{up} \otimes \mathcal{H}^{down} \equiv \mathcal{H}^{boundary} \ .
\end{equation}
Therefore, being an element in a Hilbert space, $Z$ is a ket and also a bra, of course. Therefore we can contract an upper quantum state on the boundary and a lower one on the (lower) boundary, obtaining an amplitude. So, we now have an explicit way to compute amplitudes from (a linear combination of) quantum geometrical states in LQG.
\begin{figure}[h]    
\begin{center}      
\includegraphics[width=8cm]{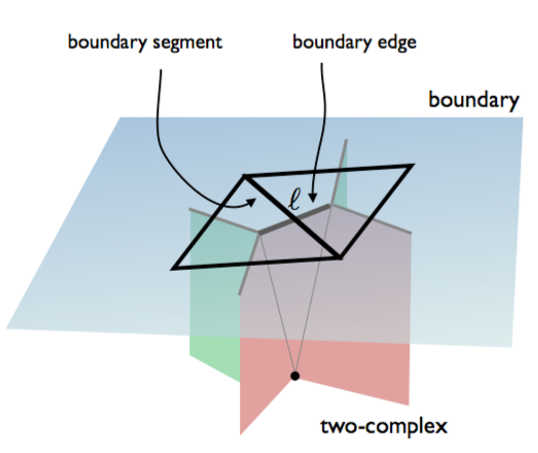}
\caption{\textit{Taken from page 96 of the book.} }
\end{center}
\end{figure}
We must remember that the above partition function depends on the chosen triangulation between the states, and it consists of a truncation, so we can explicitly denote it with $Z(h_{boundary}) \equiv Z_{\Delta}(h_{boundary})$. When we refine the triangulation, the partition function converges to a number that no longer depends on the triangulation, as in QCD. \textit{From minute 34:48 to 35:14, Rovelli briefly describes this aspect in QCD}. So, what we have defined so far are the \underline{truncated} transition amplitudes. The exact ones are defined more abstractly, namely, the value of these when we refine the triangulation. Concretely, all calculations that have been done in this formalism were performed for a fixed value of approximation in the finite triangulation. Little is still known about the actual convergence when performing this refining. There are hints and suggestions that the actual number is independent of the triangulation, but we have obtained no solid results. 

\medskip

These abstract formulas, which seem to have nothing to do with Einstein equations (in fact, we started from a generic BF action), have much to do with GR. Several theorems state that, in a suitable limit and with many caveats, the above $Z$ gives precisely the Hamilton function of GR. So, there are Einstein equations inside this construction, and we will dedicate an entire lecture to see how. We now add some details: 
\begin{itemize}
\item For each vertex, there is 5 $SU(2)$ integration since each vertex has five edges coming out of it. Studying the expression for $A_v$, we discover that, because of the gauge invariance, just four integration are independent, and the last one gives $\int_{SL(2,\mathbb{C})} dg = \infty$ (the Haar measure of a non-compact group diverges). So we integrate just four times to obtain finite quantities. 
\item These expressions are finite in the UV limit, but some infrared divergences might occur since we are summing over all possible geometries. Even if the boundary is finite, the sum can still be infinite. In the book, there is a full chapter devoted to these divergences. Here, we say that the theory with the cosmological constant gives no divergences.
\end{itemize}
The following lectures show how similar and different this formulation is compared to lattice QCD. A crucial difference is that here since GR is invariant under diffeomorphism, there is no scaling parameter!
We have seen all the treatments of LQG in the covariant formalism, writing all the basic equations. During the following lectures, we comment on the results extracting physics from that.
\section{Why there is no critical parameter in LQG}
\label{sec:Lecture_17_no_critical_parameter}
We have fully defined the theory in its wholeness. 
\subsection{Summary of the main equations}
We can summarize the main formulas of LQG:
\begin{itemize}
\item \textbf{Kinematic} \\
We work in terms of a Hilbert space associated with a graph $\mathcal{H}_{\Gamma} = L_2 \left[ SU(2)^L / SU(2)^N \right]_{\Gamma}$, in which the wavefunctions are given by $\psi(h_l)$. The most important operators are $\vec{L}_l$, which are associated with the links of the graph, satisfying the following commutation relation: $\left[ L^i_l, L^j_{l'} \right] = \delta_{l,l'} \epsilon^{ij}_{\hspace{3mm} k}L^k_{l} \ .$ 
\item \textbf{Dynamics}
In this Hilbert space, there is a vector (or, more precisely, a bra), so exists a linear map from this Hilbert space to the complex numbers, which is the amplitude, that we now rewrite as:
\begin{equation}
W_{\mathcal{C}}(h_l) = \int_{SU(2)} dh_{vf} \prod_f \delta \left( \prod_f h_{vf} \right) \prod_{v} A_{v} \left( h_{vf} \right)
\label{Amplitude LQG Lecture 17}
\end{equation}
We expressed that the amplitude depends explicitly on the 2-complex $\mathcal{C}$. The vertex amplitudes are:
\begin{equation}
A(h_f) = \int_{SL(2,\mathbb{C})} dg_e \prod_f \sum_j (2j+1) Tr_j \left[ h_f Y_{\gamma}^{\dagger} g_e g_{e'} Y_{\gamma} \right] \ .
\label{Vertex Amplitude LQG Lecture 17}
\end{equation}
\end{itemize}
\textit{From minute 4:35 to 8:45, Rovelli explains graphically what is going on and the meaning of the indices in the kinematic and the dynamical part}. 
\begin{figure}[h] 
\begin{center} 
\includegraphics[width=4cm]{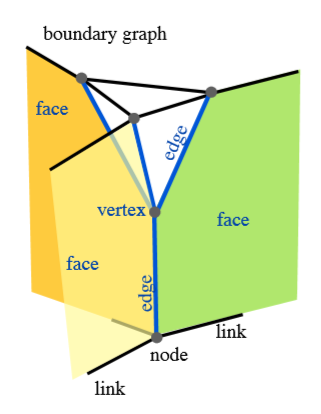}
\caption{Triangulation and two-complex terminology in 4D. \textit{Fig (7.1) of the book.}}
\end{center}
\end{figure} 
Let's do a couple of observations before going ahead. 
\subsection{Again about amplitudes}
First of all, the states $\psi(h_{h_l}) = \langle h_l | \psi \rangle $ are written based on the group elements, but we have also defined the spin network basis, where for each graph $\Gamma$ instead of the group element $h_l$ there is the basis $(j_l,v_n)$:
\begin{equation}
| \Gamma,h_l \rangle \longrightarrow | \Gamma, j_l, v_n \rangle \ .
\end{equation}
The relation between the two is:
\begin{equation}
\langle h_l | j_j, v_n \rangle \sim \bigotimes_{links} D^{j_l}(h_l) \bigotimes_{intertwiners} v_n \ .
\end{equation}
The particular graph dictates the contraction between the indices we are considering. Therefore the amplitudes \eqref{Amplitude LQG Lecture 17} and \eqref{Vertex Amplitude LQG Lecture 17} can be expressed into the spin network basis by simply inserting resolutions of identity and integrating over $SU(2)$ elements. The explicit procedure is reported in the book. We report just the final result of this process applied to the amplitude $W$:
\begin{equation}
W \left( j_l, i_n \right) = \sum_{j_f,i_e} \prod_{f}(2j_f + 1) \prod_{e}(2j_e + 1) \prod_v A_v \left( j_l, j_f, i_n , i_e \right) \ ,
\label{Ampltidue LQG con spin network Lecture 17}
\end{equation}
where the sum is over spins associated with internal faces and intertwiners associated with edges, while the vertex amplitude now turns out to be given by a function "sitting" around each vertex. If we have a 5-valent vertex, we have five nodes coming out of it. Therefore the spin network around the latter has a pentagonal shape with a star inside.
\begin{figure}[h]  
\begin{center}   
\includegraphics[width=8cm]{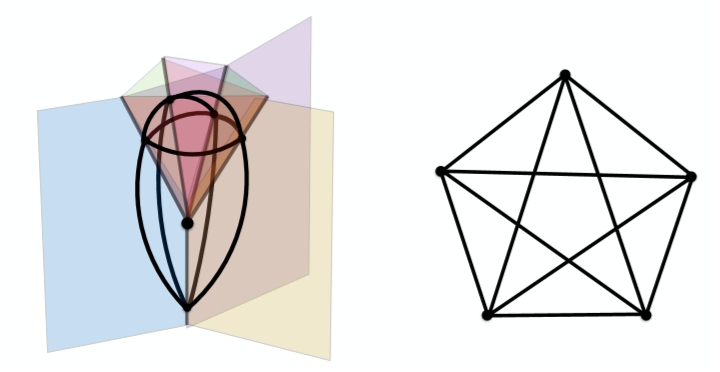}
\caption{The vertex graph in 4D. \textit{Fig (7.5) of the book. }}
\end{center}
\end{figure}
We know very well the properties of $A_v$ when $j_l$ is large (we will see this in more detail), but we don't know much about what happens for small $j_l$ except for some recent numerical works. Knowing what happens for large spins' behavior but not small ones means that we ignore what happens in the quantum regime. Generally speaking, the simplest way of taking classical limits is to take very large quantum numbers away from the "discreteness." The following is a particularly simple and useful schematic to write the vertex amplitudes. The $Y_{\gamma}$ map can be used, as we have seen, to map $SU(2)$ elements into $SL(2,\mathbb{C})$ elements, so the $SU(2)$ state $\psi$ around the vertex (in the center of the star) can be mapped into an $SL(2,\mathbb{C})$ state. This is precisely what happens inside the trace $Tr_j$ in eq. \eqref{Vertex Amplitude LQG Lecture 17}. After this mapping, we obtain an $SL(2,\mathbb{C})$ spin network, which is no longer invariant on the nodes. If we integrate over $SL(2,\mathbb{C})$ as in eq. \eqref{Vertex Amplitude LQG Lecture 17}, we are actually projecting it down to a proper $SL(2,\mathbb{C})$ invariant spin network. Therefore we can formally rewrite the integral in \eqref{Vertex Amplitude LQG Lecture 17} as:
\begin{equation}
A(\psi) = \left( P_{SL(2,\mathbb{C})} Y_{\gamma} \psi \right) \left( \mathbf{1} \right) \ .
\label{Vertex Amplitude con proiettore Lecture 17}
\end{equation} 
Namely, we compute the value of $ P_{SL(2,\mathbb{C})} Y_{\gamma} \psi$ on the identity $\mathbf{1}$ of the Lorentz group.
In a sense, the core of GR consists in taking an $SU(2)$ spin network, mapping it into an $SL(2,\mathbb{C})$ spin network, making it invariant and calculating the amplitude: this gives the Einstein-Hilbert action (in some limit). The full theory is therefore given by the equations \eqref{Amplitude LQG Lecture 17}-\eqref{Vertex Amplitude LQG Lecture 17}, where the latter can be rewritten as in eq. \eqref{Vertex Amplitude con proiettore Lecture 17}. Someone said we should never take a theory seriously if the latter cannot be written on a t-shirt. \textit{Rovelli shows his t-shirt to the class with the above equations written on it}. 
\subsection*{Some properties of this LQG formulation}
This formulation is still quite abstract, so we now state general facts about it to understand its meaning better.
\begin{itemize}
\item \textbf{Locality} \\
These fields don't live in a given spacetime. They \underline{form} a (discretized) spacetime. We can view the 2-complex as a history of spacetime quanta and a representation of spacetime. We can associate a three-geometry to the data of a spin network and a 4-geometry to the data of a spinfoam (we will treat this aspect if we have time). The total amplitude \eqref{Ampltidue LQG con spin network Lecture 17} turns out to be a sum over geometries and a configuration of something, which is a local product of amplitudes. That is precisely what we expect from the locality. We expect, in fact, the functional integral of some field(s):
\begin{equation}
Z \sim \int \mathcal{D}g e^{\frac{i}{\hbar}\int  \mathcal{L} d^4x} \ ,
\end{equation}
where the (continuous) sum in the exponential can be thought of as the (continuous) product of something: $e^{\sum x} \sim \prod e^x$. So, the product on the vertices in eq. \eqref{Ampltidue LQG con spin network Lecture 17} is like the exponential of the sum. Things in LQG interact with the next ones: there is \underline{no long range} interaction.
\item \textbf{Lorentz invariance} \\
There is explicit Lorentz invariance in LQG, introduced by the Lorentz transformations, while on the boundary, there are $SU(2)$ states written in the time gauge. But it is possible to replace this formulation with a fully manifest Lorentz invariance formalism on the boundary. So one has a proper $SL(2,\mathbb{C})$ spin network on the boundary, but there is no real gain in using this manifest covariance. 
\item \textbf{UV finite} \\
We have no UV divergences since, over the sums, we stop at a small scale since $j$ starts with $0,\frac{1}{2}$, etc., so there is no way this can reproduce a Feynman integral for high momentum, precisely because space is discrete. The entire reason of being of this construction is to have both Lorentz invariance and discreteness together. 
\item \textbf{No time} \\
As we deeply discussed, the amplitudes don't have a time variable in which they evolve since there is not supposed to be a time variable at all. We have an initial three-geometry and a probability for a final one. If we perform the classical limit, this corresponds to a four-geometry, in the middle of which we evolve proper time along the trajectory we consider.
\item \textbf{Theorems related to GR} \\
Several theorems relate this formulation to GR. Before studying them, we must treat the coherent states to which the next lecture is devoted. There are several indirect arguments showing that this is GR. We list these results (it exists great literature on them):
\begin{itemize}
\item The $n-$point function of GR has been computed starting from this construction.
\item The Friedman equation has been derived from this amplitude (in certain approximations).
\item The Hawking entropy has been derived several times and in different ways.
\item Recent works are trying to compute the tunneling probability for black holes transforming into white holes. We have a classical spacetime in which a star collapses and, at some point, a tunneling effect might occur in which classical equations are violated. These models study the possibility that something comes out of the black hole using the above LQG structure. 
\end{itemize}
\item \textbf{Variance of LQG theories} \\
We have defined LQG as a single unambiguous theory that can only be exact or wrong. Namely, there is no free parameter in it. Of course, there is a quantity $\hbar G$ "hidden" in the equations. What we have written is all dimensionless, but we can connect it to meters, seconds, etc., obtaining $\hbar G$. If we add matter, $\hbar$ and $G$ separate each other since $G$ comes out into the coupling with the matter. There is no ambiguity. However, this is not a "God-given theory," and several numbers of variants have been studied. We worked with the simplest possible variant but still do not control the full theory well enough to be sure if this is the "final" version. One variant we explicitly mentioned is that we only talked about triangulation, namely tetrahedra, and 4-simplexes. But we can write the above equations in general graphs with arbitrary valence that are not dual to any triangulation. If we have a node with more than four links, we cannot interpret it as dual to a tetrahedron. There is a theorem due to Minkowski stating that if we have a polyhedron and know the $N$ normals vectors to its faces (with length proportional to the area), the latter uniquely determines the polyhedron itself. The theorem is even stronger: if we have $N$ vectors $\vec{L}_i$ satisfying:
\begin{equation}
\sum_{i=1}^N \vec{L}_i = 0 \ .
\end{equation} 
Then there is one and only one polyhedron with the above characteristics. In the general case, the classical limit is much more difficult to understand. We can do tricky things to generalize the equations. \textit{See the lecture for further considerations}.
\item \textbf{Matter and Cosmological constant} \\
The book has a section treating the matter, even if not much literature has been (still) written on it. There is also an interesting section regarding the cosmological constant term, but we don't treat it since it would require several mathematical notions. The important result is that this makes the theory also infrared regular, namely the quantities $W\left( j_l,j_n\right)$ turn out to be mathematically finite without any physical limitations.  
\end{itemize}
Before going ahead, we clarify a crucial point: the absence of parameters. 
\subsection{Absence of a parameter}
This characteristic is what explains why things work in LQG. The striking feature of amplitude \eqref{Ampltidue LQG con spin network Lecture 17} is that we have not said a parameter is involved to its critical point. Everyone who does QCD on the lattice, treating approximation of QFT, knows that we should tune a parameter to an appropriate critical point to go to the continuous limit, but we said nothing about it. This is the true core of the LQG, i.e., how this invariance lack of background comes into the quantum theory. Let's proceed slowly, starting with the Feynman Ph.D. thesis. He had the idea that if we want to compute the propagator, we can perform a functional integral of an action:
\begin{equation}
W = \int \mathcal{D}ge^{\frac{i}{\hbar}\int dt \mathcal{L}} \ .
\end{equation}
Path integral concretely means, in QM, either a way to juggle with perturbation theory (doing equations with all terms from perturbation theory) or perform a discretization. So this is the limit of the discretized version of $W$, in which we break time into $N$ little steps:
\begin{equation}
W = \lim_{N \rightarrow \infty} \int dq_n e^{\frac{i}{\hbar} \sum_{n=1}^{N} \mathcal{L}_n[q_n]} \ .
\end{equation}
\begin{figure}[h]   
\begin{center}   
\includegraphics[width=5cm]{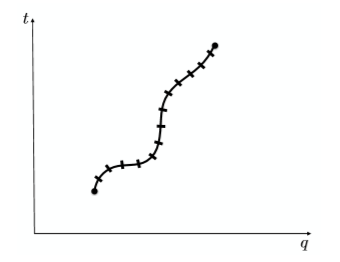}
\caption{Discretization of the parameter time. \textit{Fig. (4.1) the book.} }
\end{center}
\end{figure}
Let's do it concretely for a harmonic oscillator, which has, by definition, the following Lagrangian:
\begin{equation}
\mathcal{L} = \frac{1}{2} \left( \dot{q}^2 - \omega^2 q^2 \right) \ .
\end{equation}
What we have also to discretize is $\dot{q}$, this crucial passage is explicitly realized by:
\begin{equation}
\dot{q} \longrightarrow \frac{q_{n+1} - q_n}{\epsilon}  \ ,
\end{equation}
where $\epsilon$ is the infinitesimal interval between the discretized coordinates. Therefore the action $S^N \equiv \sum_{n=1}^{N} \mathcal{L}_n[q_n]$ is given by:
\begin{equation}
S^N = \sum_n \epsilon \left[ \left( \frac{q_{n+1} - q_n}{\epsilon} \right)^2 - \omega^2 q_n^2  \right]  \ .
\end{equation}
We rewrite it as:
\begin{equation}
S^N = \sum_n  \left[ \frac{\left( q_{n+1} - q_n \right)^2 }{\epsilon}  -  \omega^2  q_n^2 \epsilon  \right]  \ .
\end{equation}
It is convenient to get rid of the $\epsilon$ by a change of variable in the integration, rewriting the full amplitude as:
\begin{equation}
W = \lim_{\substack{N \rightarrow \infty \\ \omega_{\epsilon} \rightarrow 0}} \int dq_n e^{\frac{i}{\hbar} \sum_n^N \left[ \left( q_{n+1} - q_n \right)^2   - \omega_{\epsilon}^2 q_n^2 \right]}  \ ,
\end{equation}
where $\omega_{\epsilon} \equiv \omega \epsilon$ and the key point is precisely the limit $\omega_{\epsilon} \longrightarrow 0$. So, the path integral discretizes a continuous theory once \underline{provided that a parameter is taken to its critical value}. This happens all over discretizations and returns to the continuous limit. Doing QCD on a lattice works similarly, for example, discretizing a Yang-Mills theory on lattice links. \textit{Rovelli shortly describes discretized QCD theories on a lattice}. Let's treat the same harmonic oscillator model in a context where we have no background time, treating $q$ and $t$ on the same ground, namely both evolving in an unphysical parameter $\tau$, as we have seen in lecture 11. We pointed out several times that this is precisely how GR is formulated. We use an action that turns out to be a Lagrangian evolving in this parameter, and the latter is the same as before:
\begin{equation}
S = \int d \tau \frac{dt}{d \tau} \left[ \left( \frac{dq}{d \tau} \left( \frac{dt}{d \tau} \right)^{-1} \right)^2 - \omega^2 q^2 \right]  \ .
\end{equation}
Defining with the dot the derivatives with respect to $\tau$, we can write this in a less cumbersome notation:
\begin{equation}
S = \int d \tau \left( \frac{\dot{q}^2}{\dot{t}} - \omega^2 \dot{t} q^2 \right)  \ .
\end{equation}
Let's do a discretization of that:
\begin{equation}
S \longrightarrow S^N = \sum_n \epsilon \left[  \frac{ \left( \frac{q_{n+1} - q_n}{\epsilon} \right)^2 }{\frac{t_{n+1} - t_n}{\epsilon}} - \omega^2 \frac{t_{n+1} - t_n}{\epsilon} q_n^2 \right]  \ .
\end{equation}
The extremely remarkable feature is that the $\epsilon$ cancels, so the action does not depend on it! This is why the Regge calculus on GR has no lattice spacing, and the amplitude of LQG has no parameter inside it. The amplitude itself can be rewritten as:
\begin{equation}
W = \lim_{N \rightarrow \infty} \int dt_n dq_n e^{\frac{i}{\hbar} \sum_{n=1}^N \mathcal{L}(q_n,t_n) } \ .
\end{equation}
The lattice spacing in $\tau$ is irrelevant, as we said. We are complicating our life since the integral $\int dt_n$ is generally more challenging to compute. But the advantage is that, in a theory in which we start with a background independence, we do not have to perform a limit to a critical point.
\section{Regge calculus and intrinsic coherent states}
\label{sec:Lecture_18_Regge_calculus}
The discussion regarding the end of the previous lecture turns out to be a crucial point. Suppose we discretize and do the path integral for a theory defined and evolved on a background physical time. In that case, the continuous limit is obtained by (also) bringing a parameter to its critical value. But, when we do the same for a theory evolved in a parameter time, there is no critical value to take, while the limit is just the one in which we get the number of variables to go to infinity. So the difference is that, in the second case, we perform one single limit. This is the core of the distinction between QG and QFT or between spinfoam and lattice QCD. We must take the Yang-Mills coupling constant to its critical value in the latter. We treated a very elementary example, namely the harmonic oscillator. Let's do it for GR without going to the quantum theory. There is a very clear way of doing this, which is what Regge did in the sixties/seventies when he succeeded in realizing the first clean discretization of GR.
\subsection{Regge calculus}
We don't treat all details, but we want to see how to discretize a theory without a parameter and what it is truncation. It is essential since this gives us the intermediate step connection to perform the classical limit of LQG transition amplitude. Regge's idea was to take a Riemannian manifold and truncate/discretize it and, in terms of this discretization, write the Einstein-Hilbert action (or Einstein equations), namely in terms of the discretized variables. How he did this was the following. The most straightforward approach is to think first in 2D. We initially discretize a curved surface with triangles, assuming all triangles are flat, so we reproduce the manifold with the latter. Regge's intuition was using, as variables, the lengths of their segments. We should not call them "links" since they do not correspond at all to the links of LQG for the reason that we see in a moment. Where is curvature? If we have three links, we can compute the angles between the faces of the corresponding triangles. Therefore, the curvature is given by the fact that a complete turn around each vertex along the faces does not correspond to a $2 \pi$ angle (unless all triangles around the vertex are on the same plane).
\begin{figure}[h]  
\begin{center}
\includegraphics[width=5cm]{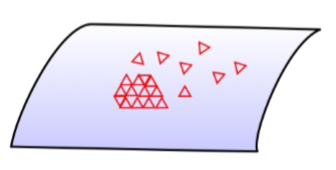}
\caption{Triangulation of a curved 2D manifold. \textit{Taken from page 88 of the book.} }
\end{center}
\end{figure}
So, we can explicitly write the angles $\theta(l)$ as a function of the three "l segments" around it (this is essentially just Euclidean geometry). We can compute the (discretized version of) the curvature $\delta$ as:
\begin{equation}
\delta = 2 \pi - \sum_n \theta_n(l) \ .
\end{equation} 
This is a crucial point. When we go to three dimensions, we have a triangulation made up of tetrahedra. We still use the lengths as variables: if we have six lengths, we have the geometry of a tetrahedron, and therefore, we can compute the angles between faces. When we have several tetrahedra attached around each segment of a tetrahedron, we have the same formula. Still, now the curvature "sits" around this segment \textit{(see the lecture from minute 5:48 to 6:23 for Rovelli explaining this by using a physical tetrahedron as support)}.
\begin{figure}[h]   
\begin{center}  
\includegraphics[width=6cm]{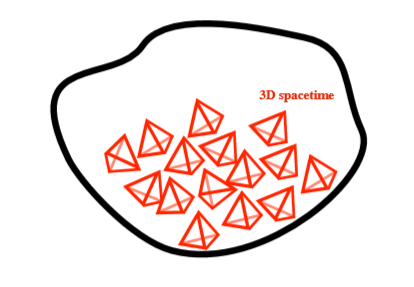}
\caption{Triangulation of a 3D region of spacetime. \textit{Taken from page 88 of the book.} }
\end{center}
\end{figure}
In 4D, we have 4-simplices, and now, each 4-simplex turns out to be bounded by five tetrahedra. \textit{Rovelli describes geometrically a 4-simplex and how to imagine the measure of the curvature of it by using a physical tetrahedron again as support, from minute 6:42 to 7:58}. Regge proved, quite remarkably, the following result. If we take the sum over all triangles in this 4D triangulation and, for each triangle $\Delta$, we consider its area (which is a function of all the lengths $l$ of the links) and finally multiply it by the curvature associated with the triangles, we get a number that up to details (we are treating only "sketches" of the full story) depend only on the $l$s:
\begin{equation}
S_R(l) = \sum_{\Delta} A_{\Delta}(l) \delta_{\Delta}(l) \ .
\end{equation}
All this $l$s approximate a geometry $g_{\mu \nu}(x)$. This $S_R(l)$ is a number. Therefore, it tells us how much curvature there is. It is local since it can be written as an integral over the manifold of something local (i.e., a Lagrangian), depending on the curvature. So, the only thing that it can be is the Einstein-Hilbert action of the metric approximated by these links:
\begin{equation}
S_R(l) \longrightarrow \int d^4x \sqrt{g} R[g] \ .
\end{equation}
Regge thought this and proved it rigorously by putting the dots in what we just said. So, the mathematical result states that this expression $S_R(l)$, if the discrete geometry approximates the continuous geometry (in a proper sense), differs from the Einstein-Hilbert action for a small quantity within the approximation. Of course, this is just a sketch of the whole story. Therefore, given a manifold with a four-dimensional metric and once fixed an $\epsilon$ of our choice, we can always find a triangulation of it. After measuring the lengths of the links, we can compute $S_R(l)$, which is "$\epsilon$ close" to the Einstein-Hilbert action. The triangulation has to be, obviously, very refine if we want to take $\epsilon$ small, therefore in physical terms, in the limit in which we refine the triangulation to infinity (infinity tetrahedra infinitely small) \underline{$S_r(l)$ converges to the E-H action}. 
\begin{figure}[h] 
\begin{center} 
\includegraphics[width=10cm]{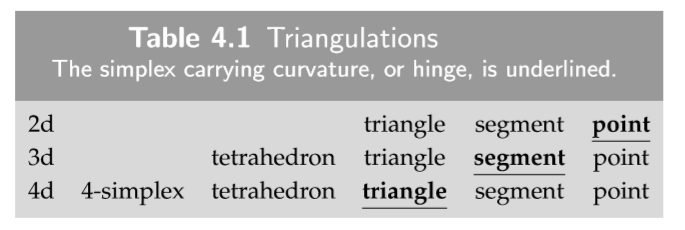}
\caption{\textit{Taken from page 88 of the book.} }
\end{center}
\end{figure}
This means that we can think of defining QG by taking a triangulation and integrating over all the metrics (namely all the $l$ variables) the exponential of the Regge action. The latter is defined for a finite triangulation, so we take the limit of the number of tetrahedra going to infinity:
\begin{equation}
Z = \lim_{N \rightarrow \infty} \int dl e^{\frac{i}{\hbar} S_R[l] } \ .
\end{equation}
We can define the theory in this manner, and we see clearly that there is no scaling parameter here for the same reason we saw during the previous lecture. The lengths of these links don't require to be taken to zero since we are integrating them. They are the (quantum) variables being considered. We are not fixing the size of the triangulation. We are just fixing several triangles and integrating all possible sizes. It is not a truncation in scale, but it is a \underline{truncation in the number of degrees of freedom} we are considering. The more degrees of freedom we choose to approximate the 4D region, the more precise it becomes the approximation. It is neither a UV nor IR cut-off since no sizes are involved. This is Regge Calculus. When we go to LQG, it is the same story: the 2-complex and the graph on the boundary involve a truncation of the number of degrees of freedom. The $j$'s can be arbitrarily large or small. They have a minimum value since they are inherently quantized. The $l$'s variables in Regge calculus are not independent, so we are integrating things whose domain of variation has a tremendously complicated shape because of the triangle inequality. There is a full literate on trying to do quantum Regge calculus, but it has never given many good results. Regge developed, together with Ponzano, a 3D version that is very much like LQG in three dimensions (with spins, intertwiners, etc.). In a sort of way, LQG can be seen as an attempt to do clear quantum Regge calculus in 4D. Let's say another thing on the Regge calculus role before going ahead. $S_R(l)$ can be written in a different way, by expressing $\delta$ as a sum over simplexes:
\begin{equation}
S_r(l) \sim \sum_{sim} \sum_{\Delta \in sim} A_{\Delta}(l) \theta_{\Delta}(l) \ .
\end{equation}
Therefore:
\begin{equation}
e^{\frac{i}{\hbar} S_R[l] } \sim \prod_{sim} e^{\frac{i}{\hbar} \sum_{\Delta}A_{\Delta} \theta_{\Delta}} \ .
\end{equation}
where $e^{\frac{i}{\hbar} \sum_{\Delta}A_{\Delta} \theta_{\Delta}}$ is the Regge action for a single 4-simplex. We remember that, in the spin network amplitude, we wrote that the total amplitude was a product of actions for an amplitude $A(j,i_n)$ concerning a single 4-simplex. Therefore, in some manner, the quantities $A(j,i_n)$ and $e^{\frac{i}{\hbar} \sum_{\Delta}A_{\Delta} \theta_{\Delta}}$ must be related. We have thus reduced a way of analyzing these amplitudes to a question about whether these vertex amplitudes are related to a single Regge action. This is the actual last thing we want to discuss. 
\subsection{Intrinsic coherent states}
There is an obstacle to overcome and a technique to develop, mainly due to Simone Speziale. Now we describe the problem and the solution (extensive literature discusses it in more detail). We return to the spin networks $| \Gamma, j_l, v_n \rangle $. These give geometrical information about tetrahedra in terms of the areas and one quantity (which can be the volume) diagonalized in the intertwiners space. But these states do not have a direct application for the following reason. If we take a wave function for a particle $\psi(x)$, the states $|x \rangle$ are, of course, states that diagonalize position, while $| p \rangle$ diagonalize the momentum. Still, if we diagonalize one set of states, then the others turn out to be spread. Do we have states in which we can represent a particle with a given position $x_0$ and a given momentum $p_0$, such that the product of the spreads $\Delta x_0 \Delta p_0$ is minimal? Yes, these are wave packets, and we know how to describe them. These are semiclassical states or wave packets coherent states. However, if we look at the spin network states, the geometry of a single tetrahedron is given by six numbers, and we can only determine five of them. Therefore, there is a quantity wholly spread. Can we write coherent states for the tetrahedron? Equivalently, can we write a linear combination of spin network states that gives me a fixed shape of the tetrahedron (regular or not), which minimizes the necessary spread in the various quantities?

\medskip

This is the first question. The answer is that this can be done with a nice and simple technique, but it took some time to figure it out. Let's be more precise. We want a state for the tetrahedron where, let's say, the \underline{four areas are sharp} but the intertwiner describes a given geometry of a tetrahedron as best as possible. The intertwiner is an element of the invariant part of the tensor product's space $I \in Inv \left( \mathcal{H}_{j_1} \otimes \mathcal{H}_{j_2} \otimes \mathcal{H}_{j_3} \otimes \mathcal{H}_{j_4} \right) $. The idea is to think later about the invariant part and write coherent states directly in the representation $\mathcal{H}_{j_1} \otimes \mathcal{H}_{j_2} \otimes \mathcal{H}_{j_3} \otimes \mathcal{H}_{j_4}$. Remember that the operators $\vec{L}$ live inside the individual Hilbert spaces and have a nice geometrical interpretation for them, namely the normal to the faces. If we express the operators $\vec{L}$ in the most semiclassical possible way (since they are three, we cannot diagonalize all of them at the same time), namely minimizing their spread within all the single Hilbert spaces, we can describe a geometry in which the four normals to the surfaces turn out to be semiclassical. Then, we can take the invariant part of the states. This last step means integrating all the rotations, which means considering these four directions to common rotations. Let's do it explicitly to see how this work. 

\medskip

We take a single Hilbert space $\mathcal{H}_j$, and we take a known basis of states, namely $| j,m \rangle$. Look at these states where the operators $\vec{L} = (L^1, L^2, L^3)$ act. These states are eigenstates of $L^3$ by definition since $m$ is a quantum number of the latter, but $L^1$ and $L^2$ are pretty much spread. If we think of spherical harmonics, we have strange shapes for a generic $m$, but we have just the sphere shape for the minimal value of $m$ (or the maximum value). Therefore, let's take the maximum value of $m$, that is, $m = j$. This corresponds to a spherical harmonic sitting on the top of the sphere. $L_z$ is sharp, and the other two variables are spread. We can compute that it is the minimal spread. So, we have at least one case in which we know a state where $L_z \equiv L^3$ has a fixed value while the others are minimally spread. We can interpret this as a semiclassical state with a fixed $z$ direction: this represents a tetrahedron with a normal to a face oriented along $z$. We don't want to be attached to this particular direction, so we want to turn it. We know how to do it since this is a representation of $SU(2)$, and rotations act on it. Thus, given any unit vector $\hat{n}$, we can always choose a rotation matrix $n$ that turns the $z$ direction into the latter (we can do this in various ways, we suppose to have chosen one particular method). Then we can rotate the state $|j,j \rangle$ from the vertical $z$ axes to the $\hat{n}$ direction by acting with a rotation matrix:
\begin{equation}
D^j(n) | j,j \rangle = | j, \hat{n} \rangle \ .
\end{equation}
\begin{figure}[h]   
\begin{center} 
\includegraphics[width=8cm]{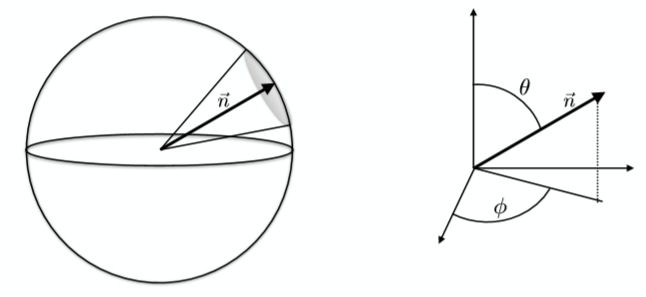}
\caption{For a generic direction  $n = (n_x,n_y,n_z)$. \textit{Fig. (8.2) of the book.} }
\end{center}
\end{figure}
The matrix corresponding to the rotation is explicitly given by:
\begin{equation}
\langle j,m | j, \hat{n} \rangle = D^j_{m,j}(n) \ .
\end{equation}
Now we have a state corresponding to a vector pointing in a direction. So we take four of them: 
$
|j_1 ,\hat{n}_1 \rangle \otimes |j_2 ,\hat{n}_2 \rangle \otimes |j_3 ,\hat{n}_3 \rangle \otimes |j_4 ,\hat{n}_4 \rangle 
$.
So we have four vectors, which means four areas. This is a state in $\mathcal{H}_{j_1} \otimes \mathcal{H}_{j_2} \otimes \mathcal{H}_{j_3} \otimes \mathcal{H}_{j_4}$, now we want to select the invariant part. To do this, we rotate it, and we integrate over all possible rotations: the result we get is certainly invariant under the common action of $SU(2)$, and it does not depend anymore on the actual four directions, but it depends on the four vectors up to a common rotation, namely on the angles between the directions.
\begin{equation}
\int_{SU(2)} dh |j_1 ,\hat{n}_1 \rangle \otimes |j_2 ,\hat{n}_2 \rangle \otimes |j_3 ,\hat{n}_3 \rangle \otimes |j_4 ,\hat{n}_4 \rangle = || j_1,j_2,j_3,j_4,\hat{n}_1,\hat{n}_2,\hat{n}_3,\hat{n}_4 \rangle \ .
\end{equation}
It does not depend on the directions but gives us information about the geometry. It depends on the four areas and the angles between them: this is the state corresponding to a given geometry of a tetrahedron. We can write this regarding the Wigner matrices since we have explicit formulas. Taking a given 4-simplex, we have five tetrahedra around it. If we fix the geometry of the 4-simplex, this determines the geometry of each tetrahedron around the latter and the area of each face. So we can write a state which has all the areas sharp and all the geometries we want just by putting in the five intertwiners space around the 4-simplex each one of the states $ || j_1,j_2,j_3,j_4,\hat{n}_1,\hat{n}_2,\hat{n}_3,\hat{n}_4 \rangle$. Therefore, we have a state $\psi$ depending on the geometry of the 4-simplex. We do not treat it explicitly, but it can be done by using the equations seen so far. This is a state on the spin network around the 4-simplex, so spinfoam amplitudes $A[ \psi ]$ are associated depending on the state, which turns out to be numbered. Has it something to do with the Regge amplitude, that is $e^{\frac{i}{\hbar} \sum_{\Delta} \theta_{\Delta}}$, in which the area of the triangles are the areas of the geometry on the 4-simplex? A theorem that Barret and his group proved states \underline{yes}: in the limit in which $j$ is large enough, there is a direct relation between the two (we will be more precise during the following lecture). This funny amplitude that seems to be defined just in terms of combinatorics calculus and groups in the large quantum number (i.e., classical) limit gives, at least locally and for one tetrahedron, the Regge action. The latter is the one that, in a suitable limit (namely, in which we refine the triangulation), gives GR. We are mentioning the sketch of what is the core of the theorem. We will be more precise regarding these coherent states in the next lecture. \textit{Rovelli answers some questions at the end of the lecture}.
\section{Extrinsic coherent states and classical limit}
\label{sec:Lecture_19_coherent_states}
We want to treat at least one application to see how the LQG works. We have no time to deal with all the details of the calculations required. Now we will be more precise concerning the previous lecture and the coherent states/Barret's theorem. Then we add some details to understand the practical application. We treat only sketches of Barret's result since it would require many mathematical notions. 

\medskip

During the last lecture, we said that, in the Hilbert space of LQG, a basis is provided by the spin network states $| \Gamma, j_l, v_n \rangle $. They diagonalize only some pieces of the 3D geometry; in particular, $j_l$ are the quantum numbers of the areas while $v_n$ are the ones for the volume. But, at a fixed volume, the angles between faces are spread. A one-parameter family of tetrahedra has all the same volume and areas but different shapes. We described a technique for writing coherent states once fixed in a given geometry. We describe a tetrahedron using eight numbers: two directions specifying the normal to each face for all four triangles, uniquely characterizing its geometry. The redundancy is given by the equivalence implied by three parameters of rotation which are supposed to define the same state. Because the sum of all the normal is zero:
\begin{equation}
\sum_{a=1}^{4} A_a \hat{n}_a = 0 \ .
\end{equation}  
Only three independent normal vectors give six numbers in total and three numbers up to the three rotation parameters. These states can be written as $| \Gamma, j_l, (\hat{n}_1,\hat{n}_2,\hat{n}_3,\hat{n}_4)_n \rangle $, where we have four unit vectors $\hat{n}$ for each node satisfying the above condition. There is an intertwiner $i^{m_1 m_2 m_3 m_4}_{\hat{n}_1 \hat{n}_2 \hat{n}_3 \hat{n}_4}$ determined by these four vectors, having four indices (each one living in a different representation $j$) and that is in $Inv \left( \mathcal{H}_{j_1} \otimes \mathcal{H}_{j_2} \otimes \mathcal{H}_{j_3} \otimes \mathcal{H}_{j_4} \right)$. To construct it, we take the state, rotate it to the direction we choose, and make everything invariant by integrating over $SU(2)$. Therefore, for each one of these vectors $\hat{n}_a$, we write it as a (rotation matrix depending upon a) group element $n_a$ acting on a unit vector $\hat{z}$ pointing in the $z$ direction:
\begin{equation}
R^i_{\hspace{1mm} j}(n_a)\hat{z}^j = \hat{n}^i_a \ .
\end{equation}
There is an arbitrariness in doing that since there are many possible rotations (we suppose to have fixed one). This partially reflects the arbitrariness in the definition of these coherent states, which require some technicalities. Finally, we integrate over $SU(2)$. We have just seen the \textbf{intrinsic coherent states} 
%
%
\subsection{Extrinsic coherent states}
These are not the full set of coherent states that we need to think to the geometry since there is also the \textbf{extrinsic coherent states}. Of course, there should be other things since the states $| \Gamma, j_l, v_n \rangle $ take into account the fact that the different components of $L^i_l$ don't commute with one another, so they are ways for writing states in which we cannot make all of them sharp, but states where the spread of the $\vec{L}_l$ is minimal. But the $\vec{L}_l$ operators are just half of the variables because there are also the $h_l$ operators. $\vec{L}_l$ are the left-invariant vector fields, while $h_l$ are the diagonal operators of a particular representation. The one in which we diagonalize part of $\vec{L}_l$, it turns out that the $h_l$ operators act in complicated manners. Therefore, the relation between $\vec{L}_l$ and $h_l$ is very much like $p$ and $x$. Can we make coherent states in which both are sharp or concentrated? These are precisely the extrinsic coherent states. There is a technique to construct them, but before examining it, we want to discuss the operators $h_l$ since we have still not treated them. We said they are associated with the links, but we have been pretty vague about the connection and the role they play between tetrahedra. We are treating this role at the end of the series of lectures, even if this connection was the story's beginning from a historical point of view. To which geometrical quantity does it correspond in GR? One is tempted to say that it is just a 3D rotation, but this is impossible since the spin connection, which knows about the 2D connection, is a function of the intrinsic geometry. It has to do with time derivatives of the three geometry, therefore with the \textbf{extrinsic geometry}: this is the reason for the name of these states. Let's be more precisely going back to classical GR that is based on the action:
\begin{equation}
S = \frac{1}{16 \pi G} \int \left( e \wedge e \right) \wedge \left( F^* + \frac{1}{\gamma}F \right) \ .
\end{equation}
The operator $\vec{L}_i$ is what describes the geometry of a 3D surface and we saw that the latter was given by $L^i = \frac{1}{2} \int_{\Delta} \epsilon^i_{\hspace{1mm}jk} e^j \wedge e^k $. When it is on a link, it consists of an integration of a 2-form over a triangle, which turns out to be dual to the link. Thus this is its space-space component, and it turns out to be the first term of the action. Its conjugate variable, which we now want to compute, is given by the derivative of the Lagrangian with respect to its time derivative. Remembering that $F = d \omega + \omega \wedge \omega$, we can forget about the last term since it does not contributes to the calculus we are interested in:
$$
S \longrightarrow \int \left( e \wedge e \right) \wedge \left(d \omega * + \frac{1}{\gamma} d \omega \right) = - \int d \left( e \wedge e \right) \wedge \left( \omega * + \frac{1}{\gamma} \omega \right)  \ .
$$
where we integrated by parts in the second passage. By changing the notation, we can rewrite the previous expression as:
\begin{equation}
- \int d_{\mu}e_{\nu}^I e_{\rho}^J \left( \omega_{\sigma}^{KL} \epsilon_{IJKL} + \frac{1}{\gamma}\omega_{\sigma}^{IJ} \right) \epsilon^{\mu \nu \rho \sigma} \ .
\end{equation} 
\textit{The indices within the parentheses should be put down, and the expression $e_{\nu}^I e_{\rho}^J$ corresponds to $\left( e \wedge e \right)^{IJ}$ in terms of flat indices}. The time derivative occurs where, in the sum, the index $\mu$ is equal to zero, while all the remaining part contains no time derivative at all. If we put $\mu = 0$, then (because of the Levi-Civita tensor) $\nu,\rho,\sigma$ must have space indices:
$$
- \int d_{0}e_{a}^I e_{b}^J \left( \omega_{c}^{KL} \epsilon_{IJKL} + \frac{1}{\gamma}\omega_{c}^{IJ} \right) \epsilon^{0 abc} \ .
$$
Remembering that:
$$
L^i = \frac{1}{2} \int_{\Delta} \epsilon^i_{\hspace{1mm}jk} e^j_a \wedge e^k_b dx^a dx^b \ ,
$$
we see that within the term $ d_{0}e_{a}^I e_{b}^J$, there is the time derivative of $L^i$, so the conjugate turns out to be precisely the term within parentheses. When we are in the time gauge, $e^0 = 0$ since the latter has the expression \eqref{Time gauge tetrad Lecture 14}, therefore $e_{a}^I e_{b}^J \rightarrow e_{a}^i e_{b}^j$ and $\frac{1}{\gamma}\omega_{c}^{IJ} \rightarrow \frac{1}{\gamma}\omega_{c}^{ij}$ but $\omega_{c}^{KL} \epsilon_{IJKL} \rightarrow \omega_{c}^{KL} \epsilon_{ijKL}$. In the latter expression, one of the indices $K,L$ must be zero and the other has to be $k$:  $ \omega_{c}^{0k} \epsilon_{ij0k} \equiv \omega_{c}^{0k} \epsilon_{ijk}$. So, the conjugate variable to the term $e_{a}^I e_{b}^J$ turns out to be a sum of two pieces, where we have two terms of the connection. The term $\omega_{c}^{ij}$ is the connection in the space direction with space-space indices. It is just the three-dimensional spin connection on the surface itself, which is the solution of the equations of motion determined by the standard equation that gives the spin connection: $de + \omega \wedge e = 0$. The term $\omega_{c}^{0k}$ has a time component and nothing to do with the three-dimensional geometry. It gives us information about the 4D geometry since it turns out to be a piece of the 4D connection. Therefore, the conjugate momentum $E^i$ to $L^i$ results in the sum of the 3D spin connection and the \textbf{Ashtekar-Barbero connection} $A^i$, which contains information on the extrinsic curvature:
\begin{equation}
E^i = \frac{1}{\gamma}\omega^i + \frac{1}{\gamma} A^i \ ,
\end{equation}
where the Ashtekar-Barbero connection is:
\begin{equation}
A^i \equiv \gamma \omega^{0i} \ .
\end{equation}
The entire game of LQG started about 35 years ago when Ashtekar realized that, by using $A^i$ as a variable and $E^i$ as its conjugate momentum, the theory simplifies a lot with respect to the Einstein formulation. It takes a form that is very much like Yang-Mills theories so that we can do several stuff on that, and these procedures led to LQG as we now know it. The point is that the group element $h_l$ is not the holonomy of the 3D spin connection. Still, it turns out to be the holonomy of the Ashtekar connection, which knows about the time derivative of the 3D geometry. 

\medskip

\begin{figure}[h]   
\centering  
\includegraphics[width=12cm]{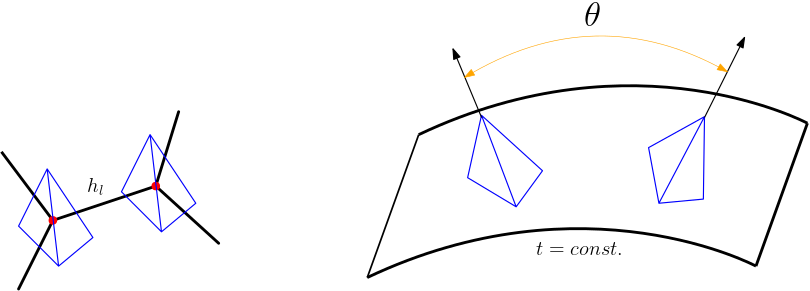}
\caption{\textit{Geometrical interpretation. On the left, we have the 3D spin network, where of course, the tetrahedra should be attached by a common triangular face. On the right, it is reported as the same picture in 4D. }}
\label{ExtCurv}
\end{figure}

Where is the extrinsic curvature? Let's suppose we have a spin network, and let's consider the links. \textit{The geometrical interpretation of all the following discussion can be visualized in Fig. (\ref{ExtCurv})}.
These links essentially connect the centers of tetrahedra. We can always choose to orient the reference frames of two tetrahedra (connected by the same link) so that there is no 3D rotation between them: we just parallel transport our reference frame from one to the other. Until we make a tour (namely, considering the curvature), we can always choose $\omega^i = 0$. Up to gauge transformations on the two nodes, we can (gauge) transform them to render the spin connection part equal to zero, but, of course, there is still the Ashtekar connection. What does it represent? It represents a part of the four-dimensional connection between the two nodes. The 4D angle $\theta$ between the normals to the tetrahedra is the \underline{discretized version of the extrinsic curvature}. It is represented by a 4D rotation, namely the exponential of the boost $\omega^{0i}$. This is the physical information contained in $h_l$, which is not disentangled from the information concerning the intrinsic geometry. If we want to reconstruct the geometry from the mathematical data, it is a bit tricky considering everything. To summarize, we have just seen that the conjugate variable is essentially the 4D angle $\theta$. 

\medskip

If we now think about the Regge action, the latter can be written as a sum over the 4-simplices of the action of a single 4-simplex:
\begin{equation}
S_R = \sum_{sim}S_{4-S} \ ,
\end{equation}
where the single action is a function of the segments $l$ of the triangulation, and it can be written as a sum over the triangles:
\begin{equation}
S_{4-S}(l) = \sum_{\Delta} A_{\Delta}(l) \theta (l) \ .
\label{Azione simplex Lecture 19}
\end{equation}
There is a tetrahedron in the boundary of each vertex, which are all connected by a common triangle face along the (dual) links associated with the common faces themselves.
\begin{figure}[h]  
\begin{center} 
\includegraphics[width=10cm]{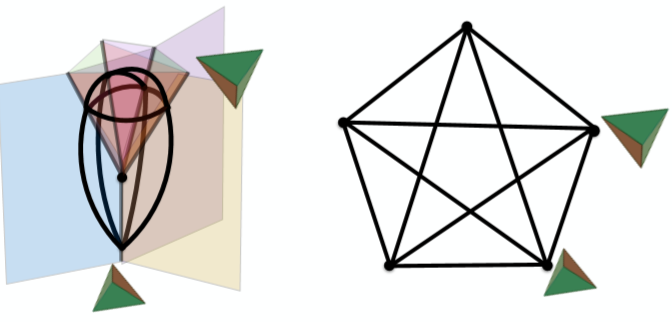}
\caption{Two tetrahedra in the boundary of a vertex are connected by a face. \textit{Fig. (8.6) of the book.} }
\end{center}
\end{figure}
The geometry is fully determined by the ten numbers $l$. In contrast, the Regge action corresponds to the area of the triangle multiplied by the 4D angle (this is what we have roughly seen during the previous lecture) between the tetrahedra in the vertices, which are captured by the variables $h_l$. These are the conjugate variables to the intrinsic geometry, whose main variable is the area. So, the structure of the Regge action turns out to be a sum over a variable and its dual variable, expressed in terms of functions of geometry: $S \sim \sum q \cdot p(q)$. If we know the ten areas of a 4-simplex, then we can fully reconstruct its geometry, which is specified by ten numbers. Therefore we can view the ten numbers $A_{\Delta}(l)$ as giving us the intrinsic geometry of the boundary and the number $\theta(l)$ providing us the conjugate variable. This is the general form of the action. \textit{Rovelli returns to the action of a free particle, illustrating that the latter can also be written using this structure. See from minute 29:28 to 33:26 for details}. In action \eqref{Azione simplex Lecture 19}, we are saying that the structure of the Hamilton function of GR, in a minimal piece of spacetime (namely a 4-simplex), is just the intrinsic geometry times the extrinsic one, as a function of the extrinsic geometry. The key is $p(q)$ as a function of $p(q_i,q_f)$. Therefore, this term contains information about dynamics. So, the Hamilton function of GR is a function of the \underline{$p$'s determined by the $q$'s}: this is what gives the dynamics. The Hamilton function \eqref{Azione simplex Lecture 19} is not a trivial expression since the angles $\theta(l)$ are expressed as a function of the areas determined by the $l$'s (or the $l$'s determined by the areas). 

\medskip

We are going to write the extrinsic coherent states explicitly. They are states giving the mean value, as like the coherent states: $\langle x|x_0, p_0 \rangle = e^{\frac{(x-x_0)^2}{2 \sigma} + p_0 \sigma}$, that are labeled by the classical quantity on which the state is build. Now we are in the condition of building states labeled by classical 3D geometry of the triangulation and classical extrinsic geometry. These are gonna be summed over spin network states with some coefficients, ultimately determined by the intrinsic and extrinsic geometry themselves:
\begin{equation}
|\text{class. 3-geometry, class.  extrinsic geom } \rangle \sim \sum_{j,v_l} |j_l, v_n \rangle C(j_l,v_l, \text{geom.}) \ .
\end{equation}
There is a construction providing these states explicitly. There is a sort of Gaussian curve around $j$ and another similar curve around the geometry, which approximates, at best it. As we can imagine, to build a quantum state (the wave packet) following the motion of a classical particle, we can take a surface in curved spacetime, triangulate it and explicitly write a quantum state that is picked around it, both in terms of intrinsic and extrinsic geometry. There are formulas to do this, but we will not write them (anyway, there is extensive literature on this topic). 
\subsection{Barret's result}
An important result obtained by Barret and his collaborators, which we have already mentioned, is the following (it is written in terms of intrinsic states). We consider a 4-simplex. We can fix the geometry (this means giving ten areas), so we write the amplitude of the intrinsic coherent states $A(j_{ab}, n_{ab})$ given by the ten areas of the tetrahedra around the 4-simplex. Let's choose an intrinsic coherent state with an arbitrary geometry for each of the five tetrahedra, specifying all the normal vectors. If we fix the areas of each tetrahedron's face around the 4-simplex, we are fixing the $n$'s as a function of the $j$'s up to rotation: $n_{ab}(j_{ab})$. If we fix the geometry, we can compute the angles $\theta$ by just elementary 4D geometry, so there are functions $\theta_{ab}(j_{ab})$. Barret's theorem states that $A(j_{ab}, n_{ab})$ is exponentially suppressed in the large $j$ limit unless the $n_{ab}$ are of the form $n_{ab}(j_{ab})$, up to rotation of all of them. So, quantum dynamics suppresses all the configurations which are not "regular." If the $n_{ab}$ are correct, the amplitude is proportional to:
\begin{equation}
A(j_{ab}, n_{ab}) \sim e^{i \sum_{}j_{ab}\theta_{ab}(j_{ab})} + \text{complex conjugate}  \ .
\end{equation}
It requires a delicate analysis, going into details of the $SL(2,\mathbb{C})$ representations. It is pretty remarkable since it seems that the theory knows about this geometry entirely, and, in turn, this geometry knows about GR.

\medskip

Now we have all the ingredients of the theory. It sounds like an abstract construction that is far away from reality. Now we use this theory to describe a gravitational phenomenon in the Universe that has to do with what happens inside a black hole, and it can be treated only with QG. \textit{Rovelli answers some questions at the end of the lecture for approximately 15 minutes}.
\section{Application: the end of black holes}
\label{sec:Lecture_20_black_holes}
We want to render this construction a little bit more concrete to show how, in a concrete application, it is possible to make some physics. Theories are good if they can be used to compute things and processes in the Universe. This is not still, however, a fully complete application, and there is still work in progress on that. We don't know, for example, if black holes continue to evolve after their "end," and there are several aspects that turn out to be still not clear yet. As we said, we now use this theory concretely. Otherwise, these remain just "juggling ideas" unless we go to some actual physics. So, where are the cases we can use the theory? Well, the ones in which quantum gravitational phenomena are not negligible. Two typical examples are the early Universe and black holes. There would be other cases, but these are the only two in which QG is manifestly required to describe what is happening. There is a vast amount of literature on applying LQG in Cosmology, both in this form and (much more) in canonical formalism. In the latter case, the kinematics is precisely the one we have seen. Still, the dynamics are described in terms of the Hamiltonian and the Wheeler-DeWitt equations rather than amplitude. There is also a vast literature on black holes from various perspectives. Using this machinery, we want to focus on one particular calculation to understand what happens inside them, i.e., in the region where we don't know what happens.
\subsection{Black Holes}
\begin{figure}[tb] 
\centering 
\includegraphics[width=4cm]{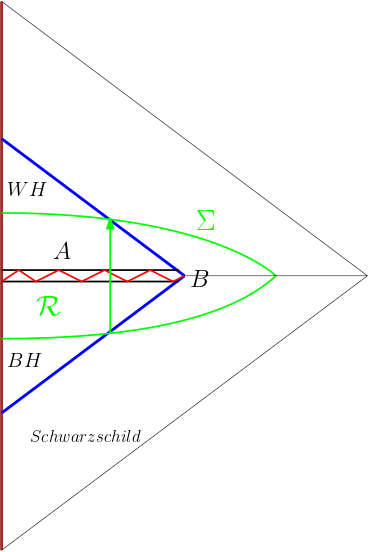}
\caption{\textit{Geometrical representation of the regions and surfaces involved, resulting from an extension of the Penrose diagram. The green arrow represents the quantum tunneling transition between the two classical states, defined on the boundary of the 3D region $\Sigma$, through the singularity of the black hole.} }
\label{BH2}
\end{figure}

Of course, we know several things about black holes since they seem perfectly described by GR as far as we can measure them, including a strong field regime. So, the horizon of a black hole turns out to be pretty well described by GR, and we have no reason to suspect that, when we go inside, GR still does not continue to describe it correctly. But, then, we get lost in the center of the black hole: we have no idea what happens for $r = 0$. The distant future of a black hole is still quite a mystery. Hawking made a very credible calculation and was repeated by other scientists in several ways, showing that the black holes shrink because of the back reaction of the Hawking radiation (a negative energy flux that falls in). The black hole should become smaller and smaller, but after this phase, nothing is still known: we need QG. 
\begin{figure}[tb]  
\centering 
\includegraphics[width=7cm]{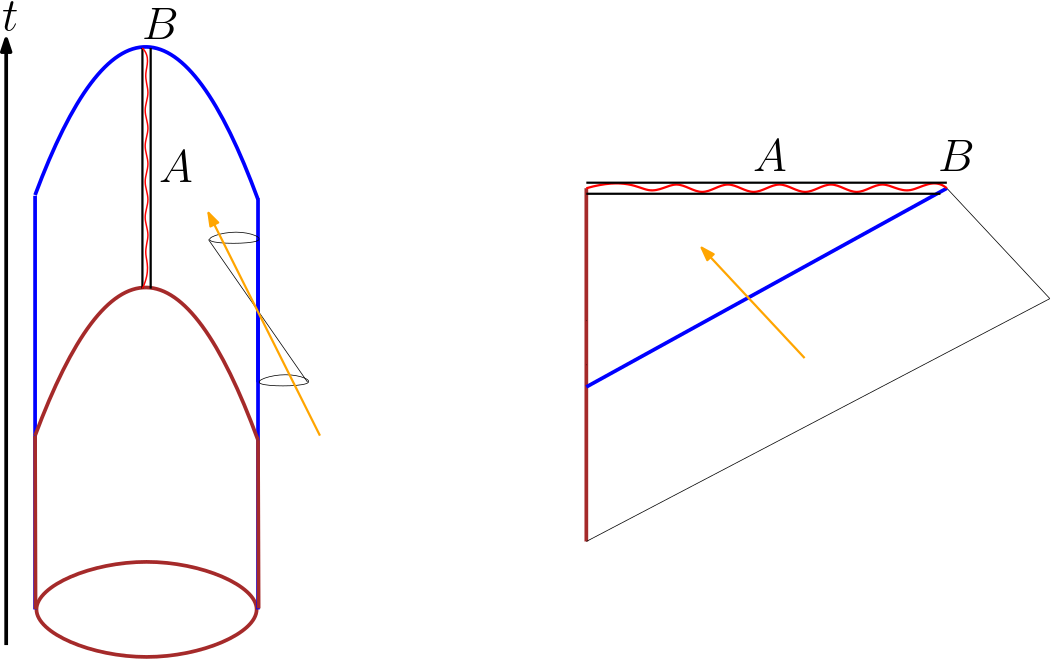}
\caption{\textit{Graphical interpretation. The blue lines denote the gravitational horizon, the brown corresponds to the surface of the collapsing star, and the orange arrow represents an ingoing photon. $B$ is the future of the black hole after the evaporation, whereas $A$ is the region around the singularity center. On the right, the Penrose spacetime diagram of the gravitational collapse is sketched.} }
\label{BH1}
\end{figure}
The questions are what happens in the region $A$ and $B$. \textit{I represented them in Fig. (\ref{BH1}), as the other surfaces involved}. We understand the physics around the black hole very well, but not in these two regions. 
Since we know nothing about what happens in $A$ and $B$, we should ignore the latter in the graphs we are used to. Whatever happens in the future, it is reasonable to expect that in a distant forward time, at some point, there is just a regular spacetime after the complete evaporation of the black hole. Therefore, another way to view the gravitational collapse is to think this is just the story's beginning. So, there is an end that turns out to be still a spacetime with a sort of standard causal structure, namely the \underline{quantum story}.
We know from classical GR that the collapsed star creates a horizon. The "cosmological censor" conjecture states that every time we have a singularity in classical GR, the latter is always "hidden" inside a horizon. It seems to be accurate up to, perhaps, some extraordinary cases. We suppose for a moment that this conjecture is true. Since GR is invariant under time reversal, the opposite is also true. If we look back from the future to go to the putative singular quantum gravity region, we are also "closed" by a horizon. We have a classical metric with a quantum transition after a certain point. We want to understand what happens. We need two things, namely, a classical description of whatever is around and a quantum one that tells us what happens inside. We notice that we need precisely what LQG gives us: we have a classical geometry closed inside a surface that can be arbitrarily chosen.
In a quantum system, we can select the boundary we want since the probability will be the same, so we can always move our boundary out. There is this freedom. Therefore we can close the region $A$ inside a boundary, namely a space-like surface with classical geometry. We need some theory that tells us how we go from two classical geometries in terms of probability amplitude, and this is precisely what LQG provides us. To study this phenomenon, we first check the metric outside this 3D space-like surface $\Sigma_+$ and then compute the metric $\Sigma_-$ inside it. We can use LQG to calculate the amplitude, and to do that (since LQG is formulated in terms of truncation), we have to truncate the theory. There is a tunneling effect between these two regions since there is no classical transition between them: the theory dies in the singularity.
Thus there is no possible classical evolution. The quantum theory gives us this tunneling, and presumably, the degrees of freedom of the latter are not arbitrarily small but are comparable to the size of the black hole (or the size of the "phenomena" we are considering). So, we want to truncate the theory by discretizing $\Sigma$ and the interior region $\mathcal{R}$ and, then, using the quantum theory. On $\Sigma$, there is a 3D metric, while $\mathcal{R}$ is the 4D region inside. Suppose we know the geometry of $\Sigma$, which means understanding the intrinsic and the extrinsic ones. In that case, we have the geometry of the discretization of $\Sigma$, and we can write an irrelevant coherent state in the Hilbert space of LQG to the truncation we have fixed. So, we write a coherent state with this geometry, and once we have that, we can compute the amplitude.
\textit{See Fig. (\ref{BH2}) for a graphical interpretation of the above discussion}.
Summarizing, we can write the following scheme:
\begin{itemize} 
\item \textbf{Geometry outside} \\
Let's start from the first point, namely the geometry outside. It can be shown that the final metric outside depends only on two parameters: the mass of the Schwarzschild black hole $m$ and the time $T$ that flows between the lower and upper regions. There are several ways to characterize this time. For example, we can choose when the horizon forms and the transition happens. The external metric can be explicitly given as a function $dS(m, T)$. The quantum theory provides the time $T$ for an object of mass $m$ to go to this transition.
\begin{figure}[tb] 
\centering  
\includegraphics[width=7cm]{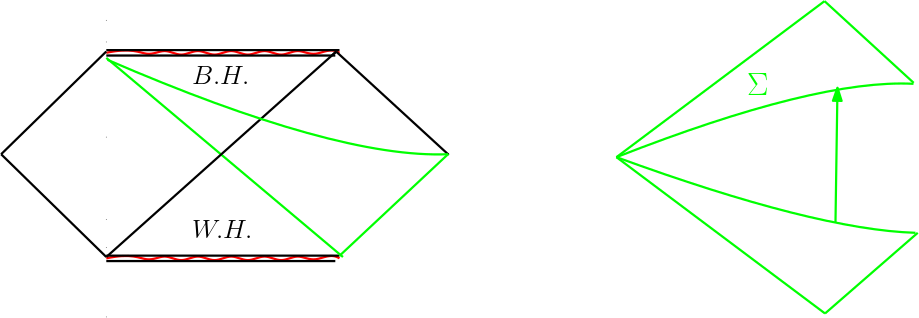}
\caption{\textit{Graphical interpretation. To construct the surface $\Sigma$, we cut the Kruskal diagram (left side). Then, we take the time reversal of the region thus obtained, computing the transition amplitude between the two (right side).} }
\label{BH4}
\end{figure}
\item \textbf{Intrinsic and extrinsic geometry of $\Sigma$ } \\
Now we have these geometries explicitly as a function of $m$ and $T$. This means we will compute an amplitude that depends on the same parameters.
\item \textbf{Discretization of $\Sigma$ and $\mathcal{R}$ } \\
Let's look at the shape of this surface. To discretize the two spheres $S_2$ attached in a point $P$ of the boundary, we could roughly choose the interior of a tetrahedron, but this would not be refined enough. Therefore, we choose four tetrahedra to triangulate the two-sphere(s). We have two kinds of these 3D figures attached along one face. The 4-valent dual graph of this triangulation is straightforward, and we call it $\Gamma$. \textit{This procedure can be visualized in Fig. (\ref{BH5})}. This is, eventually, the discretization of $\Sigma$ in spin networks. 
\begin{figure}[tb]  
\centering  
\includegraphics[width=8cm]{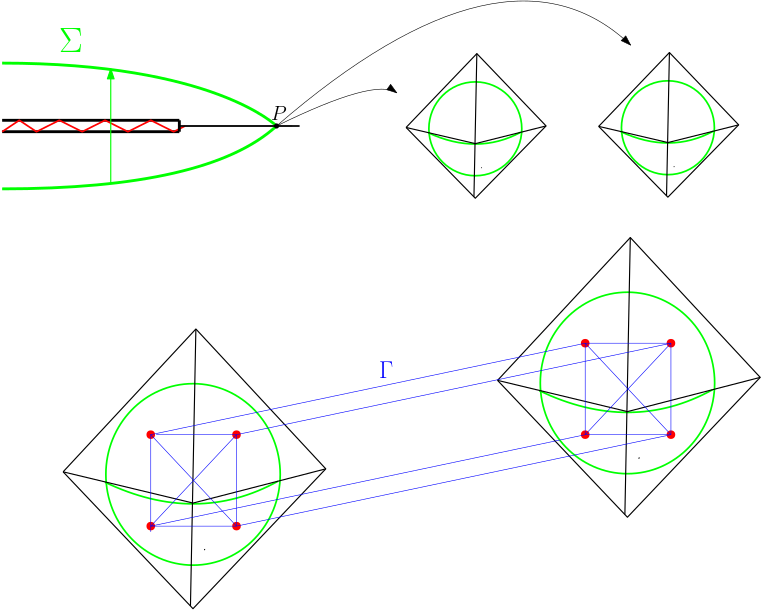}
\caption{\textit{Up I reported the two-spheres (there is one of them for each point of $\Sigma$), whose inside is triangulated in terms of four tetrahedra, joining at the point $P$ of the boundary. Down it is possible to visualize the resulting dual graph $\Gamma$. Of course, both the two $S_2$ spheres (as the four-plus four tetrahedra) should be attached.} }
\label{BH5}
\end{figure}
\item \textbf{Extrinsic coherent state } \\
We now have a discrete metric on the latter, and since we know sizes and angles, we can write down a coherent state corresponding to the metric we get from the outside. After much work, it is possible to write a state $\psi(m, T)$ in the Hilbert space of the graph, depending on the geometry.
\item \textbf{Amplitude } \\
As mentioned above, we obtain an amplitude of $W(m, T)$. Once we have a coherent state, we can compute the amplitude itself. First, we must put a 2-complex in between (a spinfoam) and use the formulas. \textit{I directly report the Fig. (\ref{figpaper}), taken from the original paper in which this calculation is performed}. Then, we choose the simplest 2-complex possible, namely a triangulation of the inside. The simplest triangulation consists of breaking the region into two 4-simplices, which are joined by a tetrahedron, and four tetrahedra enter both. So, after many stuff and technicalities, we obtain an (approximated) function of two variables.
\begin{figure}[tb] 
\centering        
\includegraphics[width=7cm]{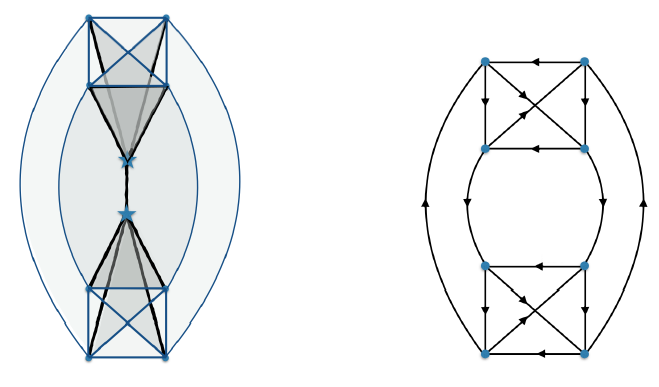}
\caption{\textit{Taken from the paper "Characteristic Time Scales for the Geometry Transition of a Black Hole to a White Hole from Spinfoams," written by Marios Christodoulou and Fabio D'Ambrosio.} }
\label{figpaper}
\end{figure}
\end{itemize}
Fabio D'Ambrosio and Mario Christodoulou have explicitly done this calculation in several papers. The square modulus of $W$ is the probability of this transition happening, for a mass $m$, in a specific time $T$. The final result turns out to be:
\begin{equation}
|W(m,T)| \sim e^{\frac{m^2}{m_P}} \ .
\end{equation}
It is very credible since this probability is suppressed for a macroscopic black hole. But, for a small black hole, this result states the opposite! Interestingly, according to this theory, there is a white hole after the evaporation of the (small) black hole, which remains a remnant. 
\end{document}